\documentclass[aps,prd,superscriptaddress,showpacs,floatfix,letterpaper,preprintnumbers,nofootinbib]{revtex4}
\usepackage{epsfig}
\usepackage{bm}
\usepackage{amsmath}
\usepackage{amstext}
\usepackage{amssymb}
\usepackage{mathrsfs}
\usepackage{color}
\usepackage{subfigure}
\usepackage{graphicx}

\begin{document}

\title{Exploring the influence of bulk viscosity of QCD on dilepton tomography}
\author{Gojko Vujanovic}
\affiliation{Department of Physics, The Ohio State University, 191 West Woodruff Avenue, Columbus, Ohio 43210, USA}
\affiliation{Department of Physics and Astronomy, Wayne State University, 666 W. Hancock St., Detroit, MI 48201, USA}
\author{Jean-Fran\c cois Paquet}
\affiliation{Department of Physics, Duke University, Durham, NC 27708, USA}
\author{Chun Shen}
\affiliation{Department of Physics and Astronomy, Wayne State University, 666 W. Hancock St., Detroit, MI 48201, USA}
\affiliation{RIKEN BNL Research Center, Brookhaven National Laboratory, Upton, NY 11973, USA}
\author{Gabriel S. Denicol}
\affiliation{Instituto de F\'{i}sica, Universidade Federal Fluminense, UFF, Niter\'{o}i, 24210-346, RJ, Brazil}
\author{Sangyong Jeon}
\affiliation{Department of Physics, McGill University, 3600 University Street, Montr\'eal, QC, H3A 2T8, Canada}
\author{Charles Gale}
\affiliation{Department of Physics, McGill University, 3600 University Street, Montr\'eal, QC, H3A 2T8, Canada}
\author{Ulrich Heinz}
\affiliation{Department of Physics, The Ohio State University, 191 West Woodruff Avenue, Columbus, Ohio 43210, USA}

\begin{abstract}
The collective behavior of hadrons and of electromagnetic radiation in heavy-ion collisions has been widely used to study the properties of the the Quark-Gluon Plasma (QGP). Indeed this collectivity, as measured by anisotropic flow coefficients, can be used to constrain the transport properties of QGP. The goal of this contribution is to investigate the influence of the specific bulk viscosity ($\zeta/s$) on dilepton production, both at Relativistic Heavy-Ion Collider (RHIC) and Large Hadron Collider (LHC) energies. We explore the sensitivity of dileptons to dynamical features that bulk viscosity induces on the evolution of a strongly-interacting medium, and highlight what makes them a valuable probe in the pursuit to also constrain $\zeta/s$.   
\end{abstract}

\maketitle
\date{\today }
\section{Introduction}

The Quark-Gluon Plasma (QGP), an intriguing state of matter composed of colored quarks and gluon degrees of freedom, is believed to have filled the universe during the first few microseconds after the Big Bang. The goal of relativistic heavy-ion collisions is to re-create and study this system of deconfined partons in the laboratory, and to analyze its properties. A fundamental property of the QGP that is vigorously sought after is its viscosity, both the bulk and shear. 

The effects of bulk viscosity ($\zeta$) alone on the elliptic flow of hadronic observables has been studied before \cite{Denicol:2009am}, which was soon after improved by simulations that include both bulk and shear ($\eta$) viscosity effects \cite{Monnai:2009ad,Song:2009rh}. These pioneering papers have highlighted some of the important features of bulk viscosity, in terms of how they affect the evolution of the hydrodynamical medium and how the distribution function of hadrons on the freeze-out hypersurface gets modified by the presence of bulk viscosity. The role of bulk and shear viscosity was further investigated through a different hadronic observable, namely  Hanbury-Brown-Twiss (HBT) correlation radii, in Ref.~\cite{Bozek:2011ua}, where it was shown that in the presence of both viscosities an improved description of the HBT radii is possible. 

Those earlier simulations however were based on Israel-Stewart hydrodynamics \cite{Israel1976310,Israel:1979wp} which did not include bulk-to-shear coupling in the equations of motion for the bulk viscous pressure and the shear stress tensor. The mathematical form of this coupling was fully derived from a microscopic theory, namely from the Boltzmann equation, in Ref.~\cite{Denicol:2012cn}, while its importance was studied in Ref.~\cite{Denicol:2014mca}. We are including bulk-to-shear couplings in our hydrodynamical equations of motion. More recent hybrid calculations composed of hydrodynamical simulation of the QGP followed by the hadronic transport evolution show that bulk viscosity is a key ingredient in improving the description of hadronic observables \cite{Ryu:2015vwa,Ryu:2017qzn}. 
 
Using the hydrodynamical simulations of Ref.~\cite{Ryu:2015vwa}, a direct photon calculation has also been carried out \cite{Paquet:2015lta}, showing that bulk viscosity in the hydrodynamical simulation affects the direct photon production at Relativistic Heavy-Ion Collider (RHIC) and Large Hadron Collider (LHC) energies. The present contribution extends the earlier work of Refs.~\cite{Ryu:2015vwa,Paquet:2015lta,Ryu:2017qzn} by looking at lepton pair (dilepton) production from the same hydrodynamical simulations. 

The dilepton invariant mass $M$, or the center of mass energy of the lepton pairs, provides them with an advantage over photons whereby low invariant mass dileptons ($M\lesssim 1.1$ GeV) are dominated by light flavor hadronic contributions (composed of $u$, $d$ and $s$ quarks), while partonic sources of lepton pair radiation become more important as the invariant mass increases. Specifically in the low invariant mass region, there are two contributions to dileptons: one from the in-medium decay of vector mesons and the other from late hadronic direct and Dalitz decays. At intermediate invariant masses ($1.1\lesssim M\lesssim 2.5$ GeV), the two main dilepton sources are direct QGP emission as well as semi-leptonic decays of open heavy flavor hadrons. Furthermore, the invariant mass degree of freedom has already been used to show that dileptons are sensitive to various aspects of the shear viscous pressure \cite{Vujanovic:2016anq,Vujanovic:2017psb}, and the present study continues in that direction by considering the effects of bulk viscous pressure on dileptons.   

Isolating the thermal contribution of dilepton production is not an easy task, but can be done, as was shown in the case of dimuons by the NA60 Collaboration at the Super Proton Synchrotron (SPS) at CERN \cite{Arnaldi:2008er,Arnaldi:2008fw,Damjanovic:2008ta}. At RHIC, the STAR Collaboration has recently acquired new dimuon data using their Muon Telescope Detector (MTD) and Heavy Flavor Tracker (HFT) simultaneously \cite{Geurts:2016}. Having both MTD and the HFT running at the same time is extremely useful as it permits to investigate the dilepton radiation coming directly from thermal radiation $1.1\lesssim M\lesssim 2.5$ GeV, after successful removal of the open heavy-flavor contribution in the region $M\lesssim2.5$ GeV. Even in the case that the open heavy flavor hadrons are not removed, one can theoretically describe their interaction with the medium and decay them semi-leptonically, allowing comparisons with experimental dilepton results (see e.g. \cite{Rapp:2009yu,Linnyk:2012pu,Vujanovic:2013jpa,Song:2018xca} and references therein). At the LHC, the upgraded ALICE experiment is planning to measure dilepton yield and anisotropic flow, with $\sim$10\% precision, for the High Luminosity LHC run after 2020 \cite{Citron:2018lsq}. Our work here assumes that the open heavy-flavor contribution is removed and considers dilepton radiation coming from QGP and light hadronic sources in the low and intermediate mass region.

This paper is organized as follows: Section \ref{sec:hydro} provides details about the dynamical modeling of the strongly-interacting medium, with an emphasis on the fluid-dynamical equations of motion used, Section \ref{sec:dilepton_prod} gives the details of dilepton production originating from both partonic and hadronic sources. Section \ref{sec:results} presents results about how bulk viscosity influences dilepton yield and anisotropic flow, while Section \ref{sec:conclusion} provides concluding remarks.

\section{Dynamical simulation}\label{sec:hydro}

The hydrodynamical equations of motion are based on the conservation laws for energy-momentum. In the fluid-dynamical model of ultrarelativistic heavy-ion collisions, one is usually probing the low longitudinal momentum fraction ($x$) region of the nuclear parton distribution functions, which are gluon-dominated \cite{Kovarik:2015cma}. Therefore, the resulting hydrodynamical equations can neglect the conservation of net quark-flavor induced charges --- i.e. net baryon number, electric charge and strangeness --- near mid-rapidity \cite{Gale:2018vuh}. Furthermore, the fact that the gluon distribution is highly populated in the low-$x$ region allows for the approximation that these gluon degrees of freedom can be dynamically evolved using classical Yang-Mills equations, which is precisely how IP-Glasma models its early-time dynamical evolution \cite{Schenke:2012wb}. Throughout this study, IP-Glasma will be used as a pre-hydrodynamical model of the strongly interacting medium, dynamically evolving the system for an assumed $\tau_0=0.4$ fm/$c$ (see \cite{Ryu:2017qzn} and references therein for details), following which we start the hydrodynamical simulation.

The hydrodynamical equations of motion include dissipation, which is described by six dissipative degrees of freedom, $\Pi$ and $\pi^{\mu\nu}$, accounting for bulk and shear viscous effects, respectively. The energy-momentum conservation equation reads:

\begin{eqnarray}
\partial_\mu T^{\mu\nu}&=&0,\nonumber\\
T^{\mu\nu}&=&T^{\mu\nu}_0+\delta T^{\mu\nu}_\Pi+\delta T^{\mu\nu}_\pi,\nonumber\\
T^{\mu\nu}_0&=&\epsilon u^\mu u^\nu - P\Delta^{\mu\nu}, \,\, \delta T^{\mu\nu}_\Pi=-\Pi\Delta^{\mu\nu}, \,\, \delta T^{\mu\nu}_\pi=\pi^{\mu\nu}
\label{eq:energy-momentum_conserv}
\end{eqnarray}  
where $\epsilon$ is the energy density, $u^\mu$ is the flow four velocity, $P$ is the thermodynamic pressure related to $\epsilon$ by the equation of state $P(\epsilon)$ \cite{Huovinen:2009yb}, $\Delta^{\mu\nu}=g^{\mu\nu}-u^\mu u^\nu$ projects on the spatial directions in the local fluid rest frame, and $g^{\mu\nu}={\rm diag}(1,-1,-1,-1)$ is the metric tensor. The dissipative degrees of freedom satisfy relaxation-type equations:
\begin{eqnarray}
\tau_\Pi \dot{\Pi}+\Pi &=& -\zeta\theta - \delta_{\Pi\Pi}\Pi\theta + \lambda_{\Pi\pi}\pi^{\alpha\beta}\sigma_{\alpha\beta}, \label{eq:bulk_relax}\\
\tau_\pi \dot{\pi}^{\langle\mu\nu\rangle}+\pi^{\mu\nu} &=& 2\eta\sigma^{\mu\nu}-\delta_{\pi\pi}\pi^{\mu\nu}\theta + \lambda_{\pi\Pi}\Pi\sigma^{\mu\nu} - \tau_{\pi\pi} \pi^{\langle\mu}_\alpha\sigma^{\nu\rangle\alpha} + \phi_7 \pi^{\langle\mu}_\alpha\pi^{\nu\rangle\alpha},
\label{eq:shear_relax}
\end{eqnarray}  
where $\dot{\Pi}\equiv u^\alpha\partial_\alpha\Pi$, $\dot{\pi}^{\langle\mu\nu\rangle}\equiv \Delta^{\mu\nu}_{\alpha\beta}u^\lambda\partial_\lambda\pi^{\alpha\beta}$, $\Delta_{\alpha\beta}^{\mu\nu}\equiv\left(\Delta_{\alpha}^{\mu}\Delta_{\beta}^{\nu}+\Delta_{\beta}^{\mu}\Delta_{\alpha}^{\nu}\right)/2-\left(\Delta_{\alpha\beta}\Delta^{\mu\nu}\right)/3$, $\theta\equiv\partial_\alpha u^\alpha$, $\sigma^{\mu\nu}\equiv\partial^{\langle\mu} u^{\nu\rangle}$, with $A^{\langle\mu\nu\rangle}\equiv\Delta^{\mu\nu}_{\alpha\beta}A^{\alpha\beta}$. Other than $\zeta$ and $\eta$, which will be discussed in a moment, the various transport coefficients present in Eqs.~(\ref{eq:bulk_relax}) and (\ref{eq:shear_relax}) were computed assuming a single component gas of constituent particles in the limit $m/T \ll 1$ \cite{Denicol:2012cn,Denicol:2014vaa}, where $m$ is their mass and $T$ the temperature, respectively. These same transport coefficients were also used in Refs. \cite{Paquet:2015lta,Ryu:2015vwa,Ryu:2017qzn}. They are summarized in Table \ref{table:bulk_shear_relax}, where $c^2_s=\partial P /\partial \epsilon$ is the speed of sound squared. The spatial resolution of the hydrodynamical simulation is $\Delta x = \Delta y = 0.17$ fm, while the temporal one is $\Delta \tau=0.015$ fm/$c$.

\begin{table}[!ht]
\caption{Transport coefficients in Eqs. (\ref{eq:bulk_relax}) and (\ref{eq:shear_relax}).} 
\centering
\begin{tabular}{c | l | c | l | l | l}
Bulk & $\tau_\Pi=\zeta\left[15(\epsilon+P)\left(\frac{1}{3}-c^2_s\right)^2\right]^{-1}$ & $\delta_{\Pi\Pi}=\frac{2}{3}\tau_\Pi$ & $\lambda_{\Pi\pi}=\frac{8}{5}\left(\frac{1}{3}-c^2_s\right)\tau_\Pi$ & & \\
\hline \hline
Shear & $\tau_\pi=5\eta\left[\epsilon+P\right]^{-1}$  & $\delta_{\pi\pi}=\frac{4}{3}\tau_\pi$ &  $\lambda_{\pi\Pi}=\frac{6}{5}\tau_\pi$ & $\tau_{\pi\pi}=\frac{10}{7}\tau_\pi$ & $\phi_7=\frac{18}{175} \frac{\tau_\pi}{\eta}$
\end{tabular}
\label{table:bulk_shear_relax} 
\end{table}

The specific shear viscosity $\eta/s$ --- where $s$ is the entropy density --- is here assumed to be temperature independent, while the specific bulk viscosity ($\zeta/s$) is assumed to exhibit a strong temperature dependence in the vicinity of the quark-hadron cross-over as shown in Refs.~\cite{Paquet:2015lta,Ryu:2015vwa,Ryu:2017qzn}. Indeed, assuming a medium that has both bulk and shear viscosity, a reasonable value of $\eta/s$ at LHC energy of $\sqrt{s_{NN}}=2.76$ TeV is $\eta/s=0.095$, obtained by fitting to hadronic observables measured by the ALICE and CMS Collaborations \cite{Ryu:2015vwa}. To narrow down some effects of bulk viscosity, two other simulations are run at LHC energy, one where bulk viscosity is removed while keeping $\eta/s=0.095$, and another where we increase shear viscosity to $\eta/s=0.16$ in order to better reproduce multiplicity as well as $v_2$ of hadrons \cite{Ryu:2015vwa}. However, this increase of $\eta/s$ comes at a price of a degraded description of the mean transverse momentum $\langle p_T\rangle$ of various hadronic species. All cases considered employ the same IP-Glasma initial conditions.

In order to fit hadronic observables with this model \cite{Ryu:2015vwa,Ryu:2017qzn}, it is necessary to modify the value of $\eta/s$ as one changes collision energy. In some sense, this is an approximate way to take into account a temperature dependence of $\eta/s$. At the top RHIC energy, $\sqrt{s_{NN}}=200$ GeV, a smaller value of $\eta/s$ (i.e. $\eta/s=0.06$) is used in order to reproduce hadronic observables measured by the STAR Collaboration \cite{Ryu:2015vwa,Ryu:2017qzn}. Given that the main goal of the present study is to explore how bulk viscosity affects dilepton production, two hydrodynamical simulations will be run at top RHIC energy, one with bulk viscosity and the other without. In both cases, shear viscosity is kept at $\eta/s=0.06$ and  the same IP-Glasma initial conditions are employed. A better description of the data with one or the other values of $\zeta/s$ does not necessarily mean that this values is preferred, because a similar quality fit could have been obtained with both values by simultaneously changing the other model parameters. Our procedure simply illustrates the effects of bulk viscosity on dilepton observables.

Lastly, hydrodynamical simulations are evolved until a switching temperature ($T_{\rm sw}$) is reached, where fluid elements are converted to hadrons. As calibration of the model is done using hadronic observables, further hadronic dynamics are performed via \texttt{UrQMD} simulations \cite{Bass:1998ca,Bleicher:1999xi}. However, no electromagnetic radiation from this stage is computed in this work. The switching temperature is also allowed to vary depending on collision energy. Within the model, the best description of the hadronic observables \cite{Ryu:2015vwa, Ryu:2017qzn} at top RHIC collision energy is reached when $T^{RHIC}_{\rm sw}=165$ MeV, whereas at LHC energy a temperature $T^{LHC}_{\rm sw}=145$ MeV is used. 

\section{Dilepton production}\label{sec:dilepton_prod}

This study considers two categories of dilepton radiation: one is the dilepton radiation from the underlying hydrodynamical simulation, which for the sake of brevity will often be called ``thermal'' dileptons, even though the medium and the dilepton rates account for non-equilibrium bulk and shear dissipation; the other source is cocktail dileptons, which consist of late Dalitz decays of pseudo-scalar mesons and vector mesons, as well as direct decays of vector mesons. 

\subsection{Thermal dileptons}  

The hydrodynamical dilepton production consists of thermal emissions from the QGP as well as in-medium decays of vector mesons in the late hadronic evolution stage. The equation of state used throughout this study \cite{Huovinen:2009yb} smoothly interpolates between lattice QCD calculations ($\ell$QCD), which contain a cross-over transition, and the hadron resonance gas (HRG) model employed at lower temperatures. In accordance with the matching between $\ell$QCD and HRG, done in the temperature region $0.184<T<0.22$ GeV \cite{Huovinen:2009yb}, we interpolate between the thermal dilepton rates as follows:
\begin{equation}
\frac{d^4 R}{d^4 q} = f_{QGP} \frac{d^4 R_{QGP}}{d^4 q} + \left(1-f_{QGP}\right) \frac{d^4 R_{HM}}{d^4 q}, 
\label{eq:f_QGP}
\end{equation}
where $\frac{d^4 R_{QGP}}{d^4 q}$ is the partonic dilepton rate, $\frac{d^4 R_{HM}}{d^4 q}$ is the dilepton rate from the hadronic medium (HM), which are both defined in the following two subsections. The QGP fraction, denoted by $f_{QGP}$, interpolates linearly between $f_{QGP}=1$ at temperature $T>0.22$ GeV, and $f_{QGP}=0$ at $T<0.184$ GeV. Thermal dilepton rates are integrated for all temperatures {\it above} the switching temperature $T_{\rm sw}$, while dileptons from the hadronic cocktail will be computed from the hypersurface of constant $T_{\rm sw}$, specified in the previous section.  

The dilepton production rate for an equilibrated system takes the following form, valid for both partonic and hadronic production sources:  
\begin{eqnarray}
\frac{d^4 R^{\ell^+\ell^-}}{d^4 q}=-\frac{L(M)}{M^2}\frac{\alpha^2_{EM}}{\pi^3 } \frac{\mathrm{Im}\left[\Pi^R_{EM}(M,|{\bf q}|;T)\right]}{e^{q\cdot u/T}-1},
\label{eq:dilep_rate}
\end{eqnarray}
where $L(M)=\left(1+\frac{2m^2_\ell}{M^2}\right)\sqrt{1-\frac{4m^2_\ell}{M^2}}$, $m_\ell$ is the lepton mass, $M^2=q_{\mu}q^{\mu}$, $q^0=\sqrt{M^2+\left\vert{\bf q}\right\vert^2}$, $\alpha_{EM}=\frac{e^2}{4\pi}\approx \frac{1}{137}$, $u^\mu$ is the local flow four velocity of the medium, while $T$ is its temperature, and $\mathrm{Im}\left[\Pi^R_{EM}\right]$ is the imaginary part of the trace of the retarded (virtual) photon self-energy. A general form of the above expression, valid off-equilibrium, uses the real time formalism where the dilepton rate is proportional to $\left[\Pi_{12}\right]^\mu_\mu$ instead of $\mathrm{Im}\left[\Pi^R_{EM}\right]/\left[e^{q\cdot u/T}-1\right]$. More theoretical details concerning off-equilibrium electromagnetic production have only been worked out for the case of real photons \cite{Shen:2014nfa,Hauksson:2017udm}.  

\subsubsection{High temperature dilepton production: partonic dilepton rates}\label{subsub_sec:qgp_dileptons}
 
Perturbative dilepton rates for an equilibrated QGP have been computed at next-to-leading order (NLO) \cite{Laine:2013vma,Ghisoiu:2014mha,Ghiglieri:2014kma}, while early lattice calculations \cite{Ding:2010ga,Ding:2013qw,Ding:2016hua} for in-equilibrium EM production have only recently been extended to include energy {\it and} three-momentum dependence of the virtual photon \cite{Ghiglieri:2016tvj,Kaczmarek:2017hfx,Jackson:2019yao}. However, lattice QCD dilepton rates are calculated on a discrete grid using the imaginary time formalism, and thus incur various uncertainties whence extracting their real-time values, limiting their phenomenological impact. Also, lattice QCD results are not yet amenable to a dissipative description of the medium. The effects of dissipation on NLO dilepton rates are yet to be fully derived. Thus, our study focuses on the QGP dilepton rate within the Born approximation where dissipative corrections are included.\footnote{Dilepton production has also been considered in out-of-equilibrium scenarios described in Refs.~\cite{Linnyk:2012pu,Song:2018dvf,Song:2018xca}. Our study focuses on dileptons production from dissipative hydrodynamics and hence we will only be focusing on this case.} The main point of this work is to explore how the presence of bulk viscous pressure affects the hydrodynamical evolution and in turn dilepton production; thus it is important to have the effects of bulk dissipation consistently included into the calculation. 

Assuming a non-dissipative fluid, the Born dilepton rate takes the following form:
\begin{eqnarray}
\frac{d^4 R_{0}}{d^4 q} &=& \int \frac{d^3 k_1}{(2\pi)^3 k^0_1} \frac{d^3 k_2}{(2\pi)^3 k^0_2} n_{0,{\bf k_1}} n_{0,{\bf k_2}} \frac{q^2}{2} \sigma \delta^4(q-k_1-k_2), \nonumber\\
                         \sigma &=& \frac{16\pi \alpha_{\rm EM}^2 \left(\sum_{f'} e^2_{f'}\right) N_c}{3 q^2},\nonumber\\
               n_{0,{\bf k_i}}&=&\left[\exp\left(k_i\cdot u/T\right)+1\right]^{-1} \quad \forall  i \in1,2 \,\, ,
\label{eq:R_born}
\end{eqnarray} 
where $u^\mu$ is the local flow velocity of the fluid, $\sigma$ is the leading-order quark-antiquark annihilation (into a lepton pair) cross section, $N_c=3$ is the number of colors in QCD, and $f'=u,d,s$ labels quarks flavors included here. Extending the isotropic dilepton rate in Eq.~(\ref{eq:R_born}) to include both bulk and shear-viscous effects amounts to modifying the quark/anti-quark Fermi-Dirac distribution functions $n_{0,\bf k}$.\footnote{Note that including the effects of dissipation in Born dilepton rates has been carried out in the context of both dissipative hydrodynamics as presented here, and anisotropic dissipative hydrodynamics (presented in Ref.~\cite{Kasmaei:2018oag} therein). The present study uses standard dissipative hydrodynamics and thus the viscous corrections to dilepton rates will be those relevant for this situation.} Using the Israel-Stewart 14-moment approximation described in \cite{Dusling:2008xj,Vujanovic:2013jpa}, the shear viscous correction to the QGP dilepton rate $\delta R_\pi$, in the local rest frame of the medium, reads
\begin{eqnarray}
\frac{d^4 \delta R_\pi}{d^4 q} & = & \frac{q^\alpha q^\beta \pi_{\alpha\beta}}{2T^2(\epsilon + P)}\left\{C_q \frac{q^2}{2} \frac{\sigma}{(2\pi)^5} \frac{T^5}{\left\vert{\bf q}\right\vert^5} \int^{\frac{E_+}{T}}_{\frac{E_-}{T}} \frac{dE_{\bf k}}{T} n_{0,\bf k} \left[1-n_{0,\bf k}\right] n_{0}\left(q^0-E_{\bf k}\right) D\right\},\nonumber\\
D&=&T^{-4}\left[(3q_0^2-|{\bf q}|^2) E^2_{\bf k}-3 q^0 E_{\bf k} q^2 + \frac{3}{4} q^4\right],
\label{eq:dR_born_shear}
\end{eqnarray} 
where $\frac{E_{\pm}}{T}=\frac{1}{2}\frac{q^0\pm |{\bf q}|}{T}$, $n_{0,\bf k}=\left[\exp\left(E_{\bf k}/T\right)+1\right]^{-1}$, $n_{0}\left(q^0-E_{\bf k}\right)=\left[\exp\left(\frac{q^0-E_{\bf k}}{T}\right)+1\right]^{-1}$, and $C_{q} = \frac{7\pi^4}{675\zeta(5)} \approx 0.97$.  The bulk viscous correction to the quark distribution function used herein is obtained by solving the effective kinetic theory of quasiparticles described in Ref.~\cite{Jeon:1995zm}, where in the relaxation time approximation the quasi-particle distribution function satisfies
\begin{eqnarray}
k^\mu \partial_\mu n_{\bf k} -	\frac{1}{2} \frac{\partial (m^2)}{\partial {\bf x}} \cdot \frac{\partial n_{\bf k}}{\partial {\bf k}} = - E_{\bf k} \frac{\delta n_{\bf k}}{\tau_R}.
\end{eqnarray}
In the local rest frame, using the Chapman-Eskog expansion in the relaxation time approximation, the leading order solution to $\delta n_{\bf k}$ reads \cite{Paquet:2015lta}
\begin{eqnarray}
\delta n_{\bf k} &=& \Pi\frac{\tau_\Pi}{\zeta} n_{0,\bf k} \left[1-n_{0,\bf k}\right]\left[\frac{E_{\bf k}}{T}-\frac{m^2_{q,\bar{q}}}{E_{\bf k}T} \right]\left(\frac{1}{3}-c^2_s\right),
\label{eq:bulk_deltan_1}
\end{eqnarray}
where $\tau_\Pi/\zeta$ is given in Table \ref{table:bulk_shear_relax}, $c^2_s$ is the speed of sound squared, and we take $m^2_{q,\bar{q}} = g^2_s T^2/3$ with $g_s=2$. Effects of the running of the couplings are not considered in the above $\delta n_{\bf k}$ \cite{Paquet:2015lta}. Using Eq.~(\ref{eq:bulk_deltan_1}) the bulk viscous correction of the QGP dilepton rate $\delta R_\Pi$ in the local rest frame is
\begin{eqnarray}
\frac{d^4 \delta R_\Pi}{d^4 q} & = & \Pi \frac{\tau_\Pi}{\zeta}\left[A\left(\frac{q^0}{T},\frac{\left\vert q\right\vert}{T}\right)+B\left(\frac{q^0}{T},\frac{\left\vert q\right\vert}{T}\right) \right]\left(\frac{1}{3}-c^2_s\right), \nonumber\\
A\left(\frac{q^0}{T},\frac{\left\vert q\right\vert}{T}\right) &=& \frac{2T}{\left\vert q\right\vert} \int^{\frac{E_+}{T}}_{\frac{E_-}{T}} \frac{dE_{\bf k}}{T} n_{0,\bf k}\left[1-n_{0,\bf k}\right] n_{0}\left(q^0-E_{\bf k}\right) \frac{E_{\bf k}}{T}, \nonumber\\
B\left(\frac{q^0}{T},\frac{\left\vert q\right\vert}{T}\right) &=& -\frac{2T}{\left\vert q\right\vert} \frac{m^2_{q,\bar{q}}}{T^2} \int^{\frac{E_+}{T}}_{\frac{E_-}{T}} \frac{dE_{\bf k}}{T} n_{0,\bf k} \left[1-n_{0,\bf k}\right] n_{0}\left(q^0-E_{\bf k}\right) \frac{T}{E_{\bf k}}.
\label{eq:dR_born_bulk}
\end{eqnarray}
The complete Born rate can therefore be expressed as $\frac{d^4 R}{d^4 q}=\frac{d^4 R_0}{d^4 q}+\frac{d^4 \delta R_\pi}{d^4 q}+\frac{d^4 \delta R_\Pi}{d^4 q}$, where the first, second, and third terms are found in Eqs.~(\ref{eq:R_born}), (\ref{eq:dR_born_shear}), and (\ref{eq:dR_born_bulk}), respectively.

\subsubsection{Low temperature dilepton production: hadronic dilepton rates}\label{subsub_sec:hm_dileptons}

In the hadronic sector, an important contribution to the dilepton production rate stems from decays of vector mesons in the QCD medium. In this work we leave out vector mesons made of charm and beauty quarks whose contribution is significant only at invariant masses beyond 2.5 GeV or so. Only the low mass vector mesons composed of up, down and strange quarks, i.e.~the $\rho$, $\omega$, and $\phi$, are included. The in-medium properties of these vector mesons are described via their spectral functions, while their connection to dilepton production is given by the Vector Dominance Model (VDM) first proposed by Sakurai \cite{Gounaris:1968mw}. Using VDM, relating the retarded virtual photon self-energy to the vector meson spectral function is done via:
\begin{eqnarray}
{\rm Im} \left[\Pi^R_{EM}\right]&=&\sum_{V=\rho,\omega,\phi}\left(\frac{m^2_V}{g_V}\right)^2 {\rm Im} \left[D^R_V\right]\nonumber\\
                                &=&\sum_{V=\rho,\omega,\phi}\left(\frac{m^2_V}{g_V}\right)^2 {\rm Im} \left[\frac{1}{M^2-m^2_V-\Pi^R_V}\right],
\end{eqnarray}
where $m_V$ is the mass of the vector meson and $g_V$ is the coupling to photons. An essential component needed to compute the spectral function ${\rm Im} \left[D^R_V\right]$ is the in-medium self-energy of the vector mesons, while the Schwinger-Dyson equation \cite{Roberts:1994dr} is used to construct the spectral function, once the self-energy is determined. The self-energy will be computed using the model first devised by Eletsky {\it et al.} \cite{Eletsky:2001bb}.\footnote{An alternative approach to calculate in-medium spectral functions relies on many-body chiral effective Lagrangians \cite{Rapp:1999ej}. This approach has not yet been extended to include dissipative corrections and hence will not be explored here.} In this model, the self-energy contains both the vacuum and medium contributions \cite{Eletsky:2001bb,Martell:2004gt,Vujanovic:2009wr} such that
\begin{eqnarray}
\Pi^R_V=\Pi^R_{V, {\rm vac}} + \Pi^R_{Va,{\rm med}},
\end{eqnarray}
where $\Pi^R_{V, {\rm vac}}$ is computed via chiral effective Lagrangians \cite{Eletsky:2001bb,Martell:2004gt,Vujanovic:2009wr} while the finite-temperature piece takes the form \cite{Eletsky:2001bb,Martell:2004gt,Vujanovic:2009wr,Vujanovic:2013jpa}
\begin{eqnarray}
\Pi^R_{Va,{\rm med}} = -4 \pi \int \frac{d^3k}{(2\pi)^3 k^0} n_{a,{\rm med}}(\omega) \sqrt{s}f_{Va}(s);
\label{eq:medium-Pi-HM}
\end{eqnarray}
here $\omega=u\cdot k$, $n_{a,{\rm med}}(\omega)$ is the distribution of the scattering partners $a$ of vector mesons $V$, while $f_{Va}(s)$ is the forward scattering amplitude of $V$ scattering onto $a$ (see Refs.~\cite{Eletsky:2001bb,Martell:2004gt,Vujanovic:2009wr} for how $f_{Va}$ is constructed). In a dissipative medium such as the one in the present study, there will be both an in-equilibrium and a dissipative contribution to $n_a$, with the latter accounting for shear and bulk viscous effects.\footnote{A different approach investigating dissipative corrections to dilepton production in the hadronic medium was explored in Ref. \cite{Schenke:2005ry} using the Kadanoff-Baym equations.} The shear viscous correction has been computed in \cite{Vujanovic:2013jpa} using the 14-moment approximation, while the bulk viscous correction obtained using the Chapman-Eskog expansion in the relaxation-time approximation \cite{Paquet:2015lta} is
\begin{eqnarray}
\delta n_{a,\Pi}=\Pi \frac{\tau_\Pi}{\zeta} n_{0,a}(\omega) \left[1\pm n_{0,a}(\omega)\right]\left[\left(\frac{1}{3}-c^2_s\right)\frac{\omega}{T} -\frac{m^2_{a}}{3 \omega T}  \right],
\label{eq:delta_n_bulk_hm}
\end{eqnarray}    
here $\tau_\Pi/\zeta$ is found in Table \ref{table:bulk_shear_relax} and $n_{0,a}$ is either a Fermi-Dirac or a Bose-Einstein distribution depending on whether $a$ is a Boson or a Fermion. Following the procedure presented in Appendix B 2 of Ref. \cite{Vujanovic:2013jpa}, substituting Eq.~(\ref{eq:delta_n_bulk_hm}) into Eq.~(\ref{eq:medium-Pi-HM}) yields a correction to the vector meson self-energy owing to the bulk-modified distribution function, which reads
\begin{eqnarray}
\delta \Pi^R_{Va,\Pi}&=&\Pi\frac{\tau_\Pi}{\zeta}\left[\left(\frac{1}{3}-c^2_s\right){\cal A} (|{\bf p}|,T)+{\cal B} (|{\bf p}|,T)\right],\nonumber\\
{\cal A} (|{\bf p}|,T)&=&-\frac{m_V m_a T}{\pi |{\bf p}|}\int^\infty_{m_a} d\omega' f^{\mathrm{a's\text{ }rest}}_{Va}\left( \frac{m_V}{m_a} \omega' \right)\times\nonumber\\
&& \negmedspace {}\times\left\{\frac{\omega_-}{T}\frac{1}{\exp\left(\frac{\omega_-}{T}\right)\pm 1}-\frac{\omega_+}{T}\frac{1}{\exp\left(\frac{\omega_+}{T}\right)\pm 1}\pm\ln\left[\frac{1\pm\exp\left(-\omega_-/T\right)}{1\pm\exp\left(-\omega_+/T\right)}\right]\right\},\nonumber\\
{\cal B} (|{\bf p}|,T)&=&\frac{m_V m^3_a}{\pi |{\bf p}| T}\int^\infty_{m_a} d\omega' f^{\mathrm{a's\text{ }rest}}_{Va}\left( \frac{m_V}{m_a} \omega' \right)\int^{\frac{\omega_+}{T}}_{\frac{\omega_-}{T}}d\zeta\frac{1}{3\zeta}\frac{\exp(\zeta)}{\left[\exp(\zeta)\pm 1\right]^2}.
\label{eq:delta_Pi_HM}
\end{eqnarray}
Here the upper (lower) signs refers to fermions (bosons), $f^{\mathrm{a's\text{ }rest}}_{Va}$ is the forward scattering amplitude evaluated in the rest frame of $a$, while $\omega_\pm=\frac{E\omega'\pm|{\bf p}||{\bf k'}|}{m_V}$, $E=\sqrt{|{\bf p}|^2+m^2_V}$, $|{\bf k'}|=\sqrt{\left(\omega'\right)^2-m^2_a}$. The shear correction to the self-energy $\delta \Pi^R_{Va,\pi}$ can be found in Appendix B 2 of Ref.~\cite{Vujanovic:2013jpa}, while the thermally equilibrated contribution $\Pi^R_{Va,0}$ is given in Appendix B 1 of Ref.~\cite{Vujanovic:2013jpa} as well as in Refs.~\cite{Eletsky:2001bb,Martell:2004gt,Vujanovic:2009wr}. 

\subsection{Cocktail dileptons}\label{sec:cocktail_dileptons}     

Once the hydrodynamical simulation reaches the switching temperature, thermal dilepton production is stopped. However, dileptons are still being radiated from late decays of hadrons, which we will refer to as the dilepton cocktail. In the low invariant mass region $0.3\lesssim M \lesssim 1.1$ GeV, this contribution mainly consists of late Dalitz decays of $\eta\rightarrow \gamma\gamma^*$, $\omega\rightarrow \pi^0\gamma^*$, $\eta'\rightarrow \gamma\gamma^*$, $\phi\rightarrow \eta\gamma^*$ mesons, as well as late direct decays of vector mesons $\rho$, $\omega$, and $\phi$. An in-depth discussion about the sources of cocktail dileptons in the context of Vector Dominance Model VDM -- followed here throughout -- is presented in Ref.~\cite{Landsberg:1986fd}. We only summarize the relevant results for our study. The invariant mass branching fraction of Dalitz decays, where the parent particle $a$ decays into the daughter particle $b$ and a virtual photon, reads \cite{Landsberg:1986fd}
\begin{eqnarray}
\frac{dB_{a\rightarrow b\gamma^*}}{d(M^2)}=N \frac{L(M)}{M^2}\left[\left(1+\frac{M^2}{m^2_a-m^2_b}\right)^2-\frac{4M^2m^2_a}{\left(m^2_a-m^2_b\right)^2}\right]^{\frac{3}{2}} \left| F_{ab}(M) \right|^2,
\label{eq:dalitz-branching}
\end{eqnarray} 
where $L(M)$ is defined below Eq.~(\ref{eq:dilep_rate}), $\left| F_{ab}(M) \right|^2$ is the form factor computed via VDM \cite{Landsberg:1986fd}, while $N$ is an overall normalization such that
\begin{eqnarray}
\int^{\left(m_{a}-m_{b}\right)^2}_{4m^2_\ell} d(M^2) \frac{dB_{a\rightarrow b\gamma^*}}{d(M^2)} = B_{a\rightarrow b\gamma^*},
\label{eq:dalitz-normalization}
\end{eqnarray}
where $B_{a\rightarrow b\gamma^*}$ is the total branching fraction measured experimentally. The kinematics of the decay being determined by Eq.~(\ref{eq:dalitz-branching}), one only needs to compute the distribution of virtual photons. In their local rest frame and at a given invariant mass $M$, the distribution of virtual photons can be obtained from the parent particle $a$ via \cite{Kapusta:1977ce}
\begin{eqnarray}
\left.\frac{d^4 N_{\gamma^*\leftarrow a}}{d^4 q}\right\vert_{a\,{\rm is\, on-shell}} = \int \frac{d\Omega '}{4\pi} \frac{m^2_a}{M^2} \frac{p'^0 d^3 N_a}{d^3 p'},
\label{eq:dalitz-distribution}
\end{eqnarray}
where primes denote quantities in the rest frame of the virtual photon, with $d\Omega'=\sin \theta' d\theta' d\phi'$ being the usual solid angle in momentum space, while $\frac{p'^0 dN_a}{d^3 p'}$ is the distribution of the meson $a$ obtained as explained in Ref.~\cite{Ryu:2017qzn}.\footnote{$\frac{d^4 N_{\gamma^*\leftarrow a}}{d^4 q}$ should be read as the virtual photon distribution obtained from decays of $a$.} Thus, the virtual photon distribution originating from Dalitz decays is
\begin{equation}
\frac{d^4 N}{d^4 p}=N \frac{L(M)}{M^2}\left[\left(1+\frac{M^2}{m^2_a-m^2_b}\right)^2-\frac{4M^2m^2_a}{\left(m^2_a-m^2_b\right)^2}\right]^{\frac{3}{2}} \left| F_{ab}(M) \right|^2 \int \frac{d\Omega '}{4\pi} \frac{m^2_a}{M^2} \frac{p'^0 d^3 N_a}{d^3 p'}.
\label{eq:dalitz_final}
\end{equation}
As far as direct decays into dileptons are concerned, the branching ratio is obtained using VDM as well
\begin{eqnarray}
M\Gamma_{V\rightarrow\gamma^*}&=& \frac{\alpha^2}{3} \frac{m^4_{V}}{g^2_V/(4\pi)}\frac{L(M)}{M^2},\nonumber\\
M\Gamma_V &=& -{\rm Im}\left[\Pi^R_V\right];
\label{eq:num_denom_V_dilep}
\end{eqnarray}
thus, the branching fraction is
\begin{eqnarray}
\frac{dB_{V\rightarrow\gamma^*}}{d(M^2)} &=& \frac{\alpha^2}{3} \frac{m^4_{V}}{g^2_V/(4\pi)}\frac{L(M)}{M^2} \frac{1}{-{\rm Im}\left[\Pi^R_V\right]}.
\label{eq:V_dilep_branching}
\end{eqnarray}
With all the various branching fractions presented, only the distribution of the original hadrons remains to be specified. Other than the $\rho$, all other mesons have a lifetime larger than the duration of the hydrodynamical simulations, thus they are treated using the Cooper-Frye (CF) prescription \cite{Cooper:1974mv} including resonance decays, which is how $\frac{p'^0 d^3 N_a}{d^3 p'}$ in Eq.~(\ref{eq:dalitz-distribution}) is obtained \cite{Ryu:2017qzn}. The Cooper-Frye prescription reads
\begin{eqnarray}
\frac{p^0 d^3 N_a}{d^3 p}&=& \int d^3 \Sigma_\mu p^\mu \frac{d_a}{(2\pi)^3}n_{a},
\label{eq:CF}
\end{eqnarray}
where the integral goes over an isothermal hypersurface of temperature $T_{\rm sw}$, $d^3 \Sigma_\mu$ is the freeze-out hypersurface element, $d_a$ is the spin degeneracy of $a$ with different isospin states being treated as separate particle species, while $n_{a}$ is a momentum distribution function. The CF distribution presented in Eq.~(\ref{eq:CF}) assumes that the spectral distribution of particles is $\delta(p^2-m^2) \Theta(p^0)$, which is valid for stable particles. To take into account short-lived particles, specifically the $\rho$ meson, the CF distribution must be generalized as follows:
\begin{eqnarray}
N_{\rho} &=& \int \frac{d^3p}{(2\pi)^3 p^0} \int d^3 \Sigma_\mu p^\mu d_\rho n_{\rho} \nonumber\\
         &=& \int \frac{d^4p}{(2\pi)^4} (2\pi) 2\delta(p^2-m^2)\Theta(p^0) \int d^3 \Sigma_\mu p^\mu d_\rho n_{\rho} \nonumber\\
		   	 &\to& \int \frac{d^4p}{(2\pi)^3} \rho_\rho(M) \int d^3 \Sigma_\mu p^\mu d_\rho n_{\rho}(M).
\label{eq:mod_CF}
\end{eqnarray}
Here $\rho_\rho(M)$ is the spectral function of $\rho$ mesons, which we write as $\rho_\rho(M)=2\frac{-{\rm Im}\left[D^R_\rho(M)\right]}{\pi}$, while $M$ is assumed to be positive semi-definite. The latter form of $\rho_\rho(M)$ reduces to the usual Breit-Wigner distribution if we assume the self-energy $\Pi^R_{\rho}(M)$ is a complex number, while $2 \delta(p^2-m^2) \Theta(p^0)$ is recovered in the limit where ${\rm Im}\left[\Pi^R_{\rho}\right]\to 0$.  

Combining Eq.~(\ref{eq:mod_CF}) and Eq.~(\ref{eq:V_dilep_branching}) yields
\begin{eqnarray}
\frac{d^4 N_{\rho\rightarrow \gamma^*}}{d^4 p} = \frac{\alpha^2}{\pi^3} \frac{m^4_{V}}{g^2_\rho}\frac{L(M)}{M^2} \left\vert D^R_\rho (M)\right\vert^2 \int d^3 \Sigma_\mu p^\mu n_{\rho}(M).
\label{eq:mod_CF_rho_dilep}
\end{eqnarray}
Note that both $\left| D^R_\rho(M)\right|^2$ and the phase-space distribution $n_{\rho}(M)$ depend on the invariant mass $M$. The latter reduces to the Bose-Einstein distribution for a medium in thermal equilibrium, while bulk and shear viscous corrections used in computing $n_{\rho}(M)$ are those of Ref.~\cite{Ryu:2017qzn}. Thus, since the $\rho$ meson is a broad resonance, its contribution from the hydrodynamical switching hypersurface will be different depending on whether its subsequent production from that hypersurface uses the vacuum version of the $\left| D^R_\rho(M)\right|^2$ distribution or its in-medium counterpart. In order to quantify these differences, we will perform three calculations: where we will (i) use the in-medium distribution of $\left| D^R_\rho(M)\right|^2$ as well as allow for $n_{\rho}(M)$ to depend on the invariant mass, or (ii) set $\left| D^R_\rho(M)\right|^2$ to its vacuum value while still letting $n_{\rho}(M)$ be invariant mass dependent, or (iii) set $\left| D^R_\rho(M)\right|^2$ to its vacuum value and evaluate $n_\rho$ on the mass shell, $n_{\rho}(M=m_\rho)$. Using method (iii) allows us to include the contribution to the multiplicity of $\rho$ mesons coming from resonances decaying into $\rho$ mesons. Indeed, if contributions from resonance decays were to be included to the $\rho$ distribution calculated via method (ii), we would have to calculate the spectral functions of all the parent resonances contributing to $\rho$-production, as well include off-shell dynamics in all decay channels, which is beyond the scope of this study.    

As far as  $\omega$ and $\phi$ vector mesons are concerned, since their vacuum lifetime is significantly larger than that of the hydrodynamical medium, we approximate their cocktail contribution to the dilepton spectrum via
\begin{eqnarray}
\frac{d^4 N_{\omega,\phi\rightarrow \gamma^*}}{d^4 p} = \frac{\alpha^2}{\pi^3} \frac{m^4_{\omega,\phi}}{g^2_{\omega,\phi}}\frac{L(M)}{M^2} \left\vert D^R_{\omega,\phi} (M)\right\vert^2 \int d^3 \Sigma_\mu p^\mu n_{\omega,\phi}(M=m_{\omega,\phi}),
\label{eq:omega_phi_dilep}
\end{eqnarray}
thus employing the $\omega,\phi$ distribution using the on-shell CF integral $\int d^3 \Sigma_\mu p^\mu n_{\omega,\phi}$, including resonance decays \cite{Ryu:2017qzn}, while keeping a non-trivial invariant mass dependence in form factor $\left\vert D^R_\omega(M)\right\vert^2$ and $\left\vert D^R_\phi(M)\right\vert^2$.

\section{Results}\label{sec:results}

Following the procedure of recent dilepton studies \cite{Vujanovic:2016anq,Vujanovic:2017psb} the scalar product method will be used to compute dilepton anisotropic flow coefficients.\footnote{In previous studies \cite{Vujanovic:2016anq,Vujanovic:2017psb} all the events in the 20-40\% centrality bin were put together in one bin, while in the present study, the 20-40\% centrality class is separated into two bins with 10\% intervals, 20-30\% and 30-40\%, which are later recombined into 20-40\%.} Within a centrality class, the anisotropic flow coefficients are computed using the method outlined in \cite{Paquet:2015lta,Vujanovic:2016anq,Vujanovic:2017psb}, namely
\begin{eqnarray}
v^{\gamma^*}_n(X)&=&\frac{\frac{1}{N_{ev}}\sum^{N_{ev}}_{i=1}v^{\gamma^{*}}_{n,i}(X)  v^{h}_{n,i} \cos \left[n\left(\Psi^{\gamma^{*}}_{n,i}(X)-\Psi^h_{n,i} \right)\right]}{\sqrt{\frac{1}{N_{ev}}\sum^{N_{ev}}_{i=1} \left(v^{h}_{n,i}\right)^2}},\nonumber\\
                &=&\frac{\left\langle v^{\gamma^{*}}_{n,i}(X)  v^{h}_{n,i} \cos \left[n\left(\Psi^{\gamma^{*}}_{n,i}(X)-\Psi^h_{n,i} \right)\right] \right\rangle_{{\rm ev}, i}}{\sqrt{\left\langle \left(v^h_{n,i}\right)^2 \right\rangle_{{\rm ev},i}}},
\label{eq:vnSP}
\end{eqnarray}
where $N_{ev}$ is the number of IP-Glasma events, $X$ is any momentum-space variable such as $M$ or $p_T$, and $\langle \ldots \rangle_{{\rm ev}, i}$ is the average over events $i$. In a single event $i$, the hadronic $v^h_{n,i}$ and $\Psi^h_{n,i}$ are given by
\begin{equation}
v^h_{n,i} e^{i n \Psi^h_{n,i}} = \frac{\int d p_T  dy d\phi p_T \left[ p^0 \frac{d^3 N^h_i}{d^3 p} \right] e^{i n\phi}}{\int d p_T dy  d\phi p_T \left[ p^0 \frac{d^3 N^h_i	}{d^3 p} \right]},
\label{eq:vn_ev}
\end{equation}
where the charged hadron distribution is integrated over the entire rapidity acceptance of the STAR detector at the RHIC and ALICE detector at the LHC, while all particles having $p_T>0.3$~GeV are used when computing $v^h_{n,i}$ and $\Psi^h_{n,i}$. The dilepton $v^{\gamma^{*}}_{n,i}$ and $\Psi^{\gamma^{*}}_{n,i}$ are computed using the same approach, with the more general distribution $\frac{d^4 N_{i}^{\gamma^{*}}}{d^4 p}$. Having computed the $v^{\gamma^*}_n(X)$ in 10\% centrality sub-bins, the latter are combined as follows:
\begin{eqnarray}
v^{\gamma^*}_n(X)[20-40\%]=\frac{\frac{dN^{\gamma^{*}}}{dX}[20-30\%]v^{\gamma^{*}}_n(X)[20-30\%]+\frac{dN^{\gamma^{*}}}{dX}[30-40\%]v^{\gamma^{*}}_n(X)[30-40\%]}{\frac{dN^{\gamma^{*}}}{dX}[20-30\%]+\frac{dN^{\gamma^{*}}}{dX}[30-40\%]}.
\label{eq:vn_bin}
\end{eqnarray}
Here $\frac{dN^{\gamma^{*}}}{dX}$ is the dilepton multiplicity in a bin, while $v^{\gamma^{*}}_n(X)$ is the corresponding anisotropic flow coefficient.
 
Results are discussed in the following subsections. We first start in Sec. \ref{sub_sec:bulk_dynamics} by inspecting the dynamics of the medium, shedding light on how bulk viscosity affects the evolution of the medium. This discussion sets the stage for investigations of bulk viscous pressure effects on thermal dilepton yield and anisotropic flow at RHIC and LHC collision energies in Sec. \ref{sub_sec:thermal_dileptons}. Novel dynamics introduced by bulk viscous pressure, which translate onto the dilepton yield and $v_2$, may be used in better constraining $\zeta/s$. The last subsection (\ref{sub_sec:cokctail_results}) calculates the dilepton cocktail, with special emphasis given to treatment of the $\rho$ on the constant temperature ($T_{\rm sw}$) switching hypersurface. This section also highlights the invariant mass region were the novel effects of bulk viscosity can be seen in our dilepton $v_2$ calculation, while  constraining $\zeta/s$ needs a measurement with a good invariant mass resolution of dilepton elliptic flow.    

\subsection{Medium dynamics under the influence of shear and bulk viscosity}\label{sub_sec:bulk_dynamics}
\begin{figure}[!h]
\begin{center}
\begin{tabular}{cc}
\includegraphics[width=0.495\textwidth]{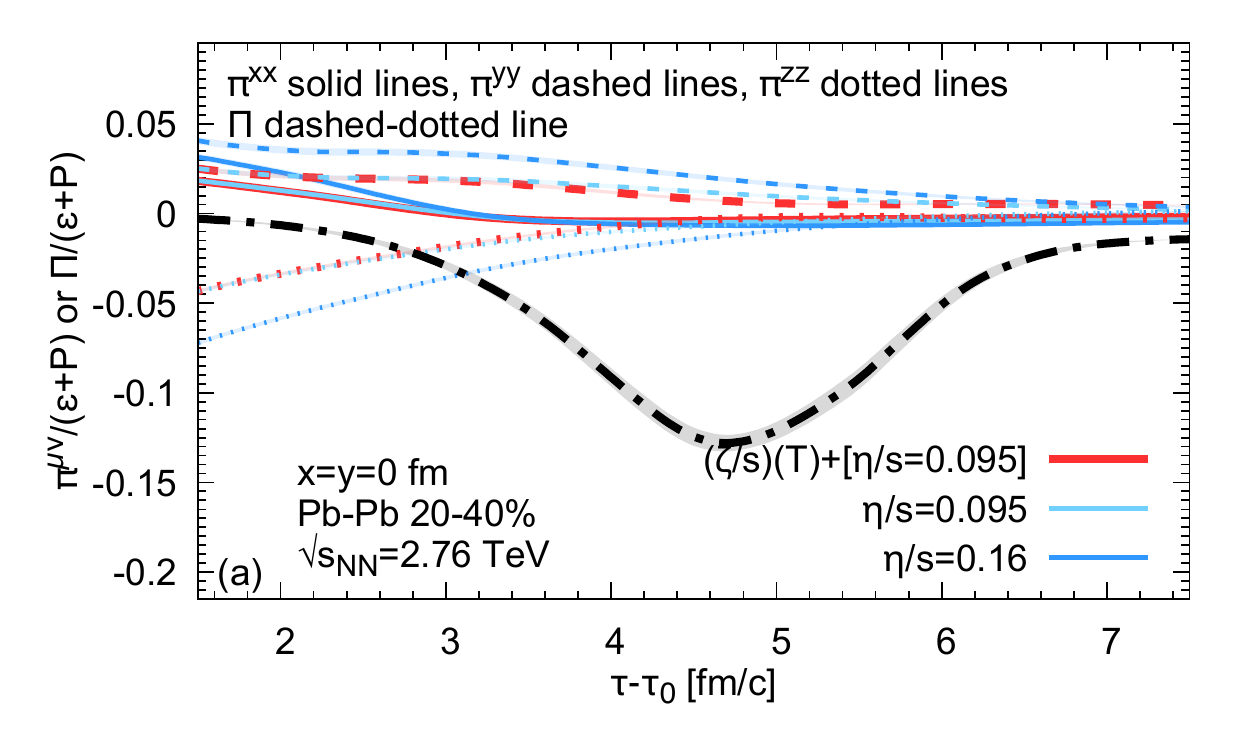} & \includegraphics[width=0.495\textwidth]{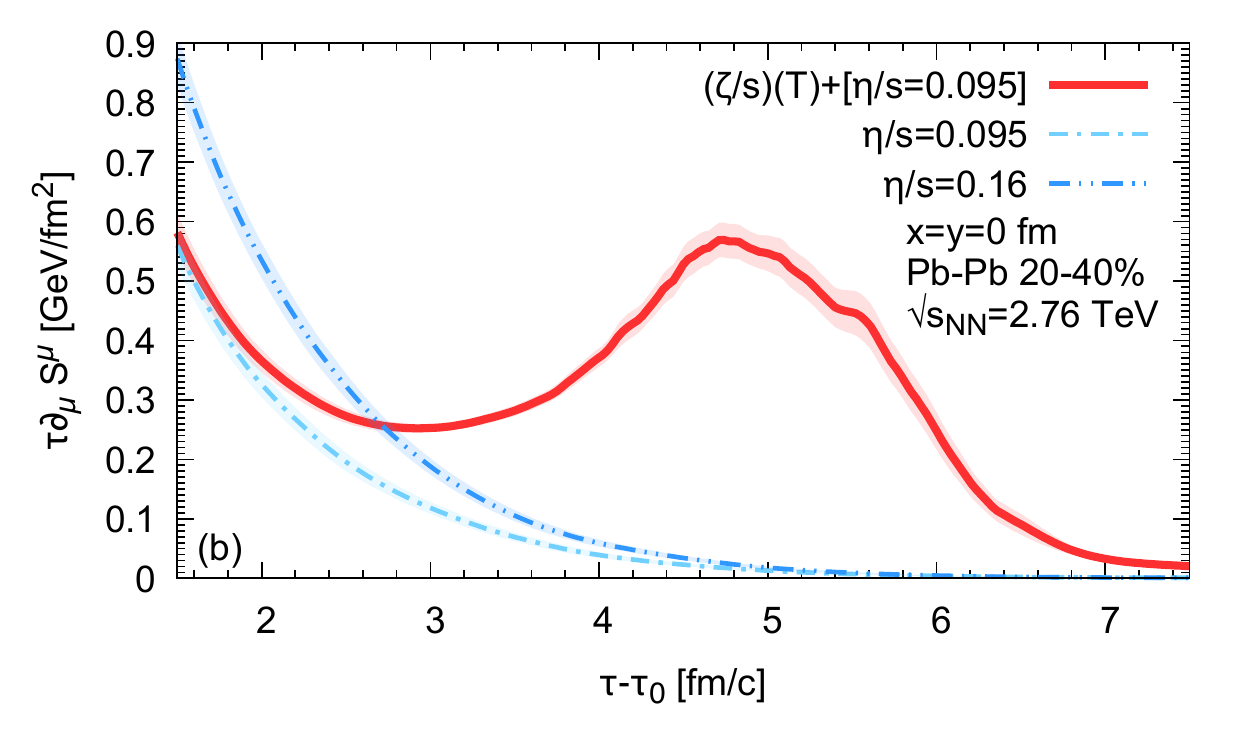}\\
\includegraphics[width=0.495\textwidth]{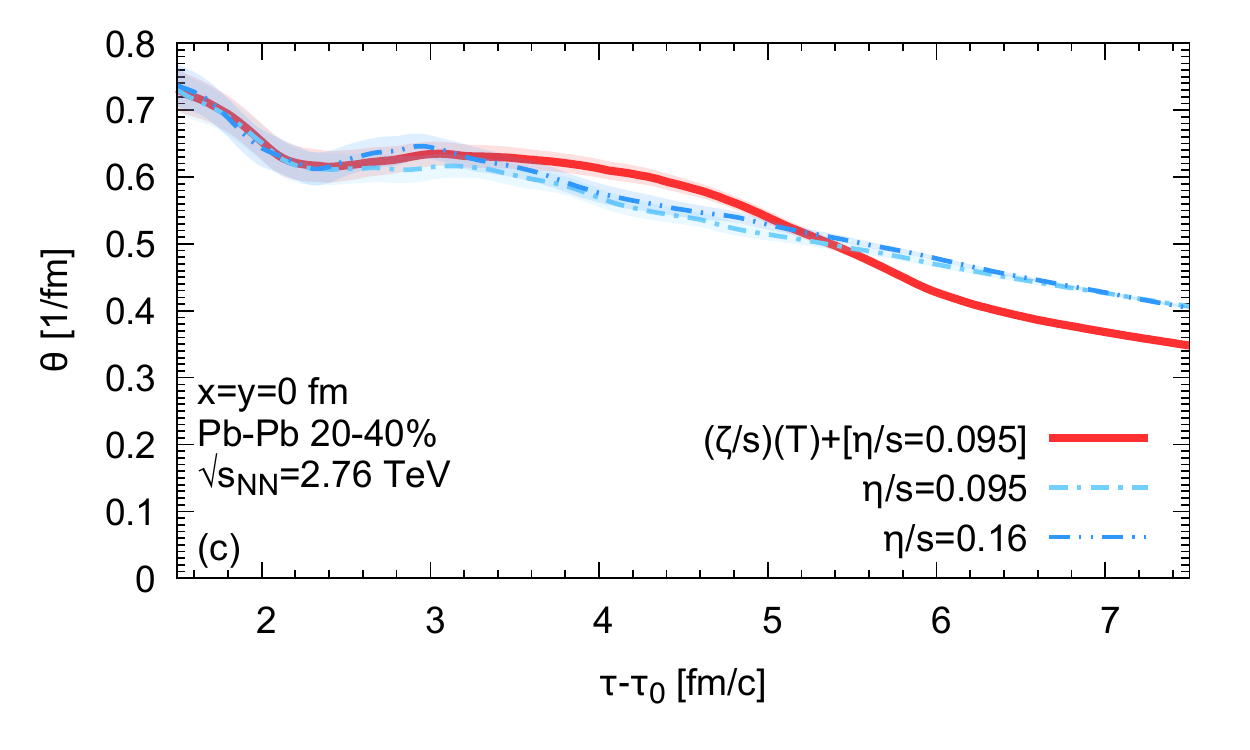} & \includegraphics[width=0.495\textwidth]{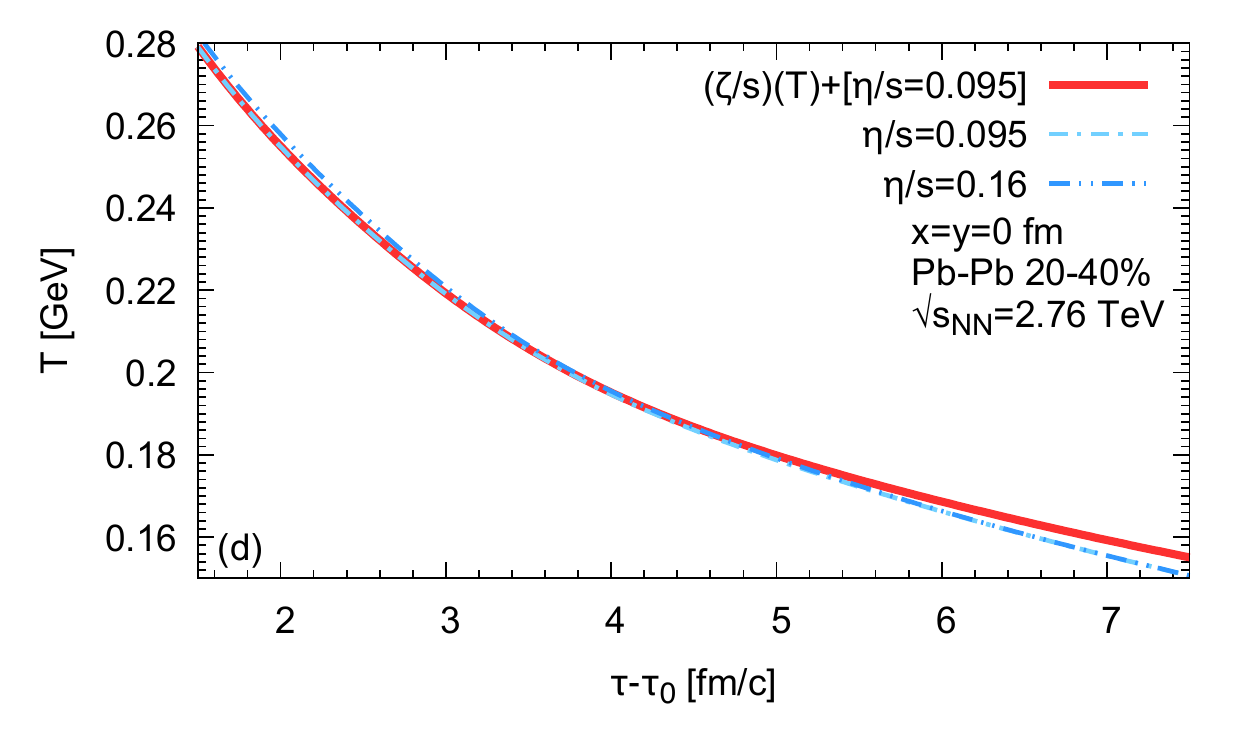}
\end{tabular}
\end{center}
\caption{(Color online) (a) Event-averaged enthalpy density normalized shear and bulk viscous pressure during $1.5\leq\tau-\tau_0\leq 7.5$ fm/$c$ of evolution. $\pi^{\mu\nu}/(\varepsilon+P)$ in the local rest frame is shown. (b) Event-averaged entropy production rate $\partial_\mu S^\mu$ rescaled by $\tau$. Note that the initial entropy density of the central cell is 44.24 GeV/fm$^2$. (c) Event-averaged expansion rate $\theta$, while (d) depicts event-averaged temperature for the central cell.}
\label{fig:later_latest_dynamics}
\end{figure}

Since effects of (shear) viscosity on the dynamics of the medium and dilepton production have been explored in the past \cite{Vujanovic:2016anq,Vujanovic:2017psb} at top RHIC collision energy, we focus on the effects of bulk viscosity on medium evolution at LHC collision energy with RHIC results presented later. In Fig. \ref{fig:later_latest_dynamics}, we explore the later-time temperature evolution of the medium, obtained through the non-trivial competition between entropy production and expansion rate. Entropy production tends to heat up the system, reducing its cooling rate; while expansion does the opposite. Focusing on a small portion of the hydrodynamical medium (with extent in the $x$ and $y$ direction of $\Delta x=\Delta y=0.17$ fm) located at the center of the simulation, we present in Fig. \ref{fig:later_latest_dynamics} the later-time dynamics of the medium, where bulk viscous pressure plays an important role. Figure \ref{fig:later_latest_dynamics}a presents the viscous portions of $T^{\mu\nu}$, sourcing the entropy production of the medium --- given by $\partial_\mu S^\mu=\frac{\pi^{\mu\nu}\pi_{\mu\nu}}{2(\eta/s)(\epsilon+P)}+\frac{\Pi^2}{(\zeta/s)(\epsilon+P)}$ --- depicted in Fig. \ref{fig:later_latest_dynamics}b. Figure \ref{fig:later_latest_dynamics}c provides the expansion rate $\theta=\partial_\mu u^\mu$ at the center of the medium, while the resulting temperature is in Fig. \ref{fig:later_latest_dynamics}d. We present in Appendix \ref{appdx:early-time} the early-time evolution of the medium.

Inspecting Fig. \ref{fig:later_latest_dynamics}a,b it is clear that the entropy production originating from bulk viscous pressure ($\Pi$) is becoming increasingly important relative to shear viscous pressure ($\pi^{\mu\nu}$), culminating in the region where $\frac{\zeta}{s}(T)$ peaks. Whether or not this entropy production is converted into a temperature increase depends on the expansion rate in Fig. \ref{fig:later_latest_dynamics}c. At first (i.e. for $1.5 \lesssim \tau-\tau_0\lesssim 4$ fm/$c$), the expansion rates are similar for all three media in Fig. \ref{fig:later_latest_dynamics}c, and the entropy production rate in Fig. \ref{fig:later_latest_dynamics}b, though significant for the medium with $\zeta/s$, is insufficient to substantially change the temperature profile in Fig. \ref{fig:later_latest_dynamics}d, partly explained by $T\propto S^{1/3}$. Later behavior (at $\tau-\tau_0\gtrsim 4$ fm/$c$) shows that the medium with $\zeta/s$ has significant entropy production, and that the expansion rate of the medium with $\zeta/s$ gradually becomes slower than the media with solely $\eta/s$. The combination of these two effects at $\tau-\tau_0\gtrsim 4$ fm/$c$ results in a slower cooling rate of the medium with specific bulk viscosity --- hence, $T_{\rm sw}$ is reached at a later time compared to media without $\zeta/s$. Similar dynamics occur at top RHIC energy. 

\begin{figure}[!h]
\begin{center}
\includegraphics[width=0.495\textwidth]{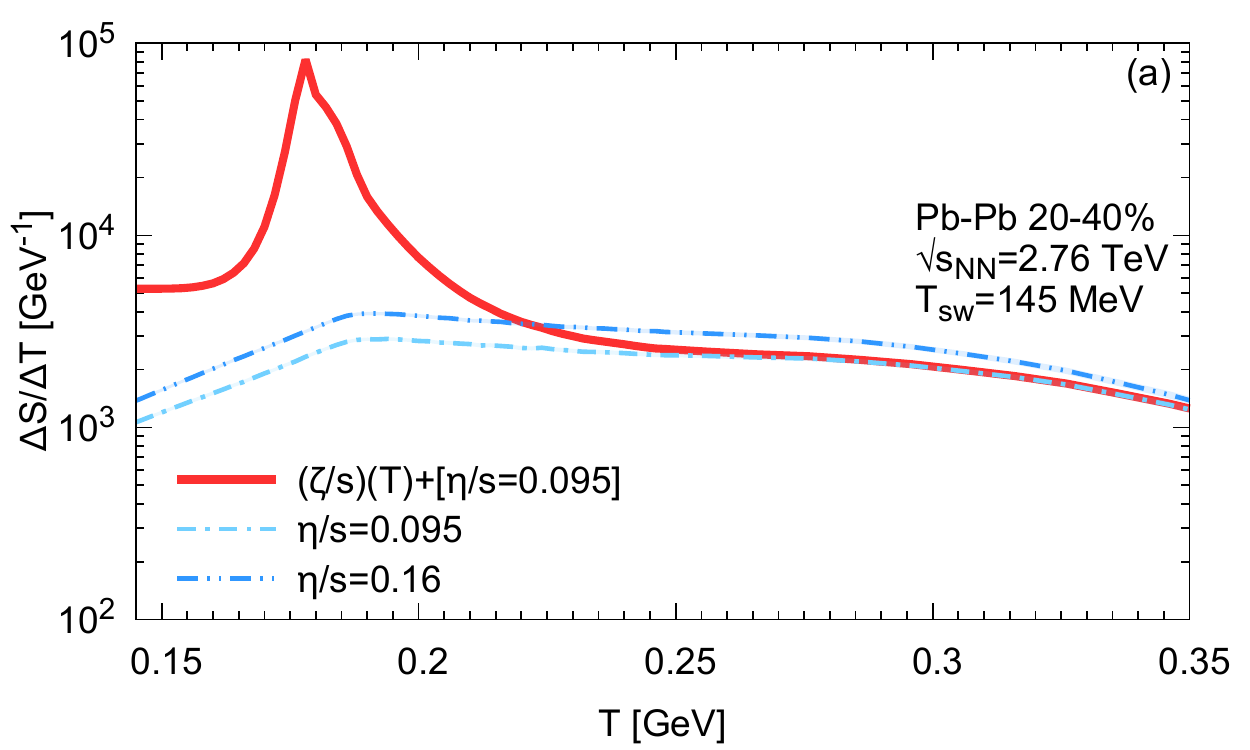}
\includegraphics[width=0.495\textwidth]{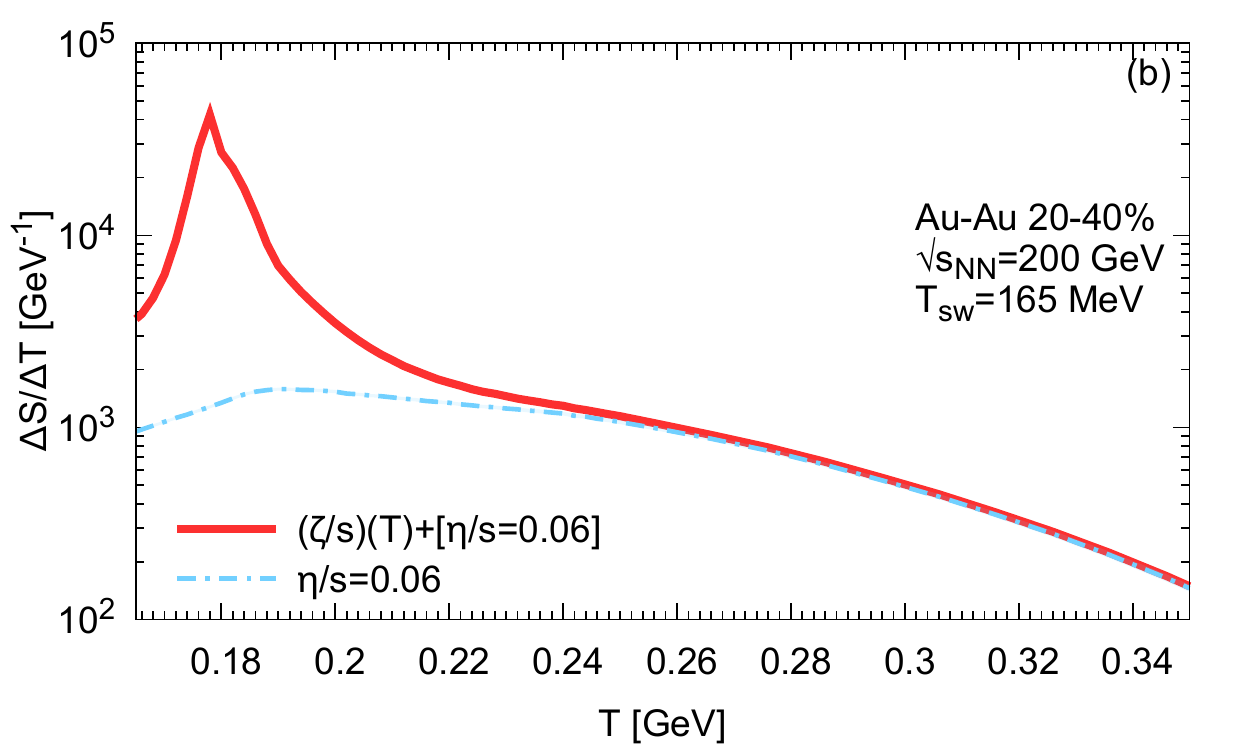}
\includegraphics[width=0.495\textwidth]{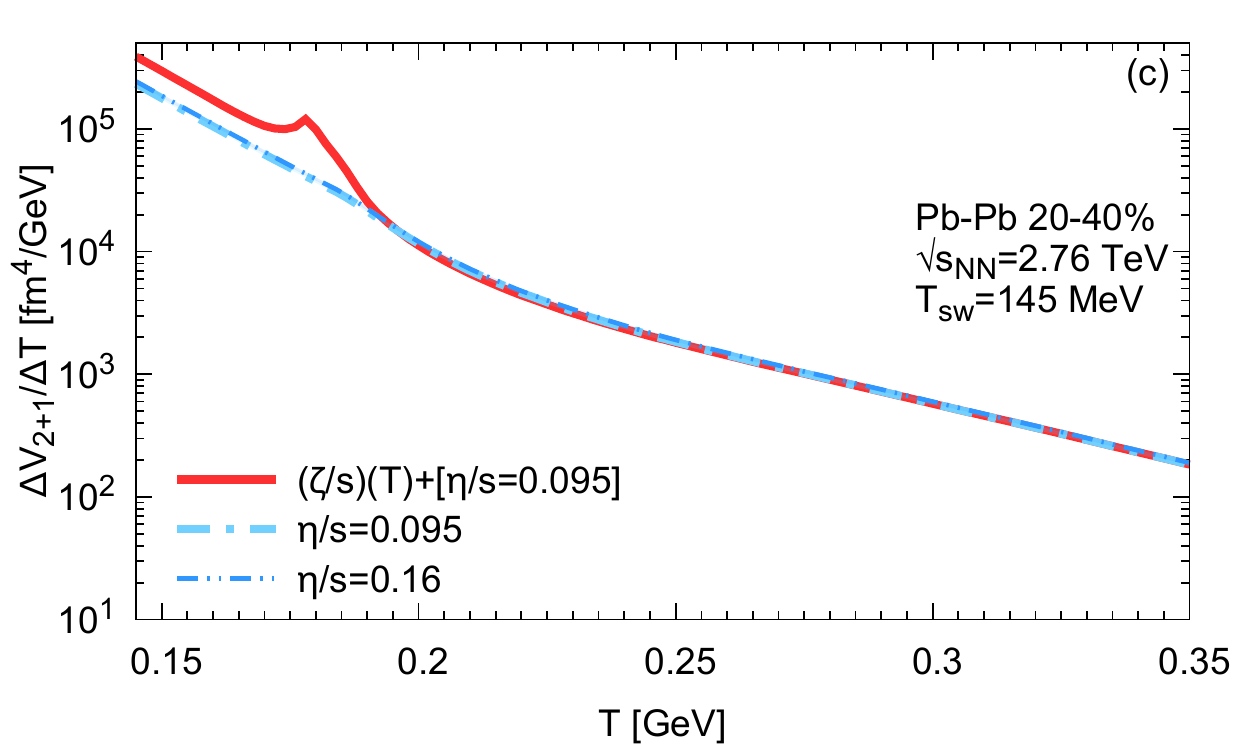}
\includegraphics[width=0.495\textwidth]{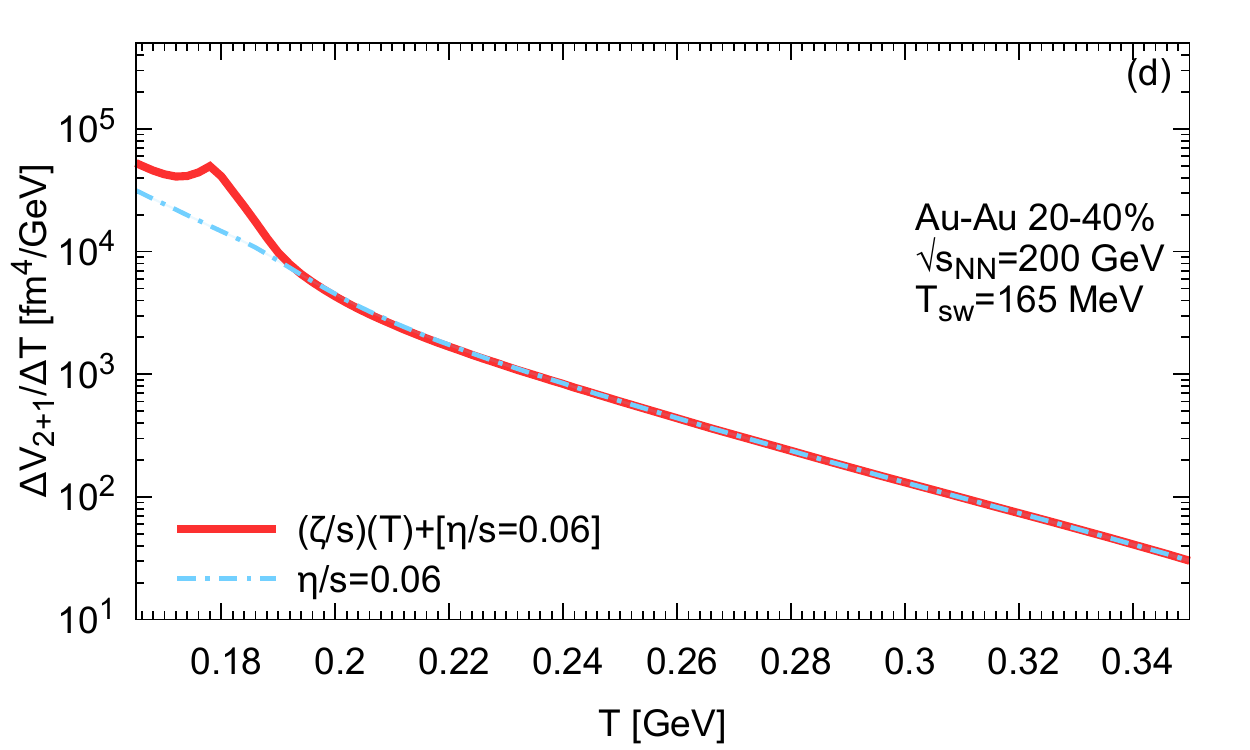}
\end{center}
\caption{(Color online) Development of the entropy within a temperature bin at LHC (a) and at RHIC (b). Hydrodynamical spacetime volume within a temperature bin at LHC (c) and at RHIC (d). The details regarding the way these quantities were computed in presented in Eq.~(\ref{eq:entropy_vol_prod_bin_T}).}
\label{fig:Ent_Vol_vs_Temp}
\end{figure}

For both RHIC and LHC hydrodynamical simulations, the presence of $\zeta/s$ in the hydrodynamical evolution, which drives larger temperatures and smaller expansion rates at late times, is also responsible for generating larger spacetime volumes at a fixed temperature $T$ --- for $T_{\rm sw}<T\lesssim 0.18$ GeV in our calculations --- as depicted in Fig. \ref{fig:Ent_Vol_vs_Temp}. Figure \ref{fig:Ent_Vol_vs_Temp} also depicts the entropy production as a function of temperature. The reduction in radial flow at $T\lesssim 0.18$ GeV shown in Ref.~\cite{Paquet:2015lta} will not be repeated here, as the same hydrodynamical simulation are employed in both calculations. The entropy production per temperature bin and the associated volume shown in Fig.~\ref{fig:Ent_Vol_vs_Temp} are computed via:
\begin{eqnarray}
\frac{\Delta S}{\Delta T} &\equiv& \frac{1}{\Delta T} \frac{1}{N_{ev}}\sum^{N_{ev}}_{i=1}\left\langle \partial_\mu S^\mu\right\rangle_{T}\nonumber\\
\frac{\Delta V_{2+1}}{\Delta T} &\equiv& \frac{1}{\Delta T} \frac{1}{N_{ev}}\sum^{N_{ev}}_{i=1}\left\langle 1 \right\rangle_{T}\nonumber\\	
\langle A \rangle_{T} &\equiv&\int \tau d\tau dy dx B(T) A\nonumber\\
B(T)&\equiv&\left\{ \begin{array}{rl}
                                  1 & T(\tau,x,y) \in [T_j-\frac{\Delta T}{2}, T_j+\frac{\Delta T}{2})\\
                                  0 & {\rm otherwise}
                               \end{array}
                        \right.
\label{eq:entropy_vol_prod_bin_T}
\end{eqnarray}
where $A$ is any quantity binned in temperature, $T_j$ is the temperature in the center of the bin $j$, with the temperature bin-width being $\Delta T$. The area under the curves in Fig. \ref{fig:Ent_Vol_vs_Temp} gives the total entropy production and volume occupied by a medium with a given $\zeta/s$ and $\eta/s$. 

Figure \ref{fig:Ent_Vol_vs_Temp}c,d depicts that the presence of bulk viscosity in the hydrodynamical medium generates a larger spacetime volume ($\Delta V_{2+1}/\Delta T$) at lower temperatures bins. This larger spacetime volume is a consequence of larger entropy production (see $\Delta S/\Delta T$ in Fig. \ref{fig:Ent_Vol_vs_Temp}a,b) and smaller expansion rate, leading to a smaller radial flow at $T_{\rm sw}<T\lesssim 0.18$ GeV as shown in Ref.~\cite{Paquet:2015lta}. The effects of bulk viscosity on the evolution presented above are concordant with the findings of Ref.~\cite{Paquet:2015lta}, where a softer photon spectrum was obtained once $\zeta/s$ was included in the hydrodynamcial evolution. This larger spacetime volume increases the dilepton invariant mass yield as seen later.\footnote{A keen reader may anticipate that the invariant mass dilepton yield, obtained from $\frac{dN}{dM}=\int dy d^2 q_\perp \int d^4 x \frac{d^4 R}{d^4 q}$, is a Lorentz invariant quantity --- and so is $\frac{dN}{dMdy}$ in our boost-invariant simulations --- and thus it will not be (directly) sensitive to radial flow, and will be far more sensitive to the spacetime volume at given temperature.}

\begin{figure}[!h]
\begin{center}
\includegraphics[width=0.495\textwidth]{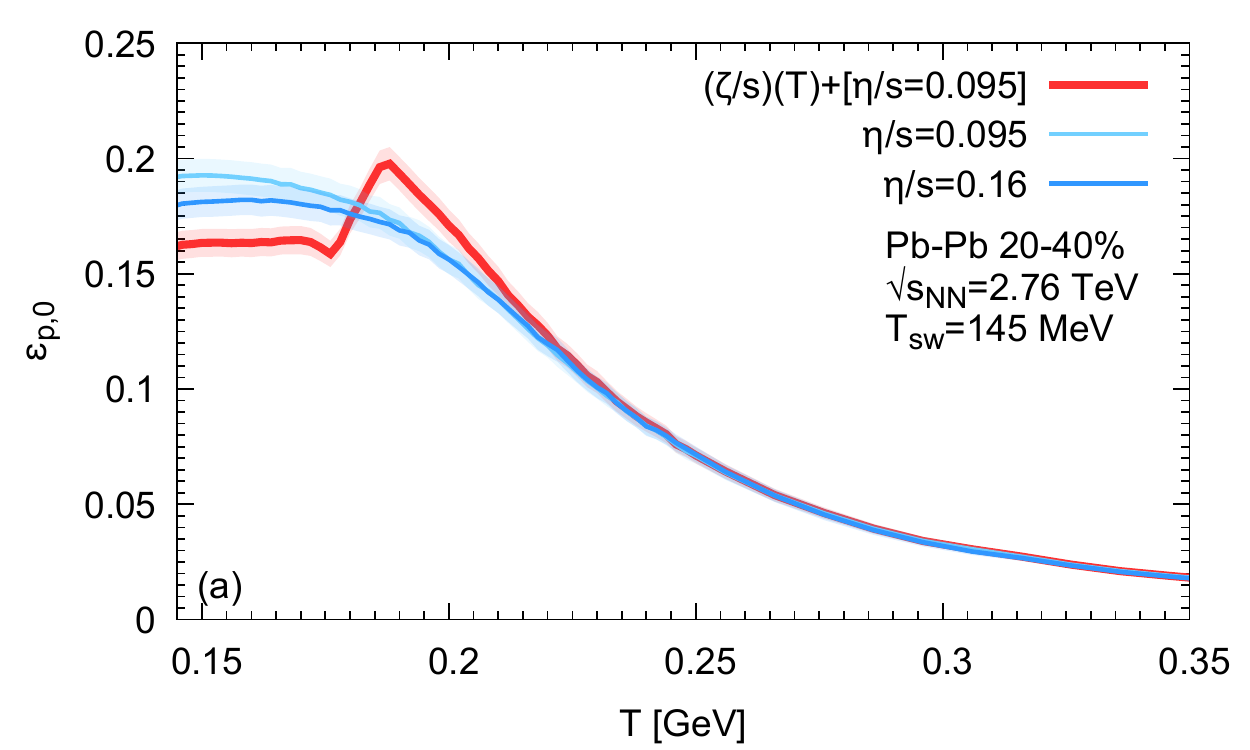}
\includegraphics[width=0.495\textwidth]{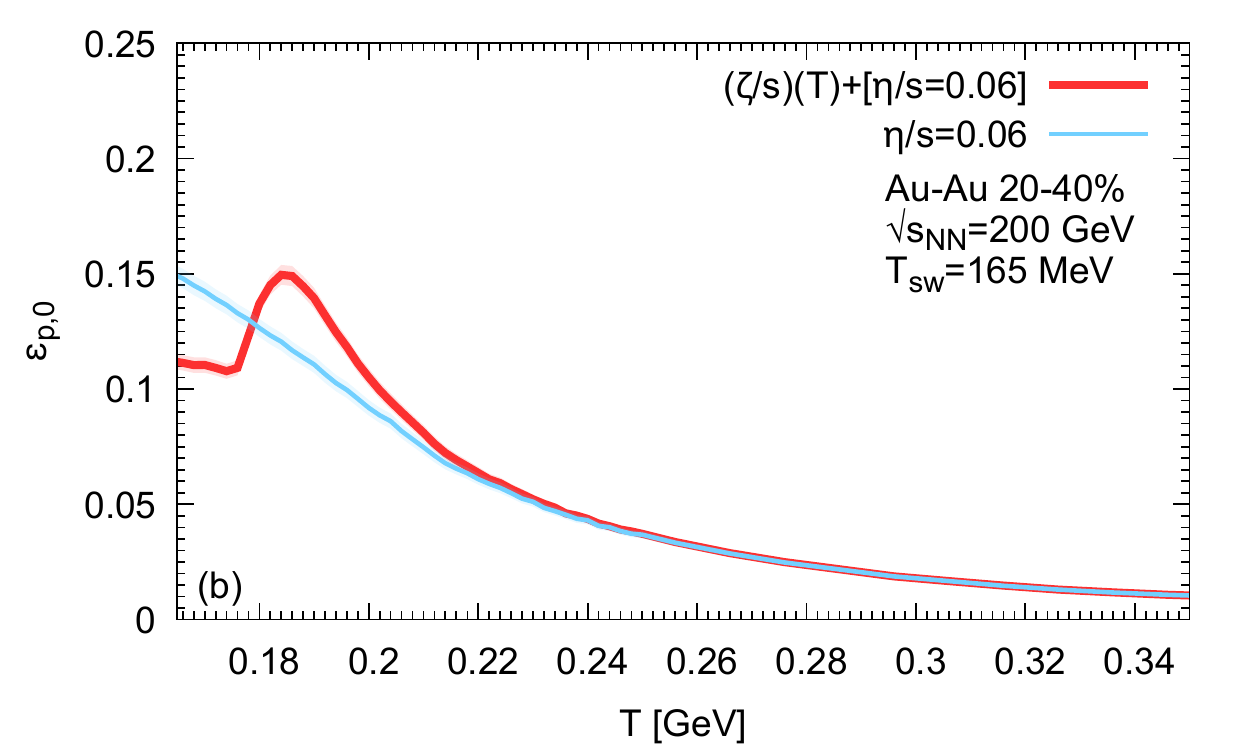}
\includegraphics[width=0.495\textwidth]{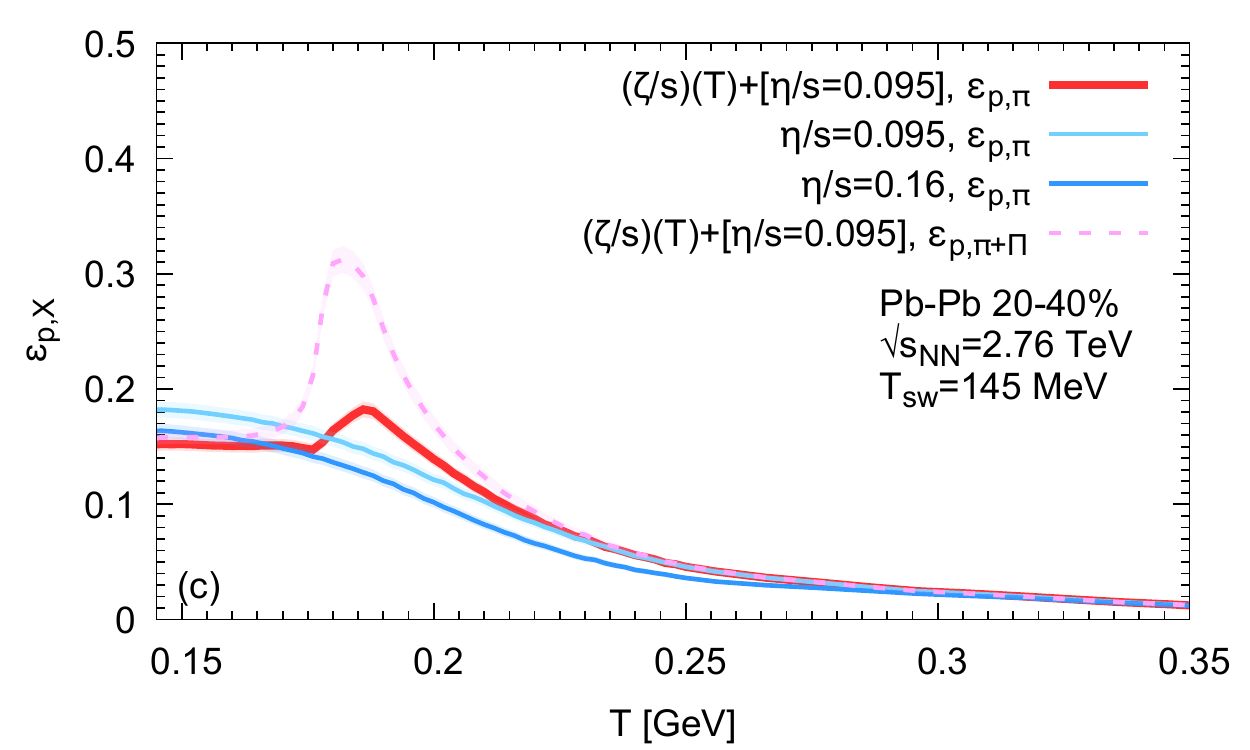}
\includegraphics[width=0.495\textwidth]{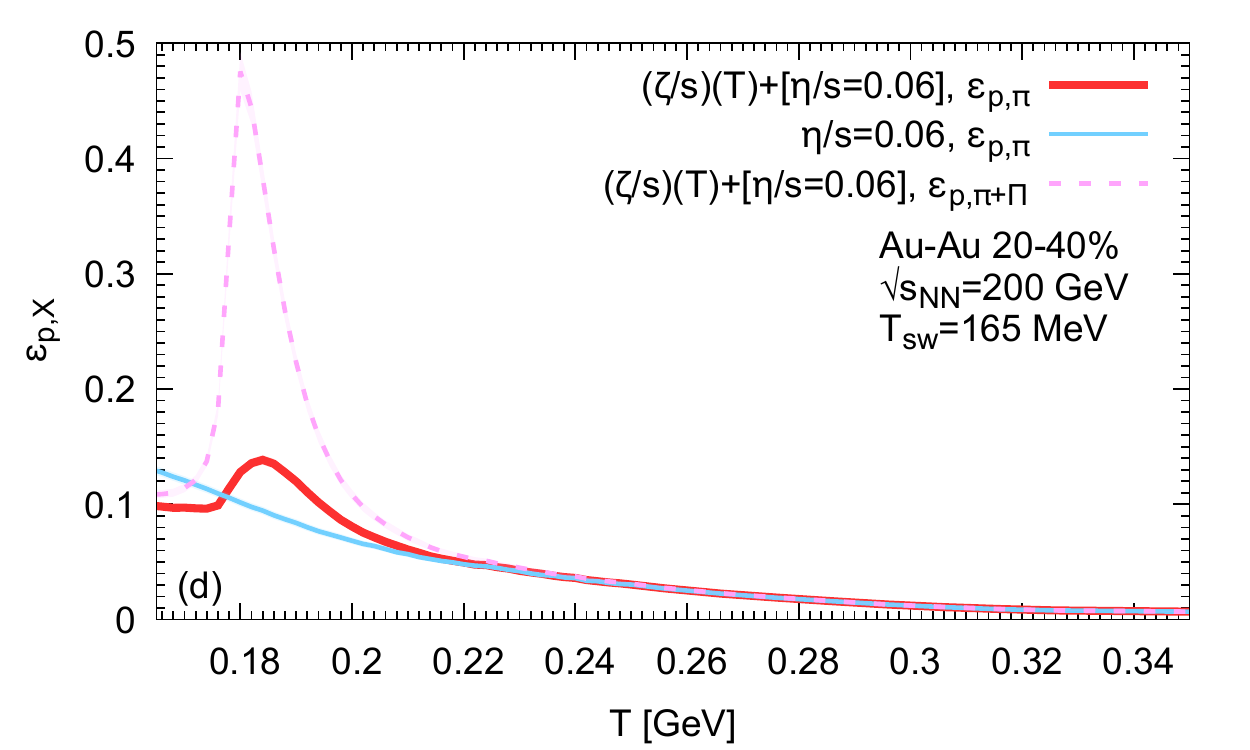}
\end{center}
\caption{(Color online) Development of the hydrodynamical momentum anisotropy as a function of temperature at LHC and at RHIC.}
\label{fig:T_munu_aniso_vs_temp}
\end{figure}

Another important aspect of the evolution is the development of the hydrodynamical momentum anisotropy, which affects $v_2$ of particle species. Figure~\ref{fig:T_munu_aniso_vs_temp} compares how the hydrodynamical momentum anisotropy is developed at RHIC and LHC energies. The hydrodynamical momentum anisotropy is computed as follows:
\begin{eqnarray}
\varepsilon_{p,X}(T)&=&\frac{1}{N_{ev}}\sum^{N_{ev}}_{i=1} \left\{\frac{\sqrt{\left[\left\langle T^{xx}_{X,i} - T^{yy}_{X,i}\right\rangle_{T}\right]^2+\left[2\left\langle T^{xy}_{X,i}\right\rangle_{T}\right]^2}}{\left\langle T^{xx}_{X,i} + T^{yy}_{X,i} \right\rangle_{T}}\right\} \nonumber\\
\end{eqnarray} 
where $\left\langle \cdot \right\rangle_T$ is defined in Eq.~(\ref{eq:entropy_vol_prod_bin_T}). $T^{\mu\nu}_X$ can be $T^{\mu\nu}_0$, $T^{\mu\nu}_{\pi}=T^{\mu\nu}_{0}+\delta T^{\mu\nu}_{\pi}$, or $T^{\mu\nu}_{\pi+\Pi}=T^{\mu\nu}_{0}+\delta T^{\mu\nu}_{\pi}+\delta T^{\mu\nu}_{\Pi}$, with $T^{\mu\nu}_0$, $\delta T^{\mu\nu}_{\pi}$, and $\delta T^{\mu\nu}_{\Pi}$ being defined in Eq.~(\ref{eq:energy-momentum_conserv}). 

Figure \ref{fig:T_munu_aniso_vs_temp} shows an enhancement in $\varepsilon_{p}$\footnote{Referring to $\varepsilon_p$ without specifying $X$ implicitly implies that the statement is valid for all $X$.} due to bulk viscous effects around the temperature where $\zeta/s$ peaks.\footnote{Note that the extra entropy production that is present near the peak of $\zeta/s$, occurring for temperatures between 175--220 MeV, is also correlated with a localized increase in anisotropic flow development in Fig.~\ref{fig:T_munu_aniso_vs_temp}. This phenomenon requires further study.} As temperatures drop, the system with bulk viscosity suppresses $\varepsilon_p$ development and expansion rate. 
There is a non-monotonic temperature profile of $\varepsilon_p$ under bulk viscous pressure shown in Fig. \ref{fig:T_munu_aniso_vs_temp}. A goal of our work is to explore dilepton's sensitivity to this profile. 

We conclude this section by giving a summary of how hadronic observables are affected by the dynamics of bulk viscosity. Computed at the surface of constant $T_{\rm sw}$, hadronic observables are only sensitive to what is happening at that temperature. Bulk viscosity acts on charged hadron $v_2$ in a similar way to shear viscosity, reducing $v_2$. A reduction in hadronic $v_2$ is a reason behind the increase in $\eta/s$ at the LHC --- from 0.095 to 0.16 \cite{Ryu:2015vwa, Ryu:2017qzn}. Furthermore, introducing $\zeta/s$ (without changing $\eta/s$), or increasing $\eta/s$ at vanishing $\zeta/s$ increases particle multiplicity owing to larger volumes at same $T_{\rm sw}$ (or larger entropy) for a medium with higher $\eta/s$, or a medium with both non-vanishing $\zeta/s$ and $\eta/s$, as shown in Refs. \cite{Ryu:2015vwa, Ryu:2017qzn}. Similarities between the effects of bulk and shear viscosity appear to be limited to the aforementioned statement, as the mean transverse momentum $\langle p_T \rangle$ of hadrons is reduced under the influence of bulk viscosity due to reduced expansion rate at late times. Increasing $\eta/s$, for $\zeta/s=0$, generates a $\langle p_T \rangle$ of hadrons that is too high compared to experimental data. Thus, to describe multiplicity, $\langle p_T \rangle$, and $v_2$ of hadrons, bulk viscosity was crucial within the present set of heavy-ion collision simulations.    

As opposed to hadronic observables, dileptons, being electromagnetic probes, should be sensitive to the entire dynamical history of the medium. As such, they may be able to probe the non-monotonic behaviour of $\varepsilon_p$ as a function of $T$ seen in Fig. \ref{fig:T_munu_aniso_vs_temp}. The latter will be discussed in the next section.

\subsection{Effects of bulk viscosity on thermal dileptons}\label{sub_sec:thermal_dileptons} 

As the effects of bulk viscosity on dilepton production are rather intricate, the discussion consists of three subsections. Subsection \ref{subsub_sec:v2_th_LHC} explores the manner in which dilepton $v_2(M)$ is affected by the presence of specific bulk viscosity at LHC collisions energy. Focus is given to the role played by the dilepton yield in obtaining the thermal dilepton $v_2(M)$ --- i.e. dileptons radiated during the hydrodynamical evolution. The thermal $v_2$ is a yield-weighted average of the individual dilepton contributions, whose production rates are described in Sec.~\ref{subsub_sec:qgp_dileptons} and Sec.~\ref{subsub_sec:hm_dileptons}. Subsection \ref{subsub_sec:dR_effects}, summarizes the effects of viscous corrections on dilepton production; a more in-depth discussion can be found in Appendix \ref{appdx:dR_pT}. Subsection \ref{subsub_sec:eps_P} discusses the effects of $\zeta/s$ on the invariant-mass-dependent dilepton yield and $v_2$ at top RHIC collision energy. To explain the results that we have found at RHIC, elements of the hydrodynamical momentum anisotropy detailed in previous section will take center stage, as we unfold their effects on dilepton $v_2(M)$. Finally, a brief study of the sensitivity of our RHIC results to the particlization temperature $T_{\rm sw}$ is presented.  

\subsubsection{Thermal dilepton $v_2(M)$ at the LHC}\label{subsub_sec:v2_th_LHC}

The dilepton production rate changes from partonic to hadronic sources as the temperature decreases. These rates are smoothly interpolated according to Eq.~(\ref{eq:f_QGP}) in the temperature interval 0.184 GeV $< T <$ 0.22 GeV. This interpolation range is chosen to yield a smooth temperature dependence; it should not be misinterpreted as suggesting the existence of hadronic matter up to temperatures above 200 MeV. The lattice-based equation of state used in our work, which controls the cooling rate and  development of hydrodynamic flow in our dynamical simulations, encodes a crossover transition from partonic to hadronic matter at a temperature $T \simeq 0.184$ GeV. For notational simplicity, the lower-temperature dileptons (emitted according to the rate based on hadronic medium sources) will be denoted by ``HM'', while the higher-temperature dileptons (emitted with a rate calculated from partonic sources will be denoted) by ``QGP''. The HM and QGP labels serve only to identify rate formula used to calculate the emissions.

\begin{figure}[!h]
\begin{center}
\includegraphics[width=0.495\textwidth]{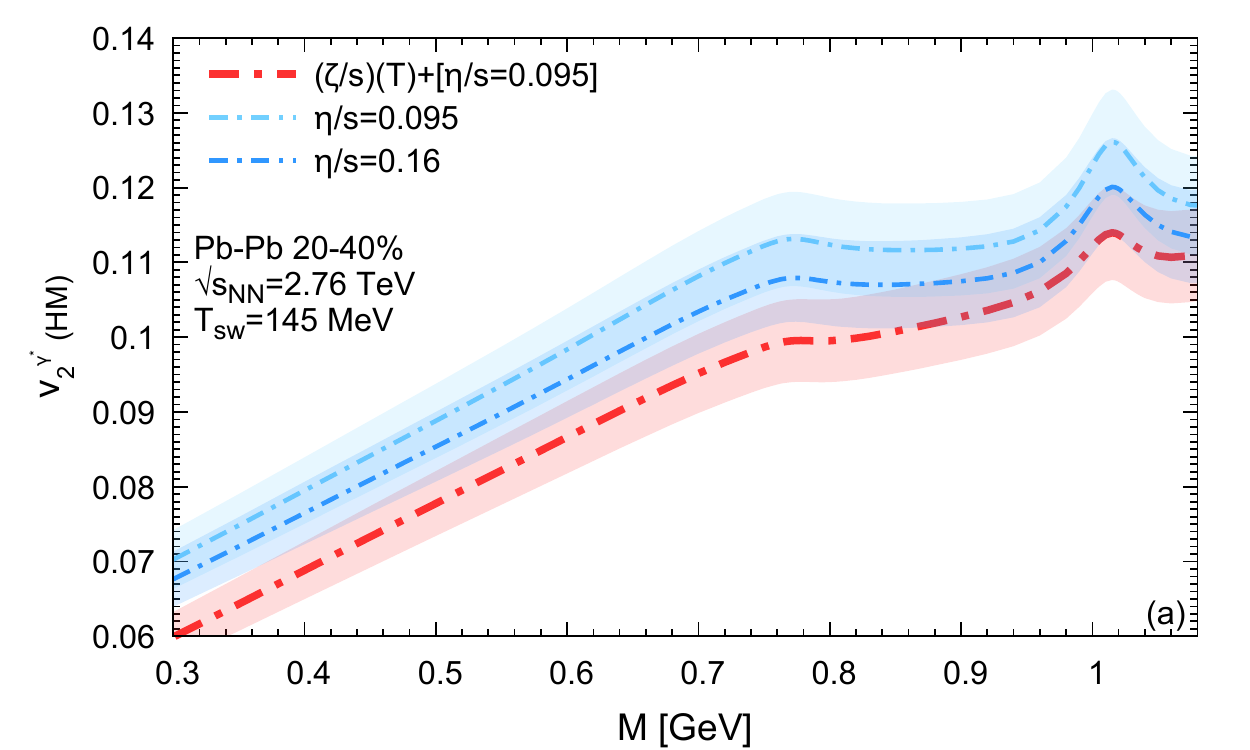}
\includegraphics[width=0.495\textwidth]{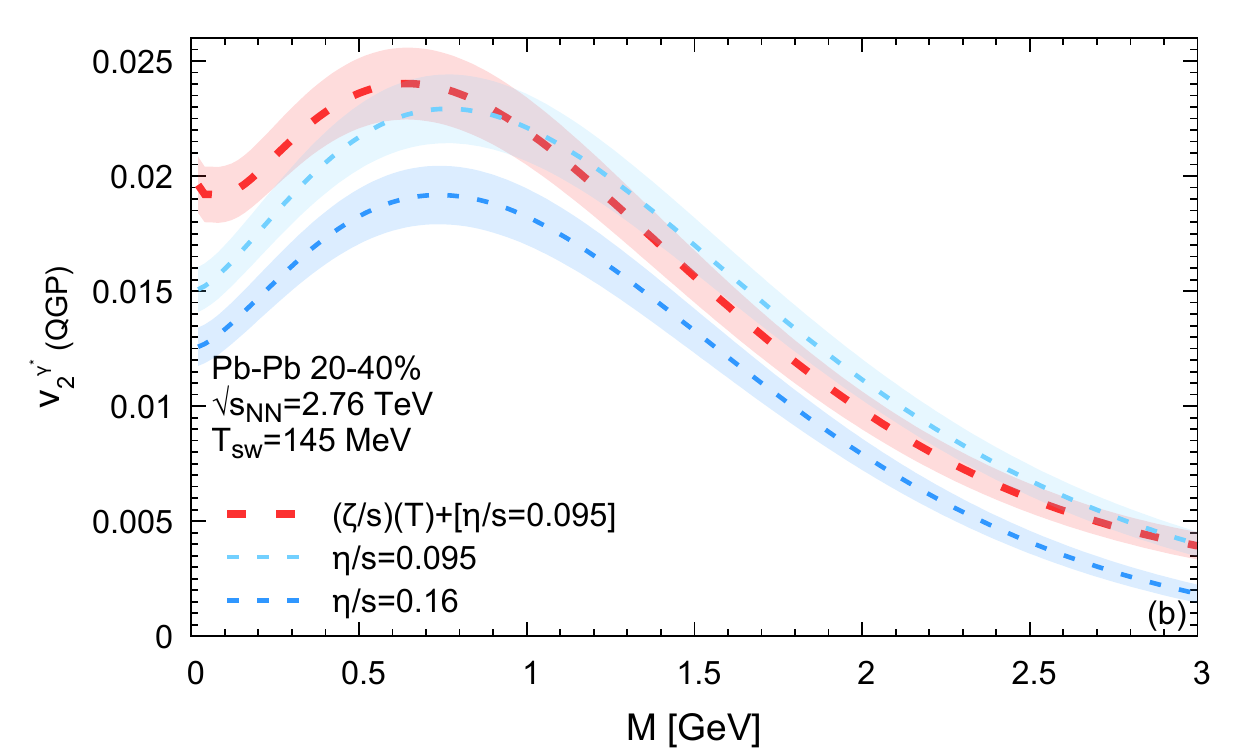}
\end{center}
\caption{(Color online) Invariant mass distribution of $v_2$ for the hadronic (a) and partonic (b) dileptons under the influence of media having different $\eta/s$ as well as a medium with non-zero values for both $\zeta/s$ and $\eta/s$. The definition of what constitutes hadronic (HM) versus partonic (QGP) dilepton radiation is presented in Eq.~(\ref{eq:f_QGP}).}
\label{fig:v2_M_HM_QGP_LHC}
\end{figure}

\begin{figure}[!h]
\begin{center}
\includegraphics[width=0.495\textwidth]{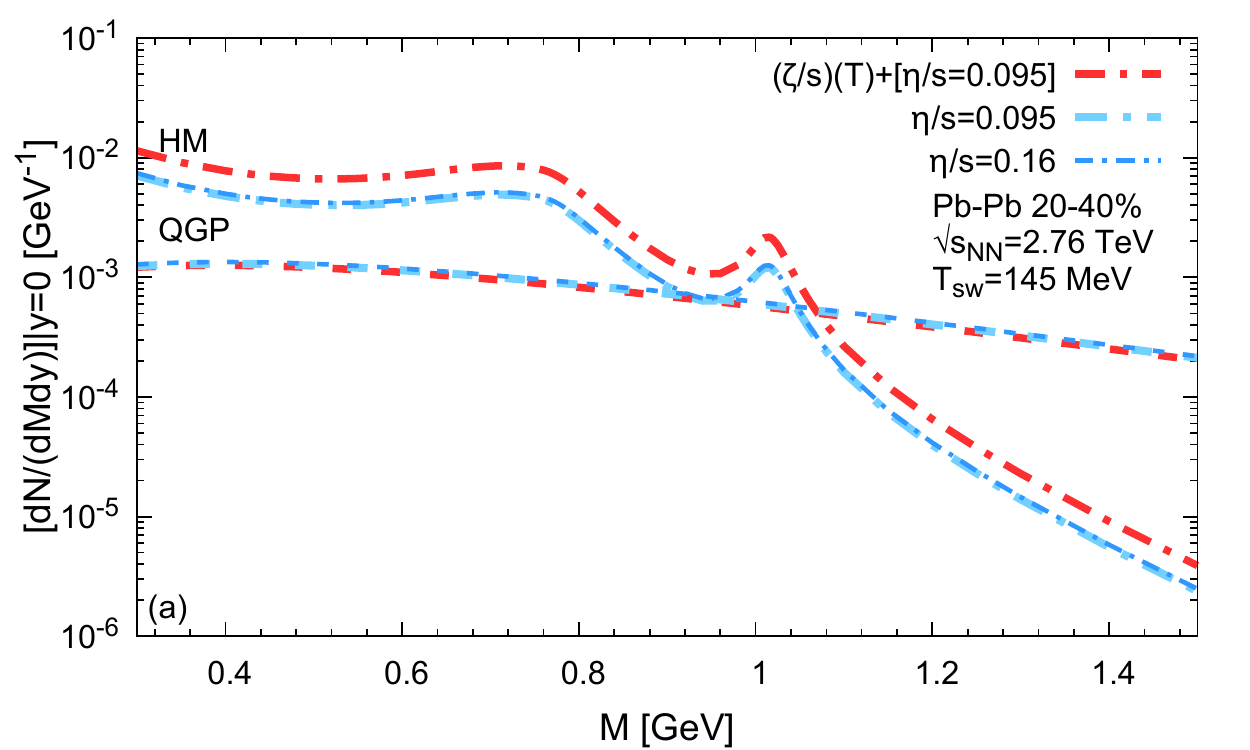}
\includegraphics[width=0.495\textwidth]{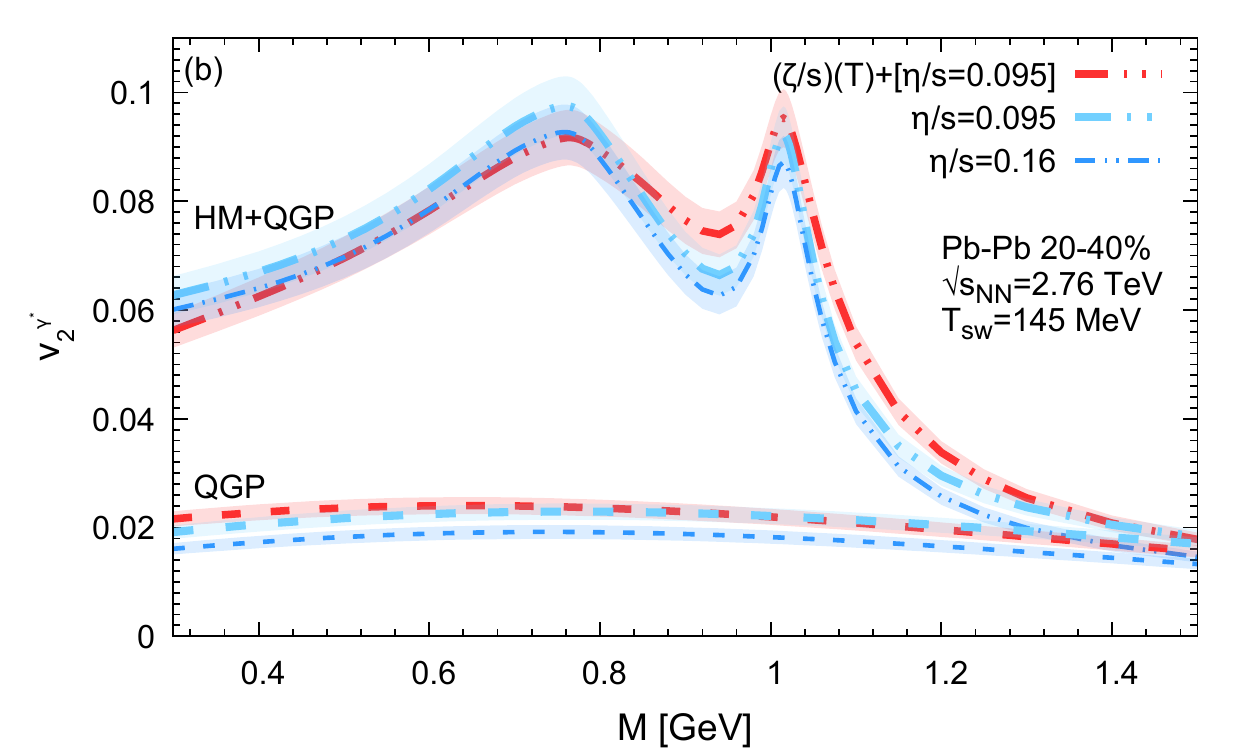}
\end{center}
\caption{(Color online) Invariant mass distribution of dilepton yield (a) and $v_2$ (b) under the influence of media having different $\eta/s$, as well as a medium with both $\zeta/s$ and $\eta/s$.}
\label{fig:yield_v2_M_thermal_LHC}
\end{figure}

Figure \ref{fig:v2_M_HM_QGP_LHC} presents the invariant mass dependence of $v_2$ for the lower (HM) and higher (QGP) temperature contributions to thermal dilepton production including all viscous corrections. In the hadronic sector depicted by Fig.~\ref{fig:v2_M_HM_QGP_LHC}a, increasing the specific shear viscosity and introducing specific bulk viscosity reduces the anisotropic flow of dileptons from the hadronic medium. In the higher temperature (QGP) sector, a more complex pattern emerges due to an interplay between the bulk viscous correction, and due to modifications to the evolution of the medium related to presence of specific bulk viscosity. The effects induced via viscous corrections will be discussed in Sec. \ref{subsub_sec:dR_effects}. Inspecting the total thermal dilepton signal in Fig.~\ref{fig:yield_v2_M_thermal_LHC}, a non-trivial invariant mass dependence in $v_2$ can be noticed. It stems from an interaction between the effects of $\zeta/s$ on dilepton yield of lower/higher temperature HM/QGP sources, and on the $v_2$ of those sources depicted in Fig.~\ref{fig:v2_M_HM_QGP_LHC}.

To understand better the $v_2(M)$ of thermal (HM+QGP) dileptons, focus should be given to the dilepton yield under the influence of various viscous effects. We first look on the $v_2$ at $M>0.8$ GeV. In that region, the dilepton yield goes from being HM dominated to being QGP dominated. Though bulk viscosity decreases the $v_2$ of lower temperature (HM) dileptons relative to any medium without $\zeta/s$, it increases the yield of those dileptons. The invariant mass yield of higher temperature (QGP) dileptons is little affected by the various values of $\eta/s$ and $\zeta/s$ explored in our study. After performing a yield-weighted average to compute the thermal $v_2(M)$ in Fig.~\ref{fig:yield_v2_M_thermal_LHC}b for $M>0.8$ GeV, the increase in the HM dilepton yield dominates over the decrease in the $v_2$ of those dileptons, thus increasing $v_2(M)$ of thermal dileptons within that invariant mass range. For $M\leq 0.8$ GeV, the lower temperature (HM) yield dominates over the higher temperature (QGP) yield; a partial cancellation between the increase in the HM yield and the reduction in the HM $v_2$ is responsible for the thermal $v_2$ result seen in this invariant mass range.

\subsubsection{Effects of viscous corrections on partonic and hadronic dilepton emissions}\label{subsub_sec:dR_effects}

Given that the effects of viscous corrections are rather intricate and that their influence of the thermal dilepton $v_2(M)$ is not large, we summarize the final results here, with details in Appendix \ref{appdx:dR_pT}. In this section the medium with $\zeta/s$ and $\eta/s$ will be used, such that various viscous corrections, presented in Eqs.~(\ref{eq:dR_born_shear},\ref{eq:dR_born_bulk},\ref{eq:delta_Pi_HM}), can be turned on or off. 

\begin{figure}[!h]
\begin{center}
\begin{tabular}{cc}
\includegraphics[width=0.495\textwidth]{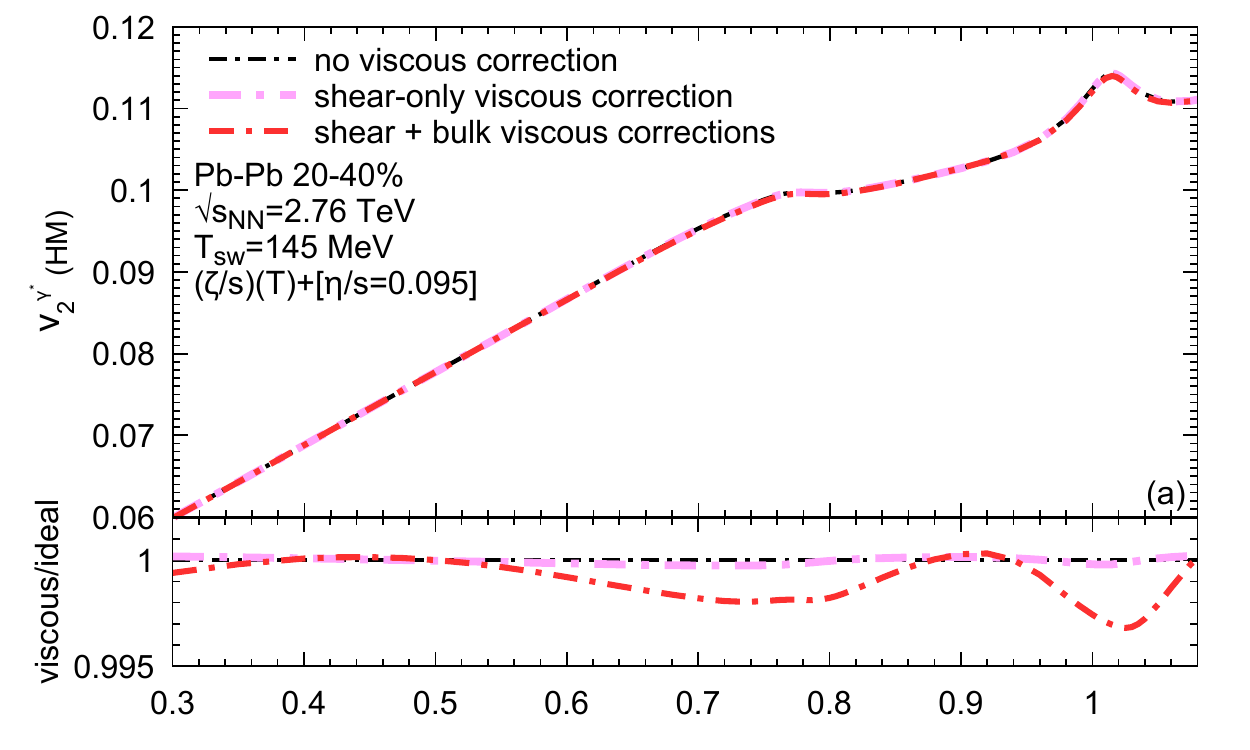} & \includegraphics[width=0.495\textwidth]{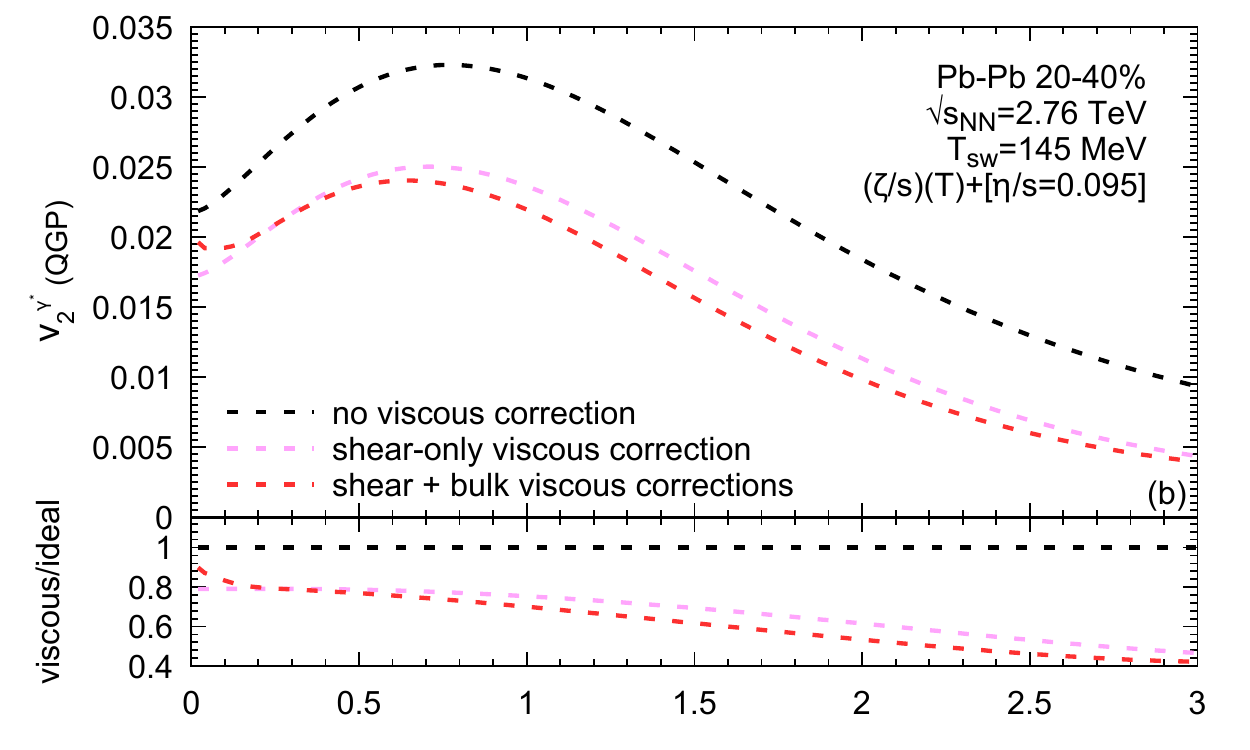}\\
\includegraphics[width=0.515\textwidth]{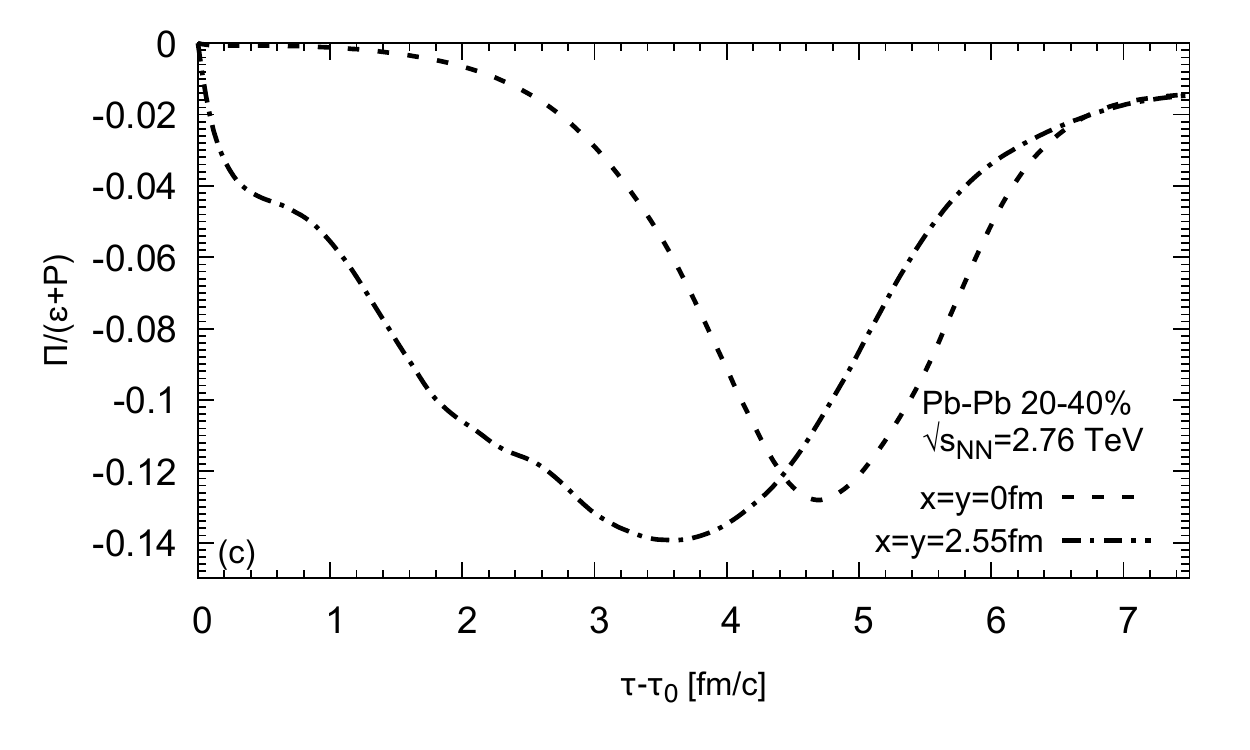} & \includegraphics[width=0.495\textwidth]{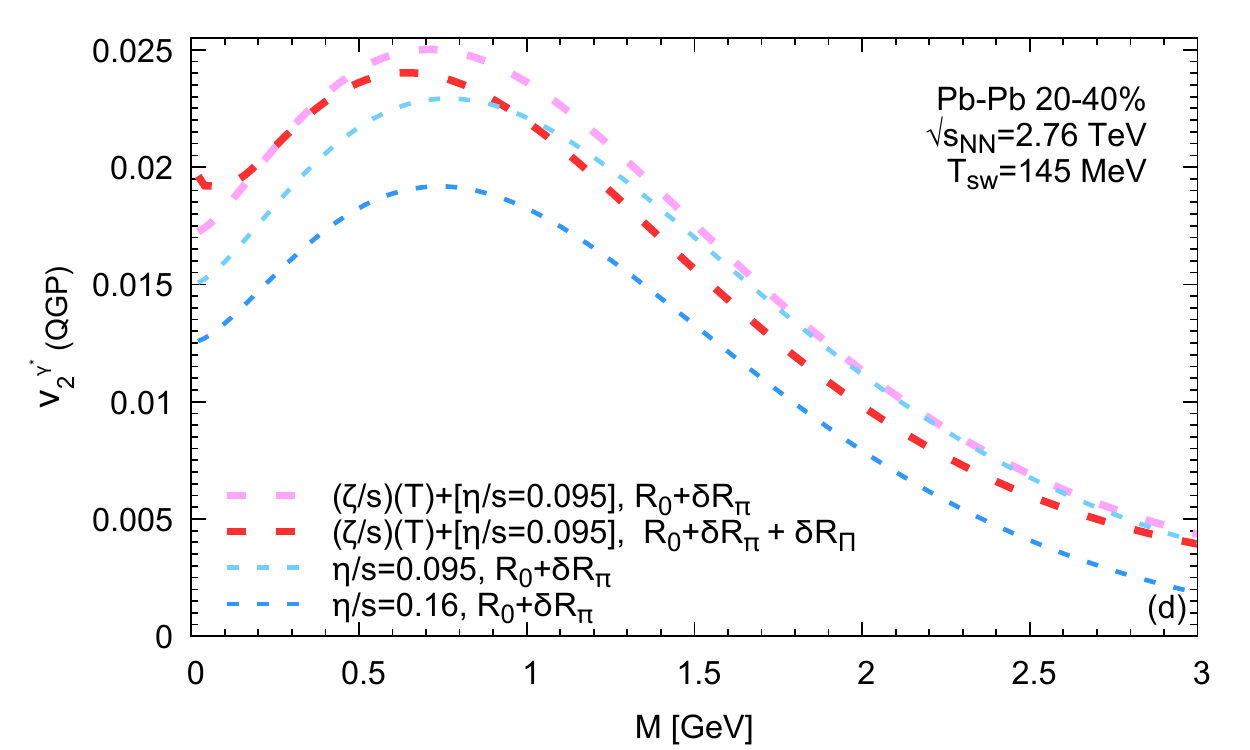}
\end{tabular}
\end{center}
\caption{(Color online) Effects of viscous corrections to the dilepton emission rate on the invariant mass distribution of $v_2$ for the hadronic (a) and partonic (b) dileptons [see Eq.~(\ref{eq:f_QGP})]. Note that in the top panel of (a), the black dash-dotted line is covered by the pink and red lines of the same type, and ones needs to look at the ratio between the viscous over the ideal HM dilepton $v_2$, presented in the bottom panel of (a), to tell those curves apart. (c) Enthalpy density normalized bulk viscous pressure at two locations in the x-y plane. (d) Comparing $v_2$ of partonic dileptons from different media with without (and with) $\delta R_{\Pi}$. }
\label{fig:dR_effects_on_M}
\end{figure}

Figure \ref{fig:dR_effects_on_M} displays the invariant mass distribution of dileptons as affected by bulk and shear viscous corrections explored herein. In Fig. \ref{fig:dR_effects_on_M}a, we show that the effects of bulk and shear viscous corrections on HM dileptons are small using the viscous corrections in Eq.~(\ref{eq:delta_Pi_HM}). A similar statement also holds for the dilepton yield. Thus, $v_2(M)$ of HM dileptons is mostly sensitive to changes in the temperature and the fluid flow $u^\mu$ profile, due to the presence of $\pi^{\mu\nu}$ and $\Pi$ within the hydrodynamical equations of motion. The HM dilepton $v_2(M)$ is not directly sensitive to the dissipative degrees of freedom themselves.   

In Fig. \ref{fig:dR_effects_on_M}b the increase in the $v_2(M)$ of QGP dileptons at low invariant masses, under the influence of bulk viscosity, mirrors what is expected from the energy dependence of the bulk correction $\delta n_{\bf k} \propto \frac{\Pi}{\varepsilon+P}\left(\frac{E_{\bf k}}{T}-\frac{m^2_{q,\bar{q}}}{E_{\bf k}T}\right)$ in Eq.~(\ref{eq:bulk_deltan_1}) which changes sign as $E_{\bf k}$ (or invariant mass) increases, while $\frac{\Pi}{\varepsilon+P}$ is typically negative\footnote{This was also shown in earlier hydrodynamical calculations, e.g.~\cite{Song:2009rh}.} as is displayed in Fig.~\ref{fig:dR_effects_on_M}c. Although  $\varepsilon_p$ increases faster for the simulation with $\zeta/s$ (relative to those without it) as temperature decreases, $v_2$ doesn't always follow this behavior, given $\delta n_{\bf k} \propto \frac{\Pi}{\varepsilon+P}\left(\frac{E_{\bf k}}{T}-\frac{m^2_{q,\bar{q}}}{E_{\bf k}T}\right) $, or equivalently $\delta R_{\Pi}$. Removing the effects of our bulk viscous correction $\delta R_{\Pi}$, we see in Fig. \ref{fig:dR_effects_on_M}d that the $v_2$ of high temperature (QGP) dileptons increases for a medium having $\zeta/s$ relative to media without it, following more closely the behavior of $\varepsilon_p$ from Fig. \ref{fig:T_munu_aniso_vs_temp}c assuming the high temperature radiation is not switched off through Eq.~(\ref{eq:f_QGP}).\footnote{The matching between dilepton $v_2(M)$ and $\varepsilon_p$, without $\delta R_{\Pi}$, also respects the assumption that high $M$ dileptons are predominatly emitted at high temperatures, thus differences in $\varepsilon_p$ --- and $v_2(M)$ --- between different media are small. As one goes to lower temperatures, thus lower $M$, a gap develops between the simulation with $\zeta/s$ and without $\zeta/s$ in both $\varepsilon_p$ (see Fig. \ref{fig:T_munu_aniso_vs_temp}c) and $v_2(M)$ (see Fig. \ref{fig:dR_effects_on_M}d).} The invariant mass yield is practically insensitive to viscous corrections.

\subsubsection{Exploring the $v_2$ at the RHIC and the LHC though the hydrodynamical momentum anisotropy}\label{subsub_sec:eps_P} 
\begin{figure}[!h]
\begin{center}
\includegraphics[width=0.495\textwidth]{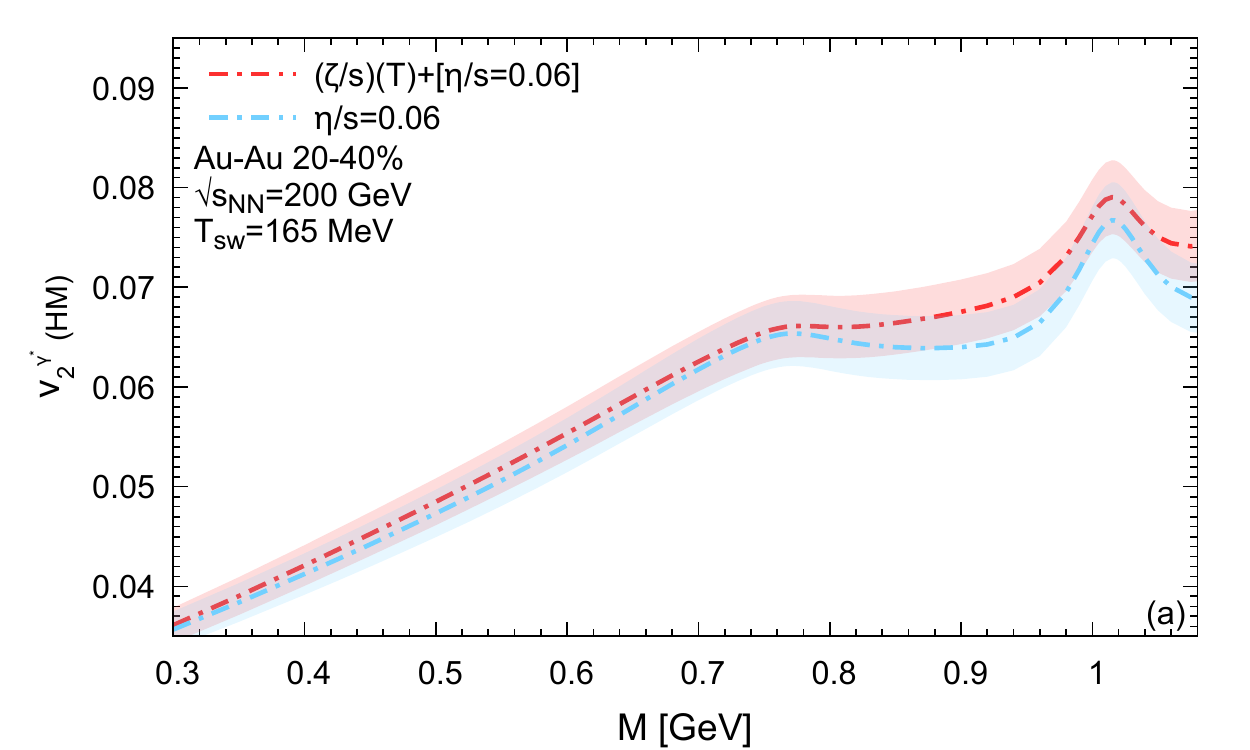}
\includegraphics[width=0.495\textwidth]{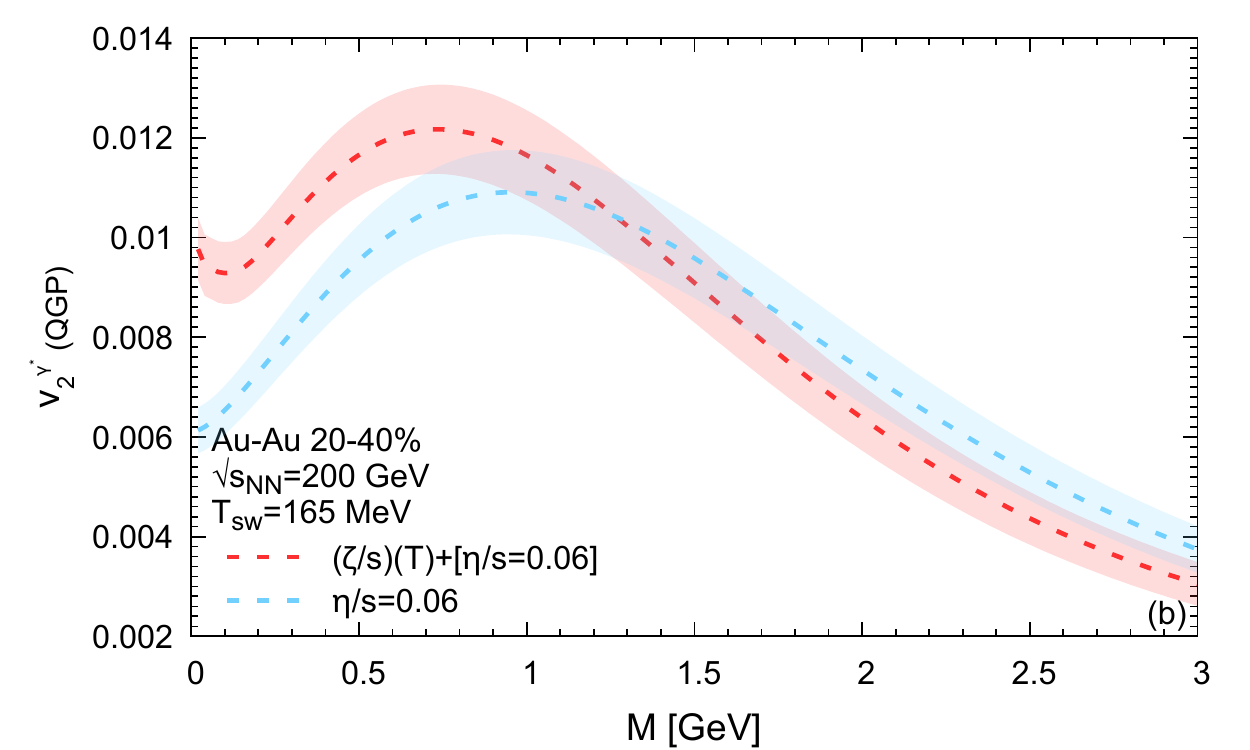}
\end{center}
\caption{(Color online) Invariant mass distribution of $v_2$ for the hadronic (a) and partonic (b) dileptons under the influence of specific bulk viscosity, keeping $\eta/s=0.06$ throughout. Equation (\ref{eq:f_QGP}) is used to distinguish between hadronic (HM) versus partonic (QGP) dilepton rates.}
\label{fig:v2_M_HM_QGP_RHIC}
\end{figure}
\begin{figure}
\begin{center}
\includegraphics[width=0.495\textwidth]{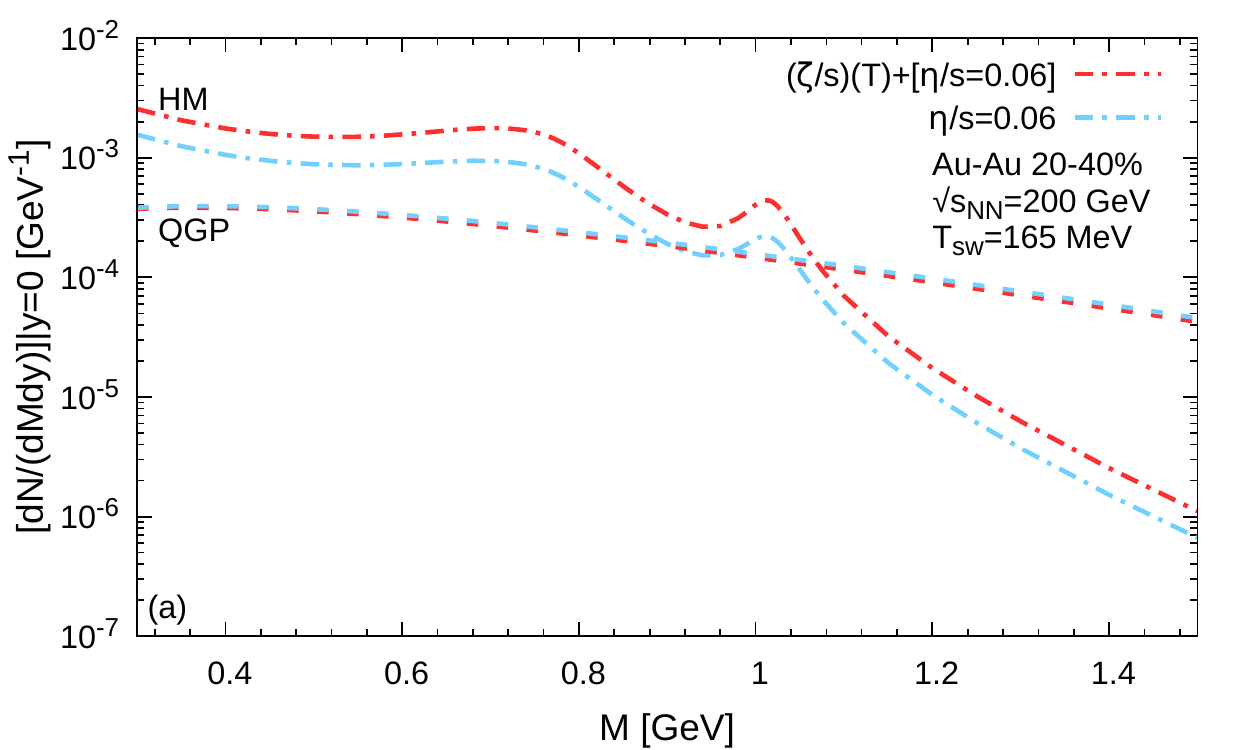}
\includegraphics[width=0.495\textwidth]{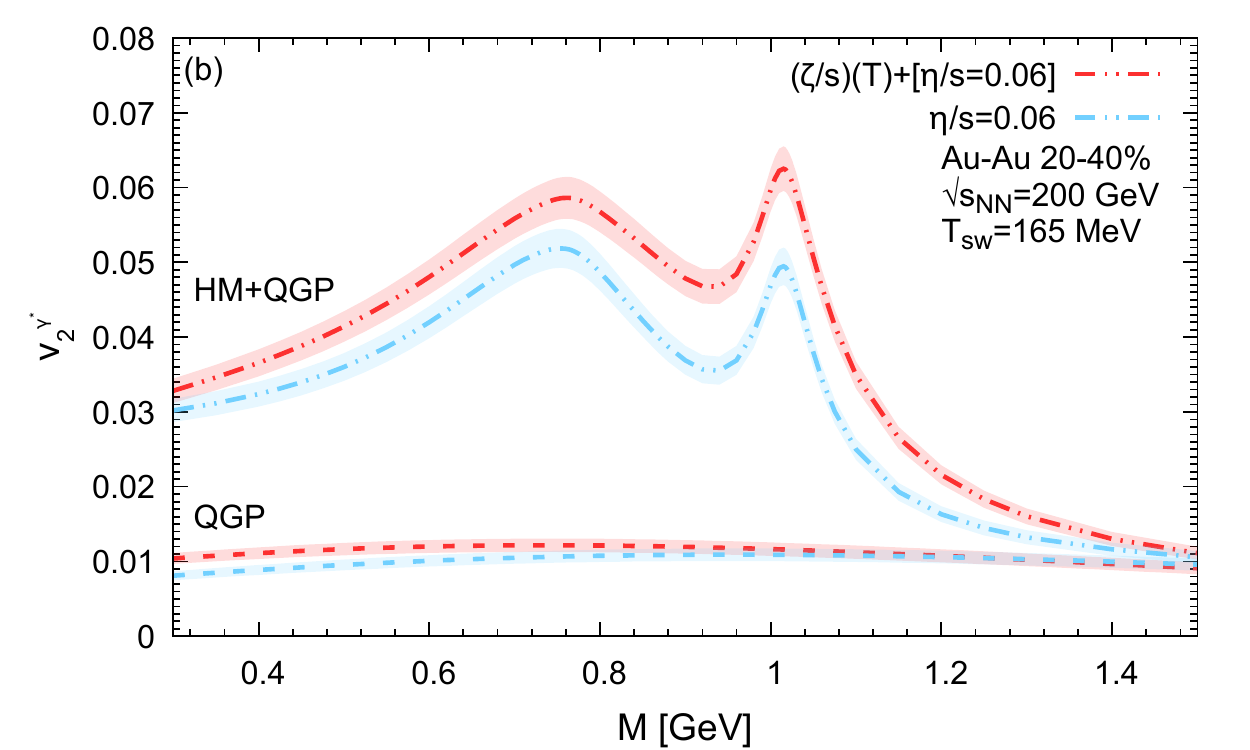}
\end{center}
\caption{(Color online) Invariant mass distribution of dilepton yield (a) and $v_2$ (b) under the influence of media with and without bulk viscosity.}
\label{fig:yield_v2_M_thermal_RHIC}
\end{figure}

The dilepton $v_2(M)$ of higher (QGP) and lower (HM) temperature sources at RHIC are shown in Fig.~\ref{fig:v2_M_HM_QGP_RHIC}, where an interesting behavior is seen. While $v_2(M)$ of higher temperature (QGP) dileptons behaves similarly across the two collision energies we have studied, the $v_2(M)$ for dileptons radiation at lower (HM) temperatures behaves differently: its anisotropic flow appears to be modestly increased under the influence of $\zeta/s$. This slight increase is enhanced once the two contributions are combined into thermal dileptons (see Fig.~\ref{fig:yield_v2_M_thermal_RHIC}b) --- for reasons given in subsection \ref{subsub_sec:v2_th_LHC}. The main origin for the increase seen in the thermal (HM+QGP) $v_2(M)$ depicted in Fig.~\ref{fig:yield_v2_M_thermal_RHIC}b comes from the increase in the yield of HM dileptons (see Fig.~\ref{fig:yield_v2_M_thermal_RHIC}a), stemming from the larger volume at $T\lesssim 0.18$ GeV seen in Fig. \ref{fig:Ent_Vol_vs_Temp}d. The latter originates from a larger entropy production of the medium with $\zeta/s$ and a reduction in expansion rate $\theta$ at late times. Therefore, even if the color order of the curves in Fig. \ref{fig:v2_M_HM_QGP_RHIC}a was inverted, the thermal $v_2(M)$ would still be increased owing to the increase in yield. With this clarification in mind, we now explore Fig. \ref{fig:v2_M_HM_QGP_RHIC} in more detail. 

As dileptons are emitted throughout the entire history of the evolution, the size of the spacetime volume (i.e. $\Delta V_{2+1}/\Delta T$ in Fig. \ref{fig:Ent_Vol_vs_Temp}d) present under the different temperature bins must be considered, to appreciate the extent to which features seen in $\varepsilon_p(T)$, e.g. the enhancement generated around the peak of $\zeta/s$ in Figs. \ref{fig:T_munu_aniso_vs_temp}a and \ref{fig:T_munu_aniso_vs_temp}b, translate onto $v_2(M)$ of lower temperature (HM) dileptons. Given that HM dileptons in our calculations are not particularly sensitive towards viscous corrections to their emission rates, we weigh each temperature bin of the inviscid $\varepsilon_{p,0}(T)$ by the volume $\Delta V_{2+1}$ under that bin. Thus, we compute the following quantity: 
\begin{eqnarray}
\varepsilon_{p,0}(\tau)&=&\frac{1}{N_{ev}}\sum^{N_{ev}}_{i=1} \left\{\frac{\sqrt{\left[\left\langle T^{xx}_{0,i} - T^{yy}_{0,i}\right\rangle_{\tau}\right]^2+\left[2\left\langle T^{xy}_{0,i}\right\rangle_{\tau}\right]^2}}{\left\langle T^{xx}_{0,i} + T^{yy}_{0,i}\right\rangle_{\tau}}\right\}\nonumber\\
\langle A\rangle_{\tau}&=&\int^{\tau}_{\tau_0} \tau' d\tau' dy dx (1-f_{QGP}) \Theta\left(T-T_{\rm sw}\right) A
\end{eqnarray} 
where $A$ is any quantity to be integrated over $\tau$, $\Theta$ is a Heaviside function, while $f_{QGP}$ is defined in Eq.~(\ref{eq:f_QGP}).
\begin{figure}[!h]
\begin{center}
\includegraphics[width=0.495\textwidth]{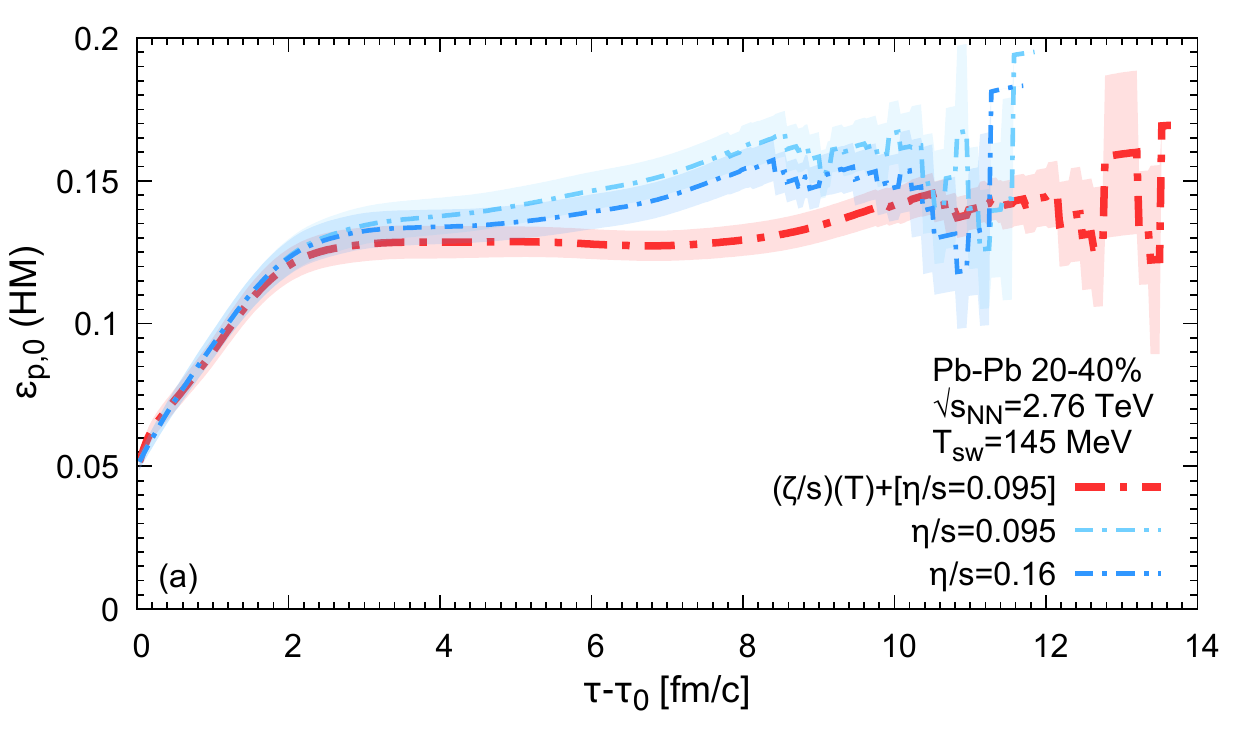}
\includegraphics[width=0.495\textwidth]{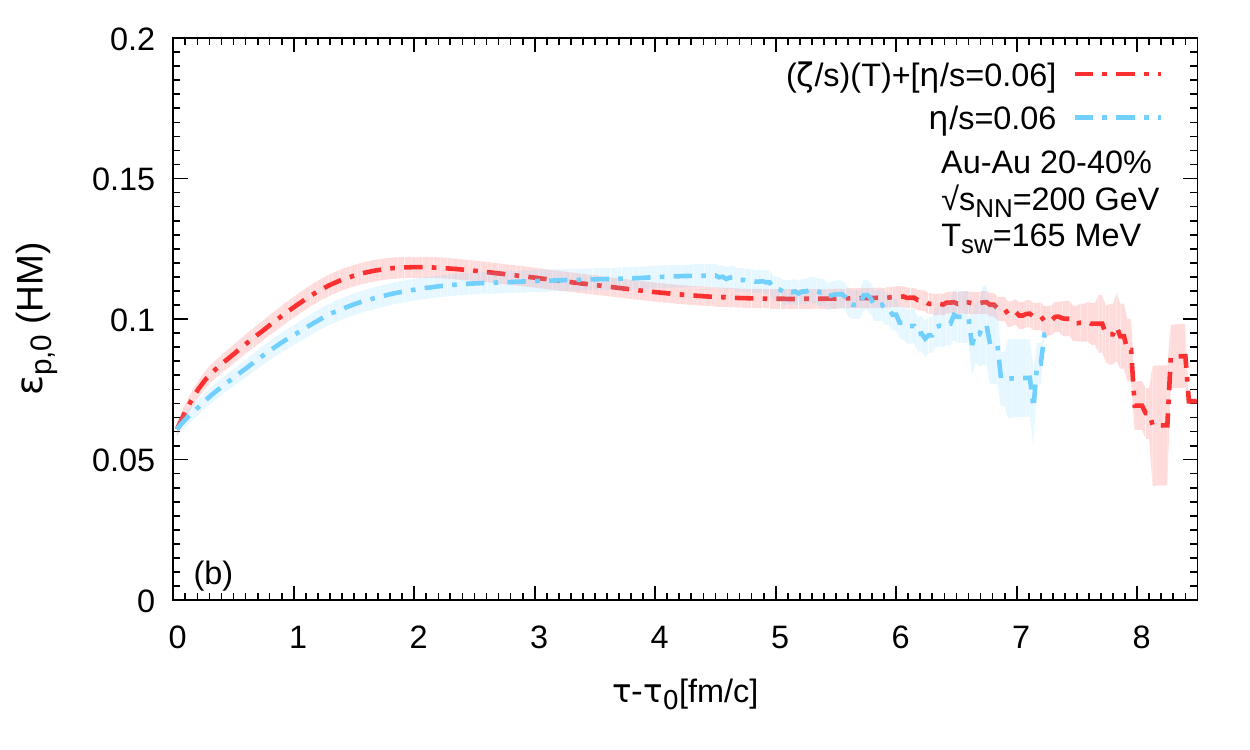}
\end{center}
\caption{(Color online) Development of the hydrodynamical momentum anisotropy in the hadronic sector [recall Eq.~(\ref{eq:f_QGP})] as a function of $\tau-\tau_0$ at LHC (a) and at RHIC (b).}
\label{fig:T_munu_aniso_vs_tau}
\end{figure}
The hydrodynamical momentum anisotropy in the temperature region where HM dileptons are emitted reveals a clearer picture [see Fig.~\ref{fig:T_munu_aniso_vs_tau}]. By looking at $\varepsilon_{p,0}(\tau)$ at late time, say $(\tau-\tau_0)\sim7.5$ fm/$c$, Fig.~\ref{fig:T_munu_aniso_vs_tau}a allows us to appreciate how much of the enhancement seen in $\varepsilon_{p,0}(T)$ translates into dilepton $v_2(M)$ at low $M$. This assumes of course that low $M$ dileptons are predominantly emitted at later times, and vice-versa for high $M$ dileptons. Near $(\tau-\tau_0)\sim7.5$ fm/$c$ in Fig.~\ref{fig:T_munu_aniso_vs_tau}a, one has effectively integrated over the entire spacetime volume $\Delta  V_{2+1}$ in the HM sector [defined in Eq.~(\ref{eq:f_QGP})], and $\varepsilon_{p,0}(\tau)$ has reached its maximal value.\footnote{Hydrodynamical events without $\zeta/s$ start to freeze-out beyond $(\tau-\tau_0)\sim7.5$ fm/$c$ at LHC collision energy, which affects the average $\varepsilon_{p,0}(\tau)$, as can be seen in Fig.~\ref{fig:T_munu_aniso_vs_tau}a. Thus comparisons between red and blue curves in Fig.~\ref{fig:T_munu_aniso_vs_tau}a become unreliable much past $(\tau-\tau_0)\sim7.5$ fm/$c$.} Comparing the three simulations in Fig.~\ref{fig:T_munu_aniso_vs_tau}a with Fig. \ref{fig:T_munu_aniso_vs_temp}a, the $\zeta/s$-induced enhancement of $\varepsilon_{p,0}(T)$ near the peak of $\frac{\zeta}{s}(T)$ does not transfer onto $\varepsilon_{p,0}(\tau)$, due to a significant amount of spacetime volume sitting away from the peak in $\varepsilon_{p,0}(T)$ (while still being above $T_{\rm sw}$). Thus, the red curve in $\varepsilon_{p,0}(\tau)$ for $(\tau-\tau_0)\sim 7.5$ fm/$c$ is smaller than the blue curves. Conversely, if there was a significant spacetime volume near the peak of $\varepsilon_{p,0}(T)$, it would show up in $\varepsilon_{p,0}(\tau)$ at $(\tau-\tau_0)\sim 7.5$ fm/$c$ by making the red curve larger than the blue curves in Fig.~\ref{fig:T_munu_aniso_vs_tau}a. However, the latter is not the case, thus the enhancement seen in $\varepsilon_{p,0}(T)$ does not translate onto the final $v_2(M)$ of lower temperature (HM) dileptons at the LHC. $\varepsilon_{p,0}(\tau)$ in Fig. \ref{fig:T_munu_aniso_vs_tau}a qualitatively mimics the behavior of $v_2(M)$ of HM dileptons in Fig. \ref{fig:v2_M_HM_QGP_LHC}a.

\begin{figure}
\begin{center}
\includegraphics[width=0.495\textwidth]{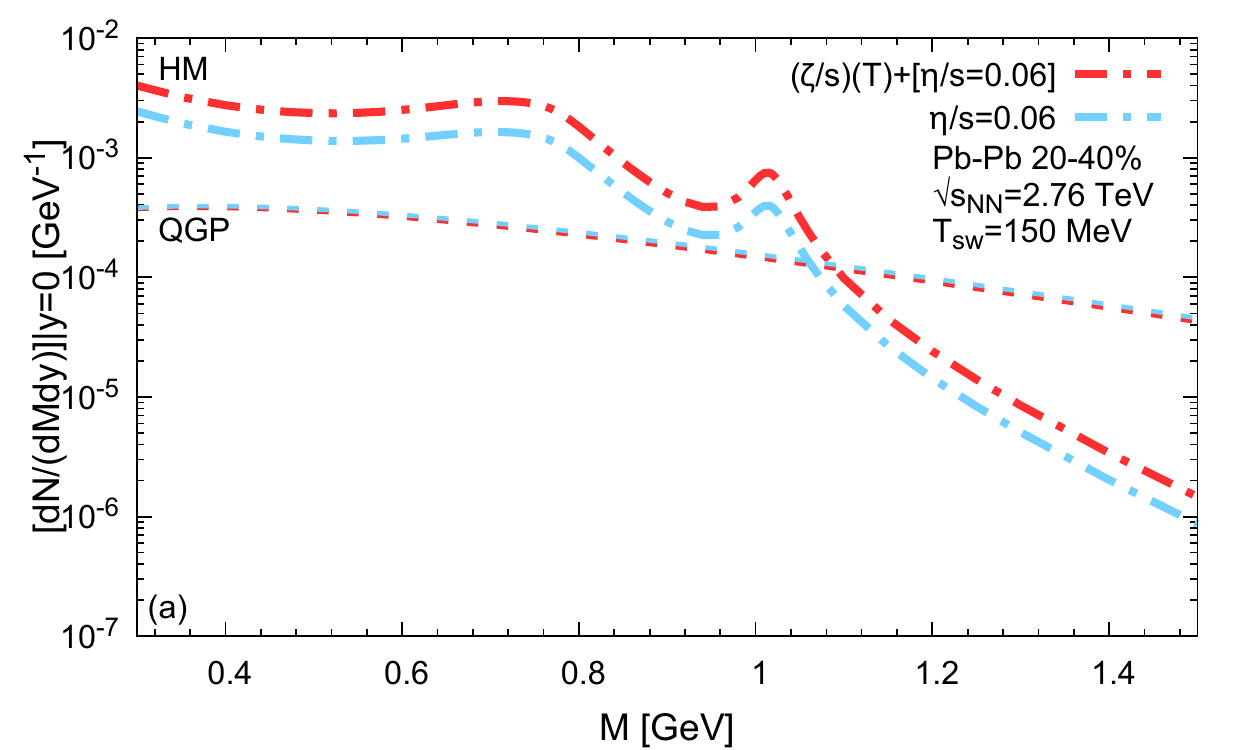}
\includegraphics[width=0.495\textwidth]{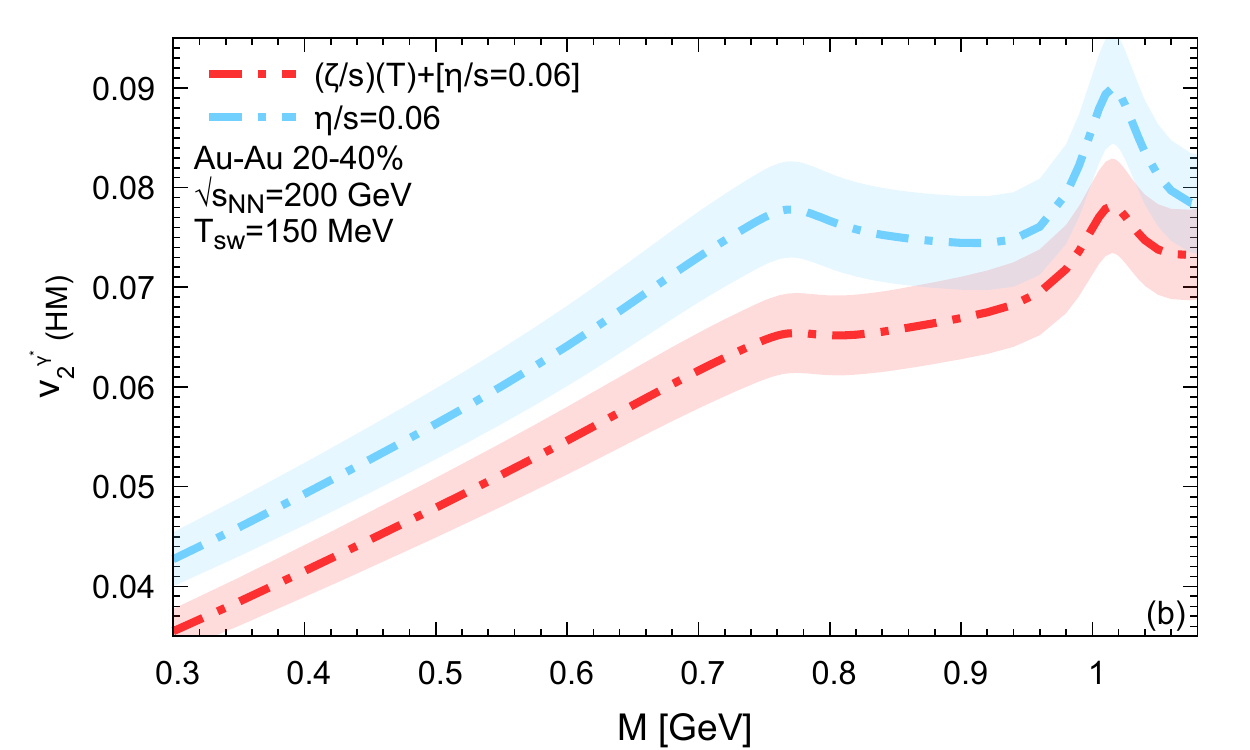}
\includegraphics[width=0.495\textwidth]{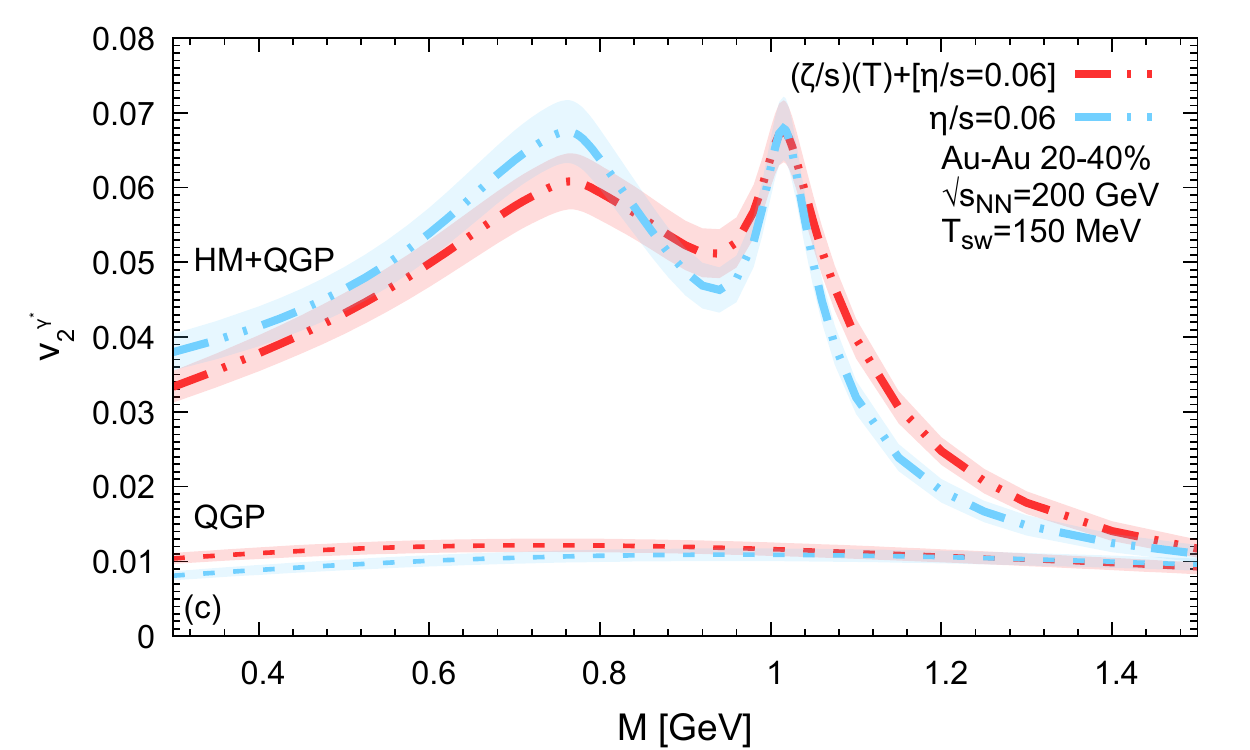}
\includegraphics[width=0.495\textwidth]{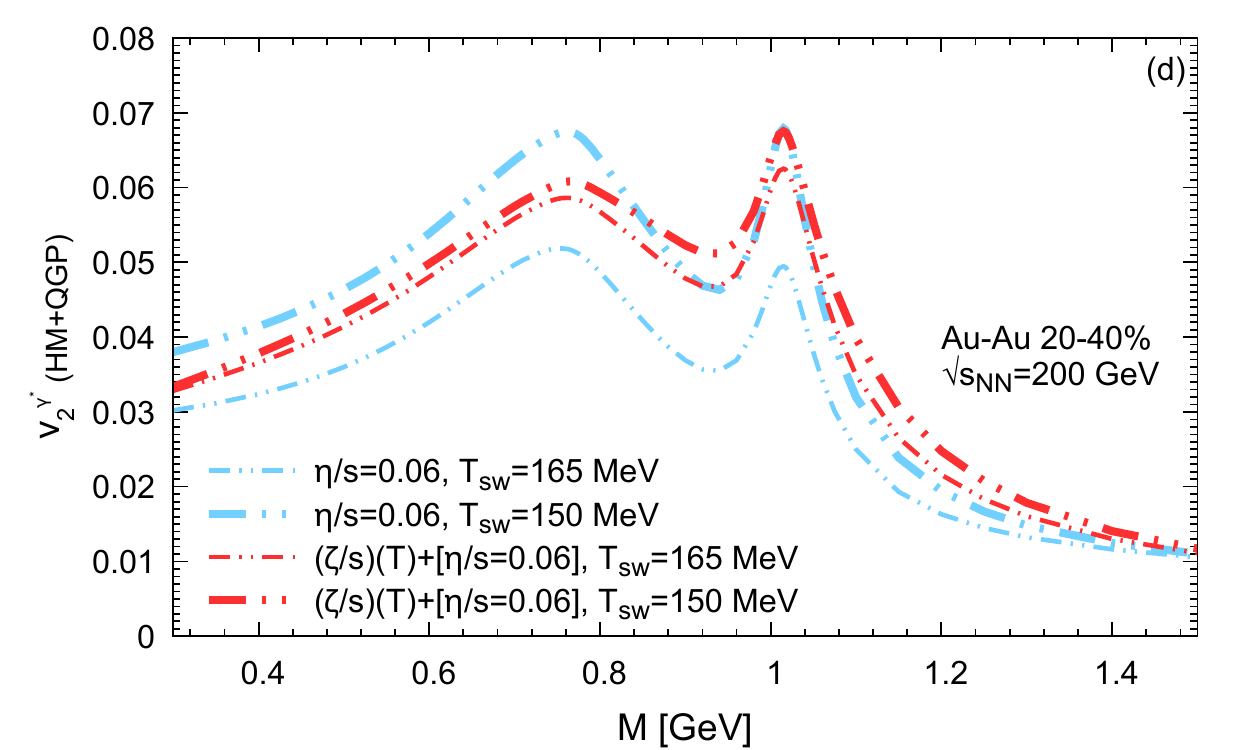}
\end{center}
\caption{(Color online)  (a) Invariant mass distribution of dilepton yield at $T_{\rm sw}=150$ MeV. $v_2$ of HM (b) and thermal (c) dileptons at $T_{\rm sw}=150$ MeV, obtained using Eq.~(\ref{eq:f_QGP}). (d) Comparison of thermal dilepton $v_2$ at $T_{\rm sw}=165$ MeV versus $T_{\rm sw}=150$ MeV.}
\label{fig:yield_v2_M_thermal_RHIC_sw150} 
\end{figure}

At RHIC on the other hand, Fig.~\ref{fig:T_munu_aniso_vs_tau}b shows that the enhancement seen in $\varepsilon_{p,0}(T)$ persists, after summing over temperature bins, showing up in $\varepsilon_{p,0}(\tau)$ for $(\tau-\tau_0)<3$ fm/$c$, where the red curve of Fig. \ref{fig:T_munu_aniso_vs_tau}b is larger than the blue curve. At $(\tau-\tau_0)\sim4$ fm/$c$, the earlier enhancement seen in the red relative to blue curves of $\varepsilon_{p,0}(\tau)$ disappears as the curves are now within uncertainty of each other.\footnote{Note that hydrodynamical simulations without $\zeta/s$ at RHIC collision energy start freezing out past $(\tau-\tau_0)\sim4$ fm/$c$ at which point comparisons between red and blue $\varepsilon_{p,0}(\tau)$ curves become less reliable.} Assuming HM dileptons with $M>0.8$ GeV are predominantly emitted at earlier times while HM dileptons with $M<0.8$ GeV are mostly emitted at later times, our calculations show that the correlation between $v_2(M)$ of hadronic dileptons and $\varepsilon_{p,0}(\tau)$ mostly holds within uncertainty. The enhancement seen in $\epsilon_{p,0}(\tau)$ at $(\tau-\tau_0)<3$ fm/$c$ has hints still present in the $v_2(M)$ of HM dileptons at $M>0.8$ GeV. However, our current uncertainties do not allow to draw more definite conclusion. The behavior of $\epsilon_{p,0}(\tau)$, and thus $v_2(M)$ of hadronic medium dileptons, is highly dependent on $T_{\rm sw}$. If $T_{\rm sw}$ is lowered far enough, the order of the $\epsilon_{p,0}(\tau)$ curves at RHIC would be the same as at the LHC, as more weight would be put to lower temperature $\varepsilon_{p,0}(T)$, where the medium without $\zeta/s$ develops more anisotropic flow than the one with $\zeta/s$. To explore this, we reduced $T_{\rm sw}$ at RHIC from 165 MeV to 150 MeV. Note that $T_{\rm sw}=165$ MeV was obtained from a tune of the hydrodynamical simulations with \texttt{UrQMD} to hadronic observables presented in Refs.~\cite{Ryu:2015vwa, Ryu:2017qzn}. To remain within $\sim 5$\% agreement with the best fit (i.e. the one including bulk viscosity) obtained at 165 MeV \cite{Ryu:2015vwa, Ryu:2017qzn}, we couldn't lower $T_{\rm sw}$ below 150 MeV. The corresponding dilepton yields and $v_2$ are presented in Fig.~\ref{fig:yield_v2_M_thermal_RHIC_sw150}. 
At the lower switching temperature for RHIC collisions, the $v_2(M)$ of HM dileptons in Fig.~\ref{fig:yield_v2_M_thermal_RHIC_sw150}b follows a similar pattern as at the LHC energy in Fig.~\ref{fig:v2_M_HM_QGP_LHC}a. After performing a yield-weighted average, the thermal dilepton $v_2$ shown in Fig.~\ref{fig:yield_v2_M_thermal_RHIC_sw150}c still displays an inversion in the ordering between the different runs of $v_2(M)$ around $M\sim 0.9$ GeV and $M\gtrsim 1.1$ GeV, similarly to what was seen at the LHC. We also notice in Fig. \ref{fig:yield_v2_M_thermal_RHIC_sw150}d that bulk viscous pressure slows down the anisotropic expansion of the medium at lower temperatures/late times. This effect is directly seen in the red curves of Fig.  \ref{fig:yield_v2_M_thermal_RHIC_sw150}d, where $v_2(M)$ of dileptons increases less going from $T_{\rm sw}=165$ MeV to $T_{\rm sw}=150$ MeV for the medium with $\zeta/s$ relative to the one without. The sizeable increase in dilepton $v_2(M)$ present for the medium without $\zeta/s$, as depicted by the blue curves in Fig. \ref{fig:yield_v2_M_thermal_RHIC_sw150}d, is caused by a decrease in $T_{\rm sw}$.  

\subsection{Cocktail dileptons and the $\rho$ spectral function}\label{sub_sec:cokctail_results}
To ascertain whether the bulk viscosity-induced increase in thermal dilepton $v_2(M)$ at RHIC may be observed experimentally, cocktail dileptons should be included. The latter have a sizable contribution to total dilepton yield and $v_2$. A crude estimate of the cocktail dilepton production may be obtained by letting hydrodynamics evolve to a lower switching temperature, while using the same thermal dilepton rates. However, such an approach is flawed since there are additional dilepton production channels that are not accounted for within the thermal dilepton emission rates (integrated over the hydrodynamical evolution). Indeed, the lifetime of some parent hadrons decaying into dileptons (e.g. via the Dalitz channel) is much longer than the average time the medium spends evolving hydrodynamically. Thus, cocktail dileptons should not be estimated by running hydrodynamics to a lower $T_{\rm sw}$. Instead, a different approach will be taken is to compute cocktail dileptons from $T_{\rm sw}$ freeze-out surface, as detailed below. 

The most important cocktail dilepton channels have been discussed in Sec. \ref{sec:cocktail_dileptons} and will all be considered here. Given that the lifetime of all mesons contributing to our dilepton cocktail is large (except for the $\rho$), while their branching fraction to dileptons is small, a portion of the decays of cocktail mesons will happen during the later stages of a hadronic transport simulation, which is well captured by free-streaming. Present work will calculate cocktail dileptons using the free-streaming assumption. A more complete calculation that includes dynamical dilepton production from hadronic transport will follow in an upcoming publication, where SMASH\footnote{SMASH stands for Simulating Many Accelerated Strongly-interacting Hadrons.} \cite{Staudenmaier:2017vtq,Oliinychenko:2017tzr,Staudenmaier:2016hmh,Weil:2016fxr} will be used to calculate dilepton generation from the hadronic cascade.\footnote{How well this free-streaming approximation holds will be revisited in an upcoming publication where a comparison between free-streaming and SMASH will be done.}  

\begin{figure}[!h]
\begin{center}
\includegraphics[width=0.495\textwidth]{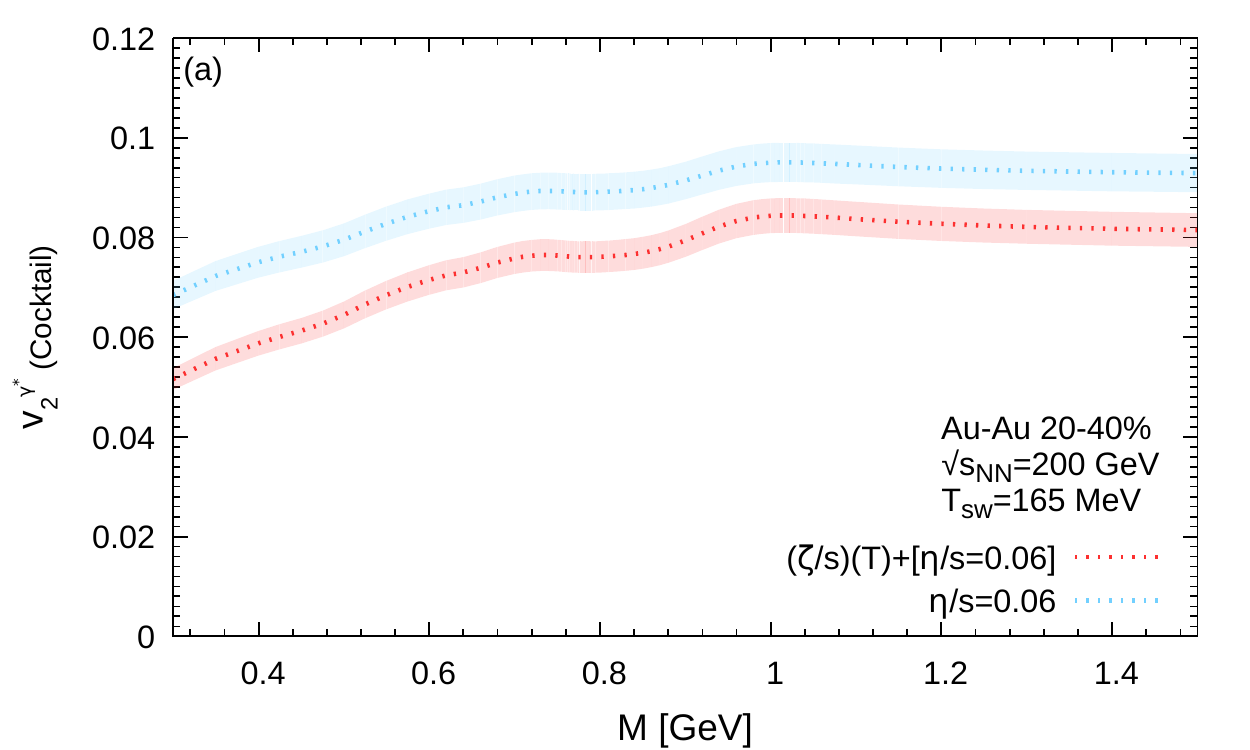}
\includegraphics[width=0.495\textwidth]{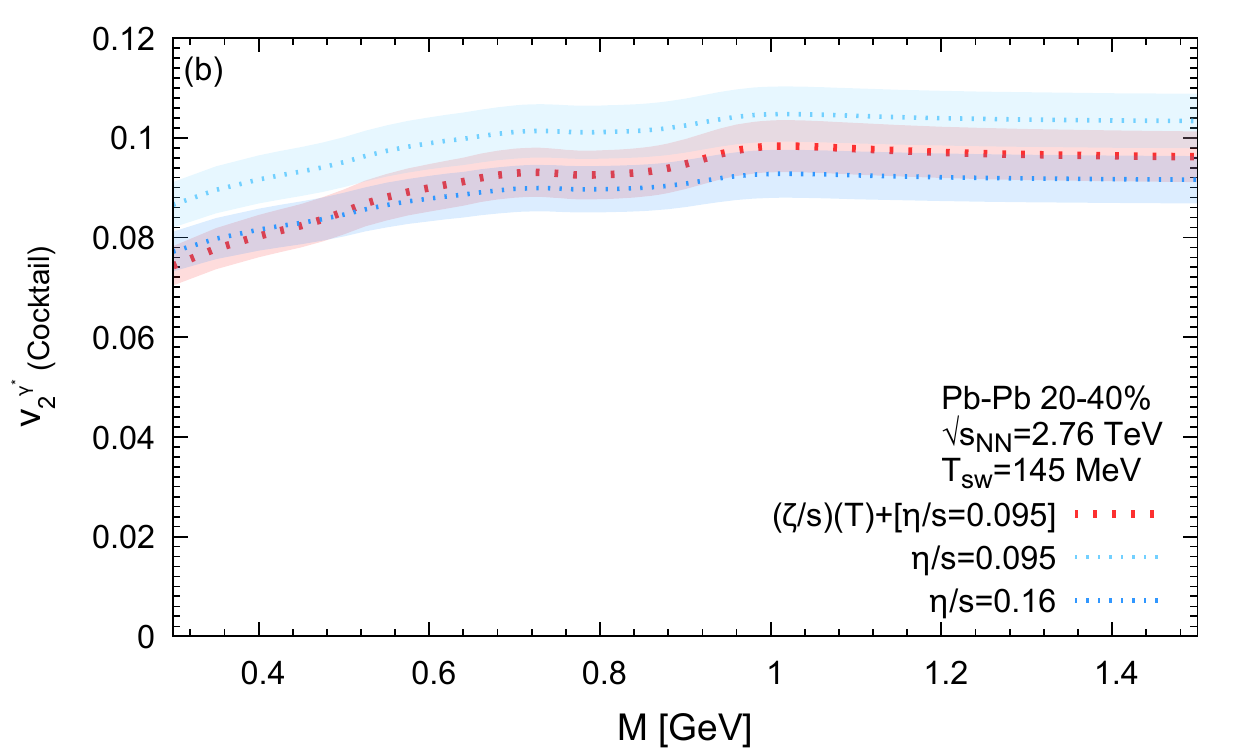}
\end{center}
\caption{(Color online) Invariant mass distribution of dilepton cocktail $v_2$ at RHIC (a) and LHC (b) collision energies under the influence of bulk and shear viscosities without any contribution from $\rho$ mesons.}
\label{fig:v2_M_cocktail_wo_rho_RHIC_LHC}
\end{figure} 
Figure \ref{fig:v2_M_cocktail_wo_rho_RHIC_LHC} presents the first event-by-event calculation, based on a realistically expanding medium, of cocktail dilepton $v_2$ at RHIC and LHC collision energies, excluding the contribution from the $\rho$. The cocktail dilepton $v_2$ at RHIC and LHC collisions energies behaves similarly. These results include direct decays of vector mesons as well as late Dalitz decays of both pseudoscalar and $\omega$ and $\phi$ vector mesons. Except the $\rho$ which will be taken into account later, all other mesons tallied in our calculation of the dilepton cocktail have a long lifetime, i.e. they are narrow resonances. Therefore, their $\frac{p^0 d^3 N}{d^3 p}$ distribution was obtained from the Cooper-Frye formula, including resonance decays, using the on-shell approximation as detailed in Ref.~\cite{Ryu:2017qzn}. Their subsequent decays into dileptons was computed through Dalitz decays as prescribed in Eq.~(\ref{eq:dalitz_final}), while direct vector meson decays follows Eq.~(\ref{eq:omega_phi_dilep}).  

\begin{figure}[!h]
\begin{center}
\includegraphics[width=0.495\textwidth]{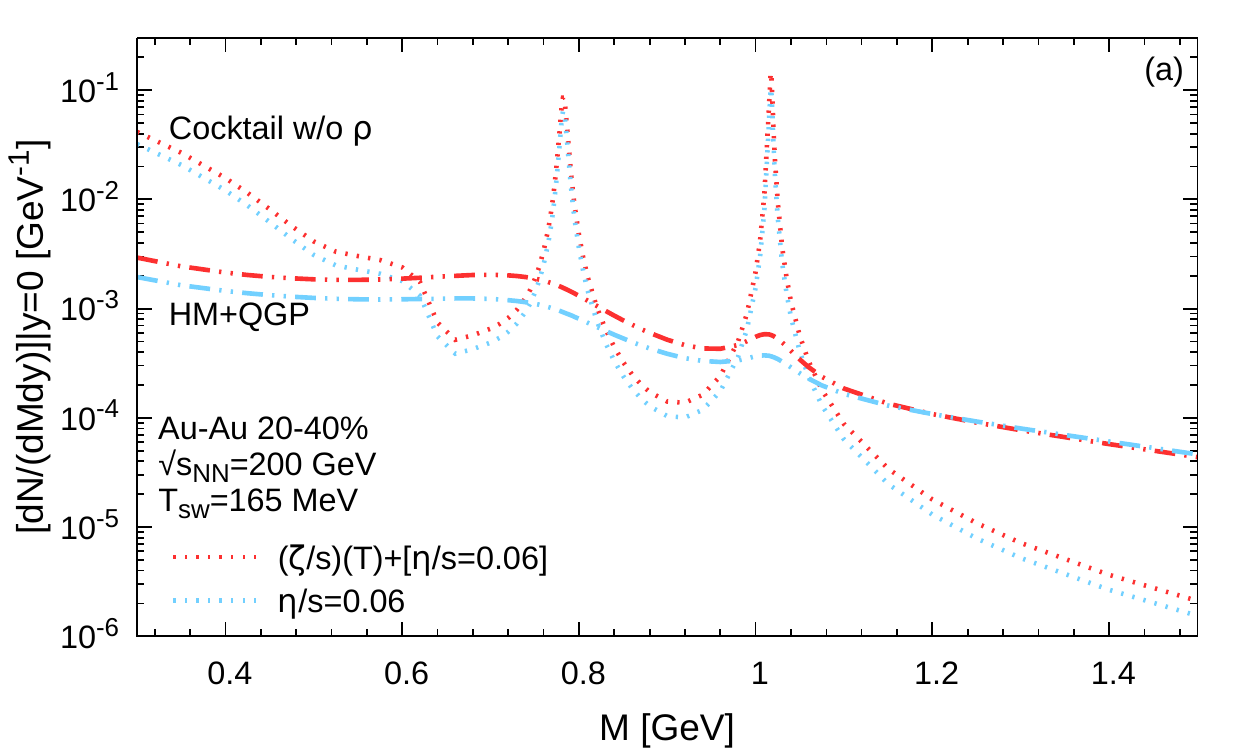}
\includegraphics[width=0.495\textwidth]{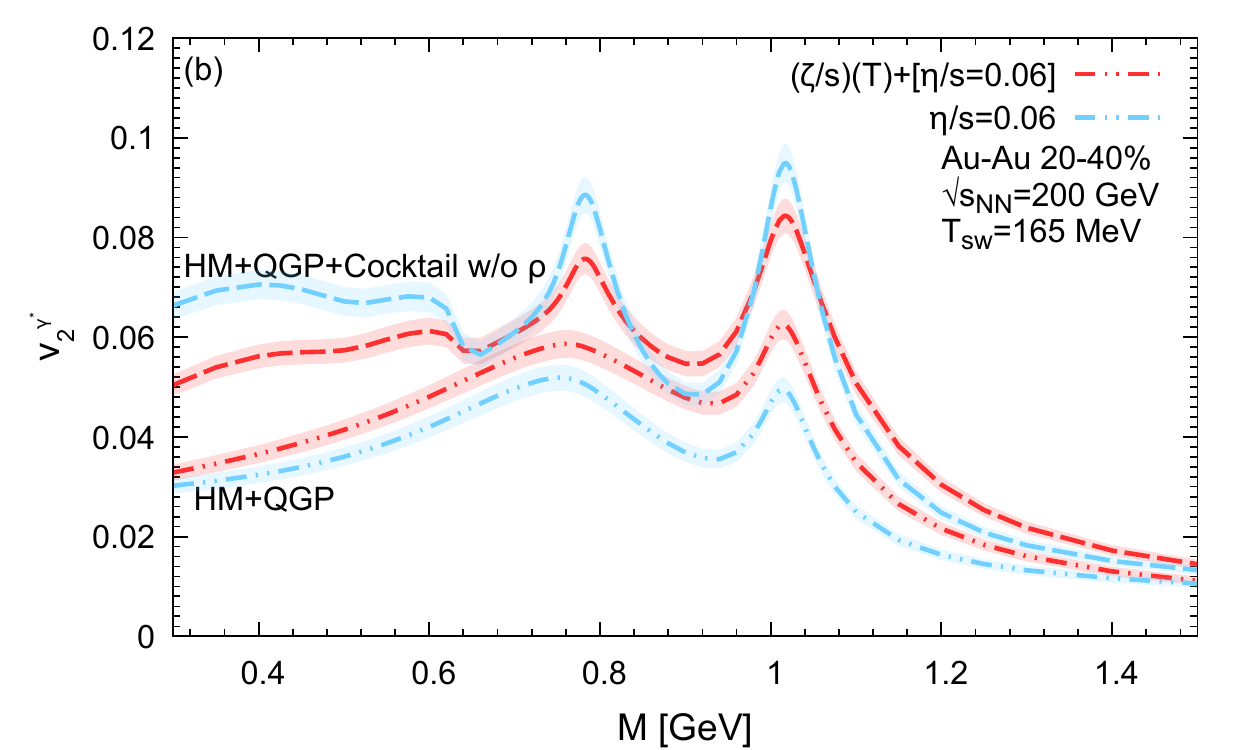}
\includegraphics[width=0.495\textwidth]{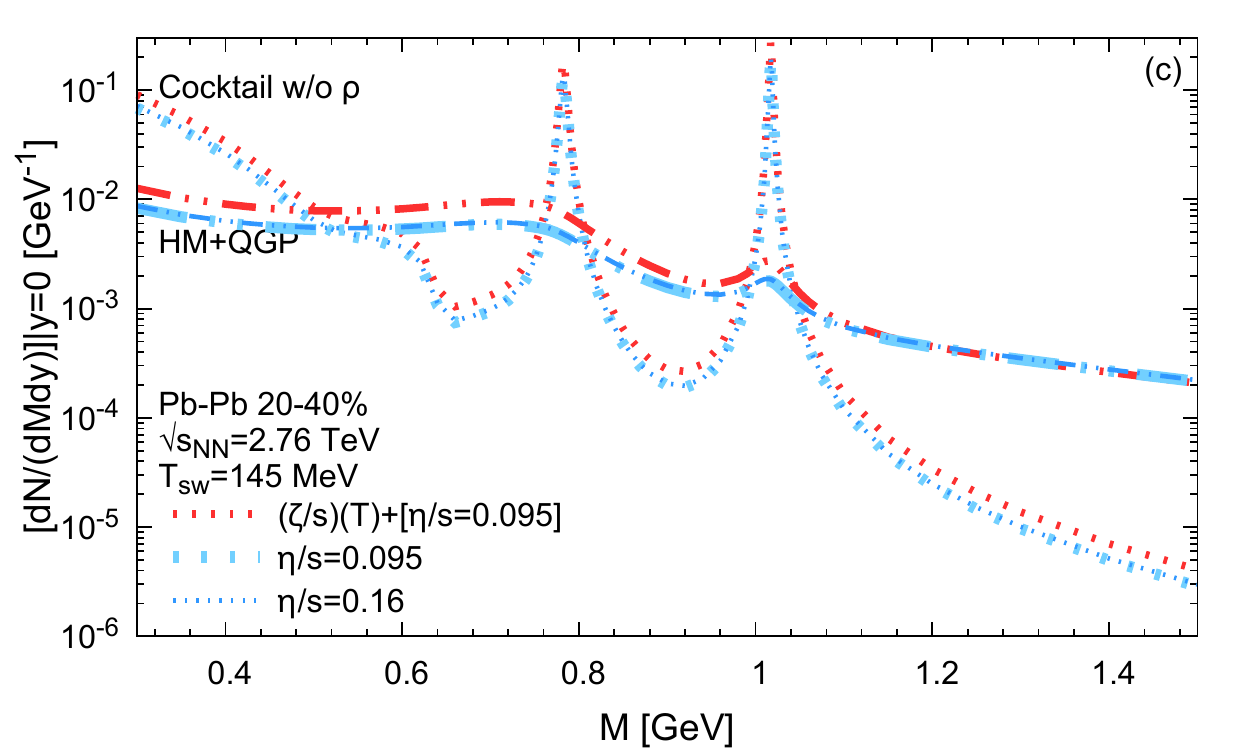}
\includegraphics[width=0.495\textwidth]{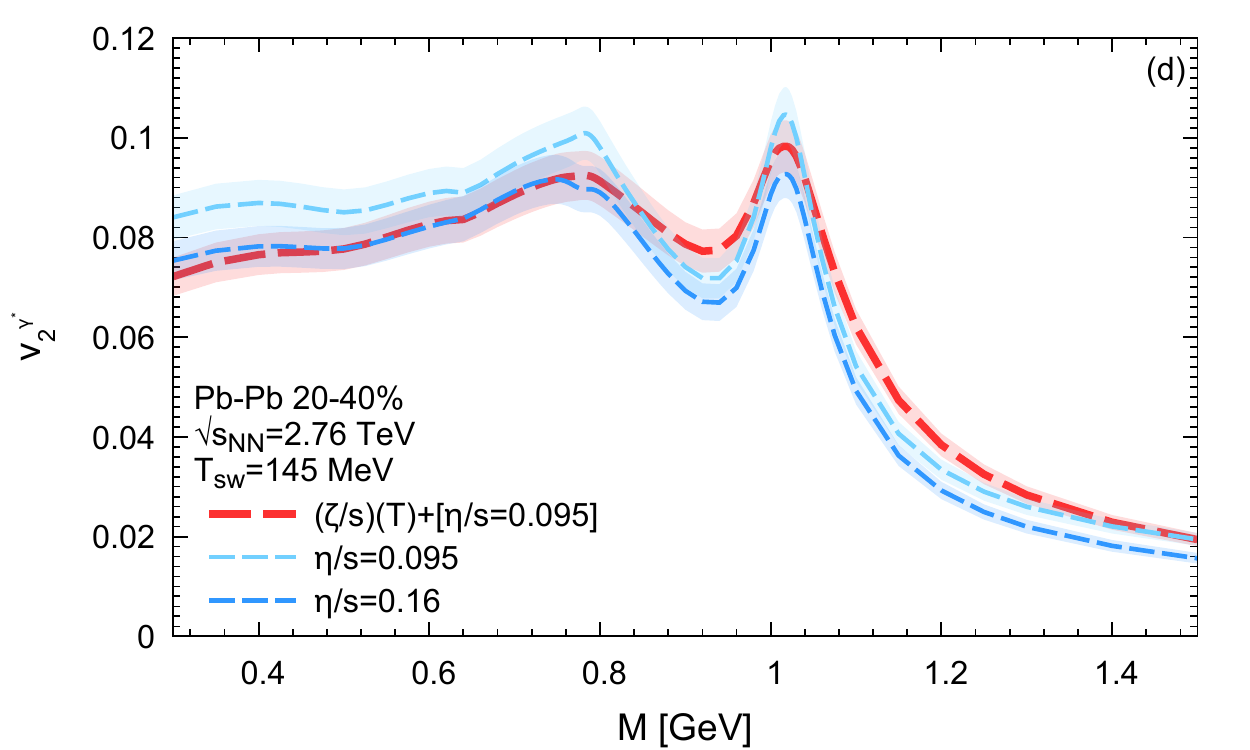}
\end{center}
\caption{(Color online) All panels in this figure exclude the contribution from the cocktail $\rho$. (a) Invariant mass distribution of dilepton yield at RHIC for thermal (HM+QGP) dileptons as well as cocktail dileptons. (b) Invariant mass distribution of dilepton $v_2$ at RHIC for thermal (HM+QGP) and all sources in (a). (c) Invariant mass distribution of dilepton yield at LHC for thermal (HM+QGP) dileptons as well as cocktail dileptons. (d) Invariant mass distribution of dilepton $v_2$ at LHC combining all sources in (c).}
\label{fig:v2_M_cocktail_wo_rho_thermal_tot}
\end{figure}

A combination of cocktail and thermal dileptons is presented in Fig.~\ref{fig:v2_M_cocktail_wo_rho_thermal_tot}, allowing to investigate how much of the bulk viscosity-induced effects seen in thermal dileptons shows up in the total $v_2$, and thus may have experimental signatures.

Focusing first on the results at RHIC in Fig.~\ref{fig:v2_M_cocktail_wo_rho_thermal_tot}b and comparing them to those of Fig.~\ref{fig:yield_v2_M_thermal_RHIC_sw150}c, one can see that after combining cocktail and thermal (HM+QGP) dileptons using the yield in Fig.~\ref{fig:v2_M_cocktail_wo_rho_thermal_tot}a, the inversion in the order of the red versus blue curves persists around the same invariant masses, i.e. $M\sim 0.9$ GeV and $M\gtrsim 1.1$ GeV. The origin of the inversion in Fig.~\ref{fig:v2_M_cocktail_wo_rho_thermal_tot}b is mostly driven by yield effects, as before. Combining the thermal and cocktail dileptons at the LHC collision energy in Fig.~\ref{fig:v2_M_cocktail_wo_rho_thermal_tot}d (using the yield in Fig.~\ref{fig:v2_M_cocktail_wo_rho_thermal_tot}c) generates a similar pattern as in Fig.~\ref{fig:v2_M_cocktail_wo_rho_thermal_tot}b. Thus, the effects of $\zeta/s$ in our calculations can be seen via  the ratio $\frac{v_2(M=0.9\, \mathrm{GeV})}{v_2(M=0.3\, \mathrm{GeV})}$ at both collision energies. In our study, the presence of bulk viscosity, in addition to shear viscosity, yields $\frac{v_2(M=0.9\, \mathrm{GeV})}{v_2(M=0.3\, \mathrm{GeV})}>1$ while shear viscosity alone makes it less than 1 [see Table \ref{table:v2_diff_mass_wo_rho} for details]. 
\begin{table}[!h]
\caption{The $\frac{v_2(M=0.9\, \mathrm{GeV})}{v_2(M=0.3\, \mathrm{GeV})}$ ratio as a tool to measure the effects of bulk viscosity (excluding the $\rho$ contribution to the dilepton cocktail)} 
\centering
\begin{tabular}{c | c | c | c || c | c | c}
LHC & $(\zeta/s)(T)+[\eta/s=0.095]$ & $\eta/s=0.095$ & $\eta/s=0.16$ & RHIC & $(\zeta/s)(T)+[\eta/s=0.06]$ & $\eta/s=0/06$ \\   
\hline
 & 1.09 & 0.881 & 0.920 & & 1.09 & 0.733 
\end{tabular}
\label{table:v2_diff_mass_wo_rho} 
\end{table}

To complete the investigation of the influence bulk viscosity has on dilepton $v_2$, the contribution of the $\rho$ will be incorporated in two steps. The first computes the production of $\rho$ mesons on the switching hypersurface while the second includes $\rho$s generated by resonance decays. 

\begin{figure}[!h]
\begin{center}
\includegraphics[width=0.495\textwidth]{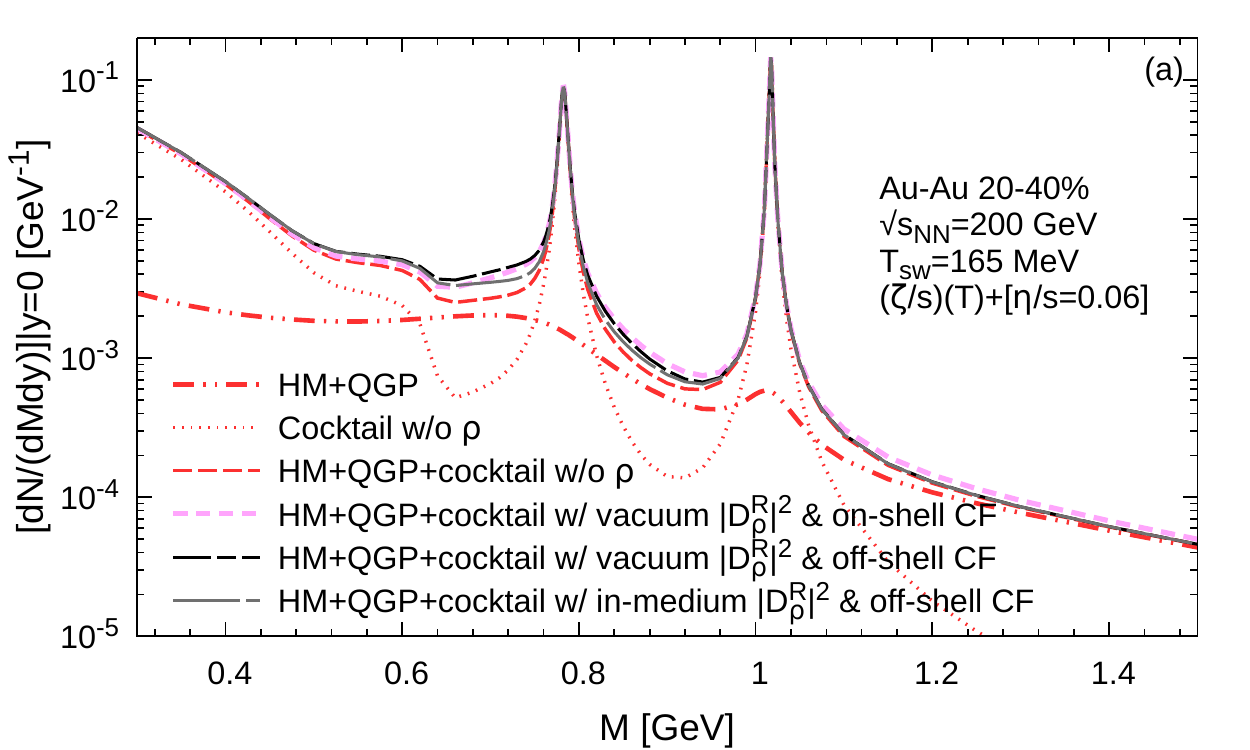}
\includegraphics[width=0.495\textwidth]{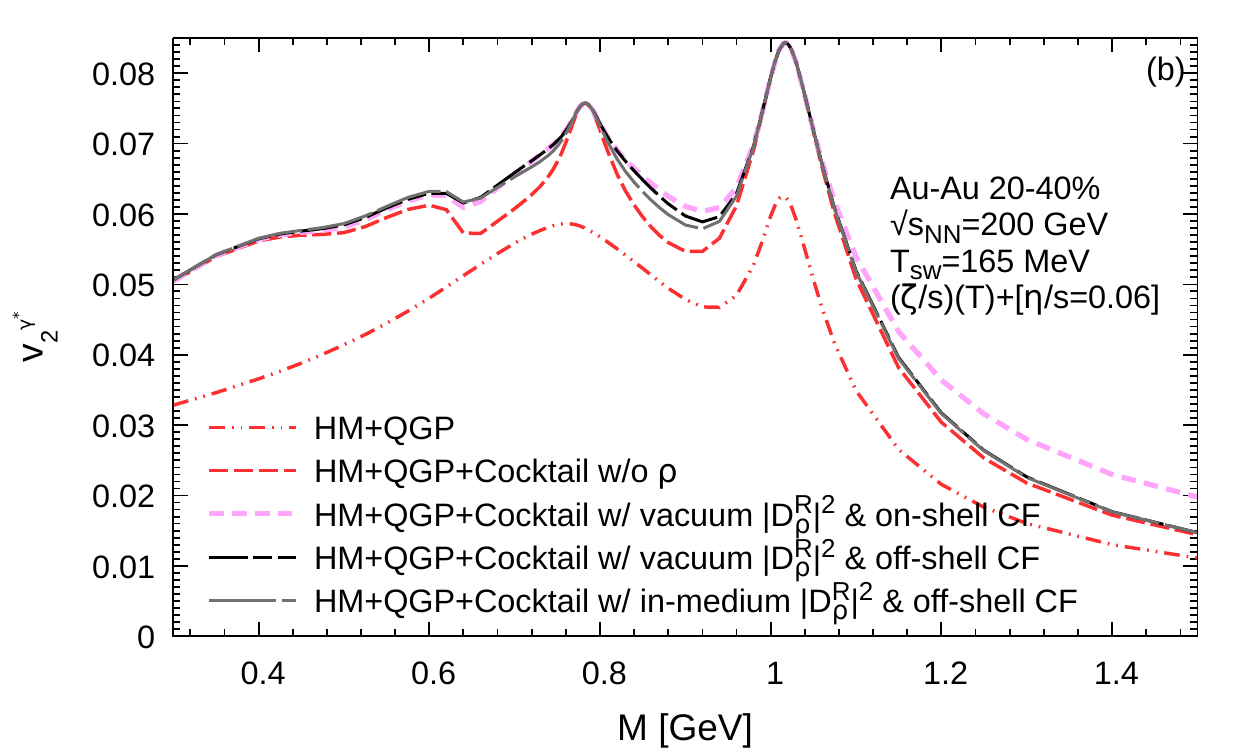}
\includegraphics[width=0.495\textwidth]{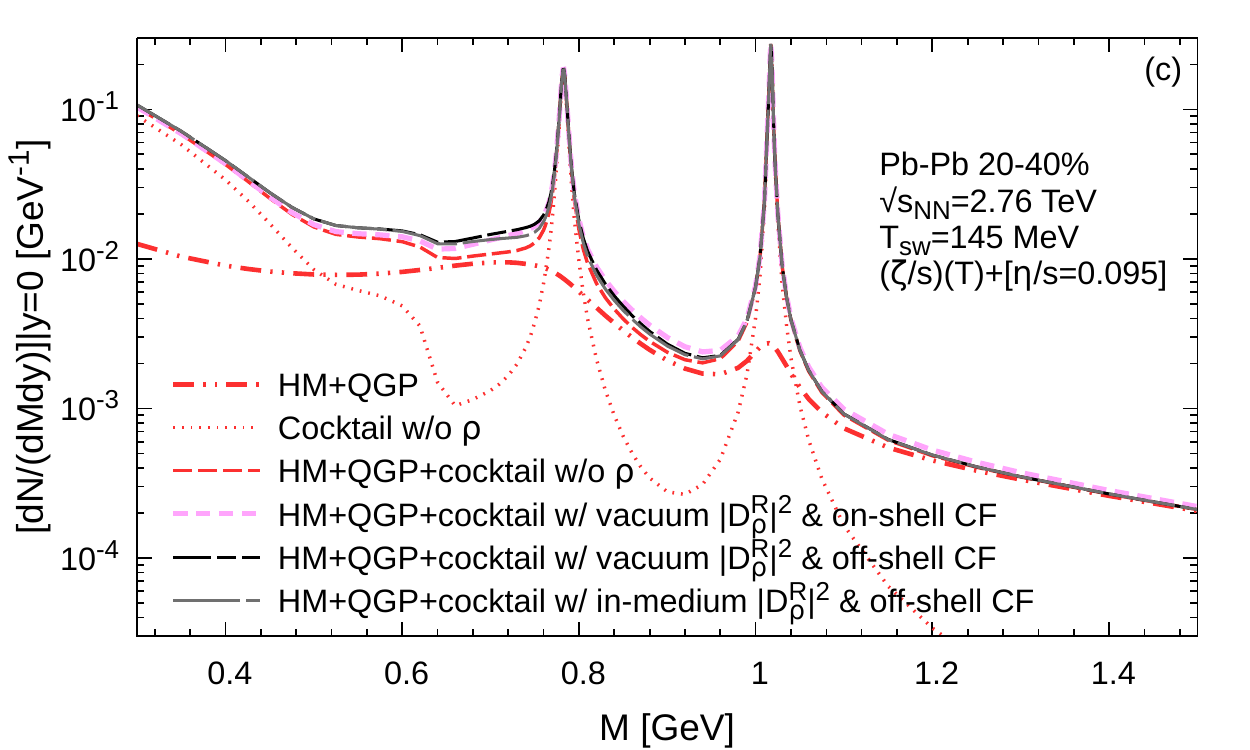}
\includegraphics[width=0.495\textwidth]{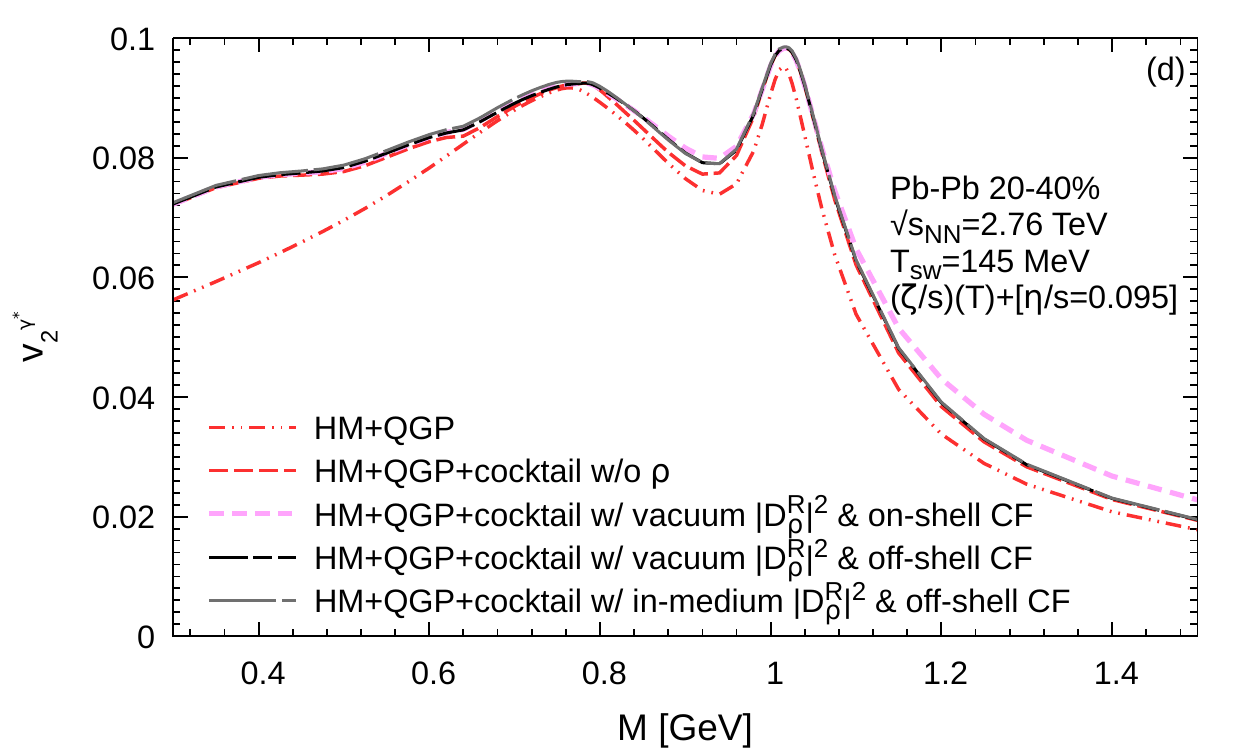}
\end{center}
\caption{(Color online) All panels in this figure include $\rho$ from the hypersurface of constant $T_{\rm sw}$, but neglect resonance decays. (a) Invariant mass distribution of dilepton yield at RHIC for thermal (HM+QGP) dileptons as well as cocktail dileptons.\\ 
(b) Invariant mass distribution of dilepton $v_2$ at RHIC for the sources presented in (a). (c) Invariant mass distribution of dilepton yield at LHC for thermal (HM+QGP) dileptons as well as cocktail dileptons. (d) Invariant mass distribution of dilepton $v_2$ at LHC for the same sources as in (c).}
\label{fig:v2_M_cocktail_w_rho_thermal_tot}
\end{figure}

There are three different approaches to calculate the contribution of the $\rho$ on the switching hypersurface. As outlined in subsection \ref{sec:cocktail_dileptons}, the first approach consists of assuming that the $\rho$ meson width is broadened on the switching hypersurface compared to its vacuum value, thus will employ the in-medium $\rho$ distribution $\left| D^R_\rho \right|^2$ in Eq.~(\ref{eq:mod_CF_rho_dilep}), while using the invariant mass dependent version of the Cooper-Frye (CF) integral, namely $\int d^3 \Sigma^\mu p_\mu\, n_{\rho}(M)$. Note that apart from the invariant mass dependence, $n_\rho$ is otherwise the same as in Ref.~\cite{Ryu:2017qzn}. The second option uses the vacuum description for the $\rho$ meson, i.e. neglecting in-medium contributions to $\left| D^R_\rho\right|^2$, while still computing the Cooper-Frye integral with a $\rho$ meson density $n_\rho(M)$ that varies with the invariant mass of the dilepton. The last option employs the vacuum description of $\left| D^R_\rho \right|^2$, and also enforces the on-shell condition in the CF integral, namely $\int d^3 \Sigma^\mu p_\mu\, n_{\rho}(M=m_\rho)$. The results of these three prescriptions are compared in Fig.~\ref{fig:v2_M_cocktail_w_rho_thermal_tot} where, for the moment, all contributions to the $\rho$ coming from resonance decays are neglected. 

Since Fig.~\ref{fig:v2_M_cocktail_w_rho_thermal_tot} overlays the various cocktail sources at the same time, a general pattern can be noticed: the total dilepton $v_2$ at RHIC (see Fig.~\ref{fig:v2_M_cocktail_w_rho_thermal_tot}b) is more affected by the dilepton cocktail than at the LHC (displayed in Fig.~\ref{fig:v2_M_cocktail_w_rho_thermal_tot}d). The reason for this is two-fold: first, the $v_2$ of the cocktail at RHIC is much larger than the thermal (HM+QGP) $v_2$ across all $M$.\footnote{At the LHC, the contribution of the cocktail is more pronounced compared to thermal (HM+QGP) $v_2$ once $M\lesssim 0.65$ GeV.} Second, at LHC collision energy, the thermal (HM+QGP) dilepton yield is larger than the cocktail dilepton yield over a wider range of invariant masses (see Fig.~\ref{fig:v2_M_cocktail_w_rho_thermal_tot}c) compared to RHIC in Fig.~\ref{fig:v2_M_cocktail_w_rho_thermal_tot}a. This is expected, as the higher collision energy at the LHC produces a larger spacetime volume of the (hydrodynamical) medium compared to RHIC.\footnote{It may also produce a larger spacetime volume for the late hadronic rescattering stage but since in this work we do not follow that stage dynamically, rather letting the hadrons freeze-out kinetically directly on the switching surface with $T_{\rm sw}$, it is premature to discuss about dileptons emitted during hadronic rescattering.} The combination of these two effects explains why the total dilepton $v_2$ at RHIC is more sensitive to the dilepton cocktail than it is at the LHC.   

Focusing specifically on the $\rho$ along constant $T_{\rm sw}$ hypersurface, its contribution to the total dilepton $v_2$ is more significant at RHIC than at the LHC. This is especially seen for $M\lesssim 1.1$ GeV, where it can contribute as much as $\sim 10$\% to the total $v_2(M)$ at RHIC, with the LHC being much smaller in that invariant mass region. Moving to higher invariant masses $M\gtrsim 1.1$ GeV, the $v_2$ shown in Figs.~\ref{fig:v2_M_cocktail_w_rho_thermal_tot}b and ~\ref{fig:v2_M_cocktail_w_rho_thermal_tot}d is sensitive to the approach used in computing the $\rho$ on the switching hypersurface, both RHIC and LHC collision energy. Of primary importance is the fact that the $\rho$ meson must be treated as an off-shell particle; specifically, the $\rho$ must have a mass distribution in the Cooper-Frye integral $\int d^3 \Sigma^\mu p_\mu\, n_{\rho}(M)$ which enters through a mass-dependent density $n_\rho(M)$.\footnote{Whether or not the width of the $\rho$ is broadened along the hypersurface of constant $T_{\rm sw}$ is not something that can be easily distinguished in our calculations, as can be seen from the thin black and gray lines in Fig. \ref{fig:v2_M_cocktail_w_rho_thermal_tot}.} Indeed since $\int d^3 \Sigma^\mu p_\mu\, n_{\rho}(M)\propto \exp\left(-M/T\right)$, the exponential suppression in invariant mass controls the convergence of the  total dilepton $v_2$ towards the thermal $v_2$ as $M$ increases, more so than its form factor $\left| D^R_\rho\right|^2$. This is illustrated in Figs. \ref{fig:v2_M_cocktail_w_rho_thermal_tot}b and \ref{fig:v2_M_cocktail_w_rho_thermal_tot}d. By comparing Figs. \ref{fig:v2_M_cocktail_w_rho_thermal_tot}a,c with Figs. \ref{fig:v2_M_cocktail_w_rho_thermal_tot}b,d we see that the precise way we do or do not include the medium effects on the spectral function of the $\rho$ has much smaller effects on the dilepton mass spectra than on their $v_2$ in the high-mass region $M\gtrsim 1.1$ GeV. Thus, evaluating the contribution from $\rho$ meson decays to cocktail dileptons must properly account for the exponential suppression $\propto \exp(-M/T)$ to the number of contributing $\rho$ mesons, especially to avoid overpredicting the dilepton $v_2$ for invariant masses above about 1 GeV.  

\begin{figure}[!h]
\begin{center}
\includegraphics[width=0.495\textwidth]{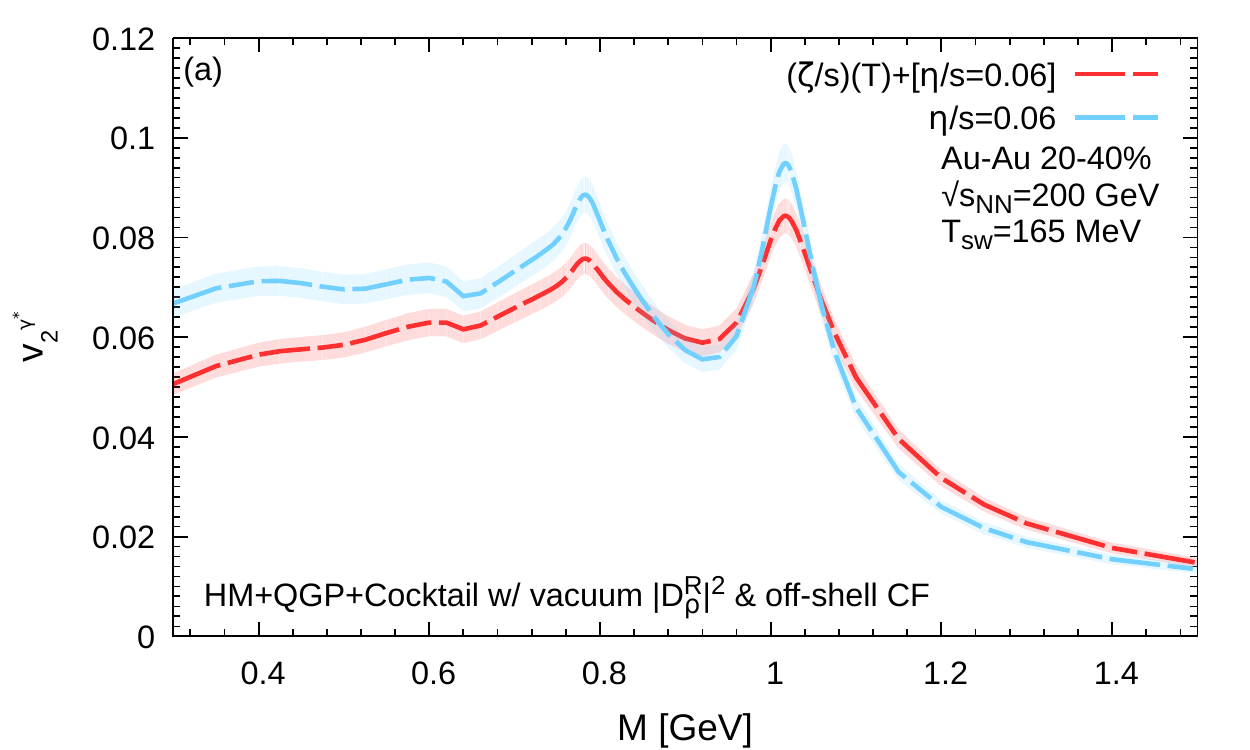}
\includegraphics[width=0.495\textwidth]{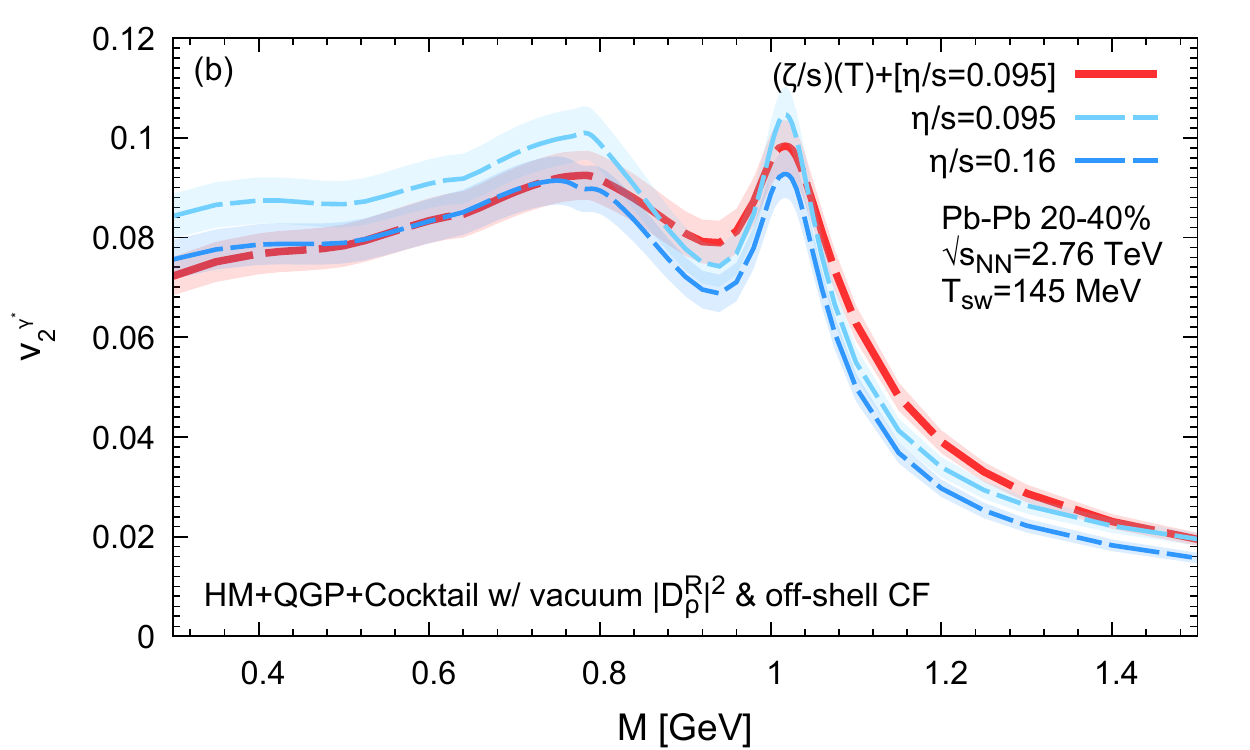}
\includegraphics[width=0.495\textwidth]{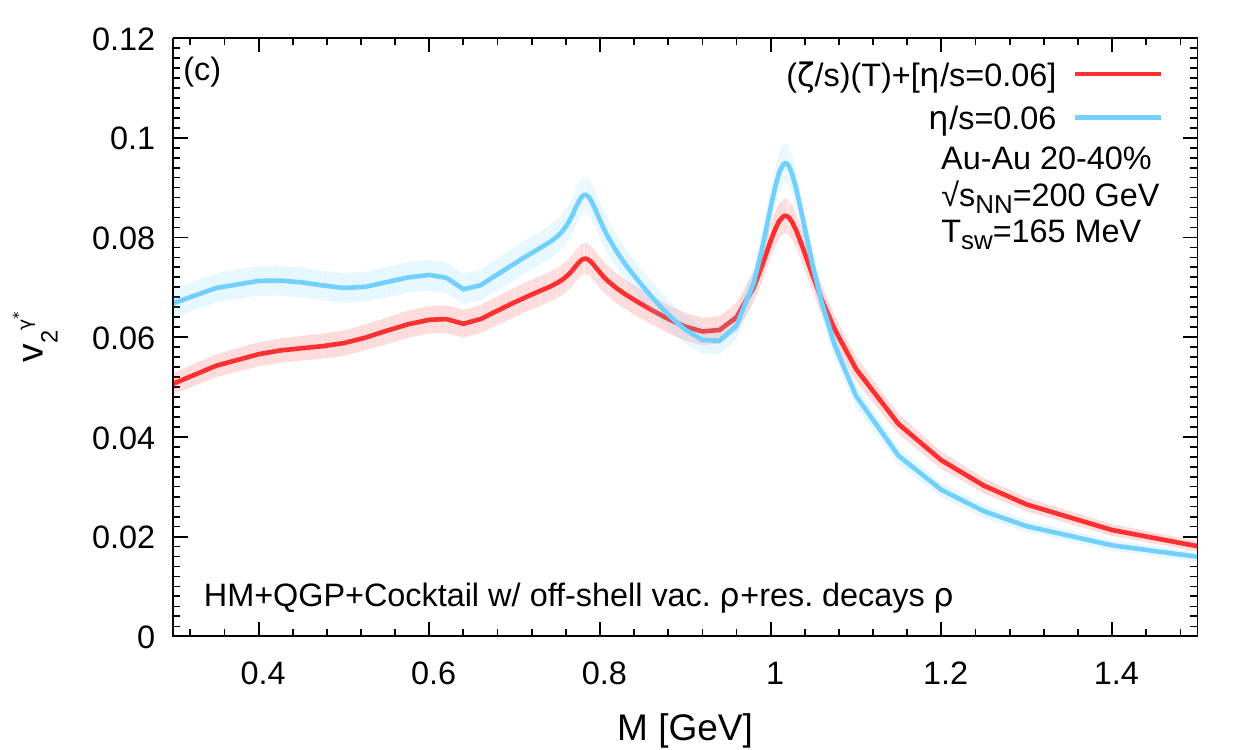}
\includegraphics[width=0.495\textwidth]{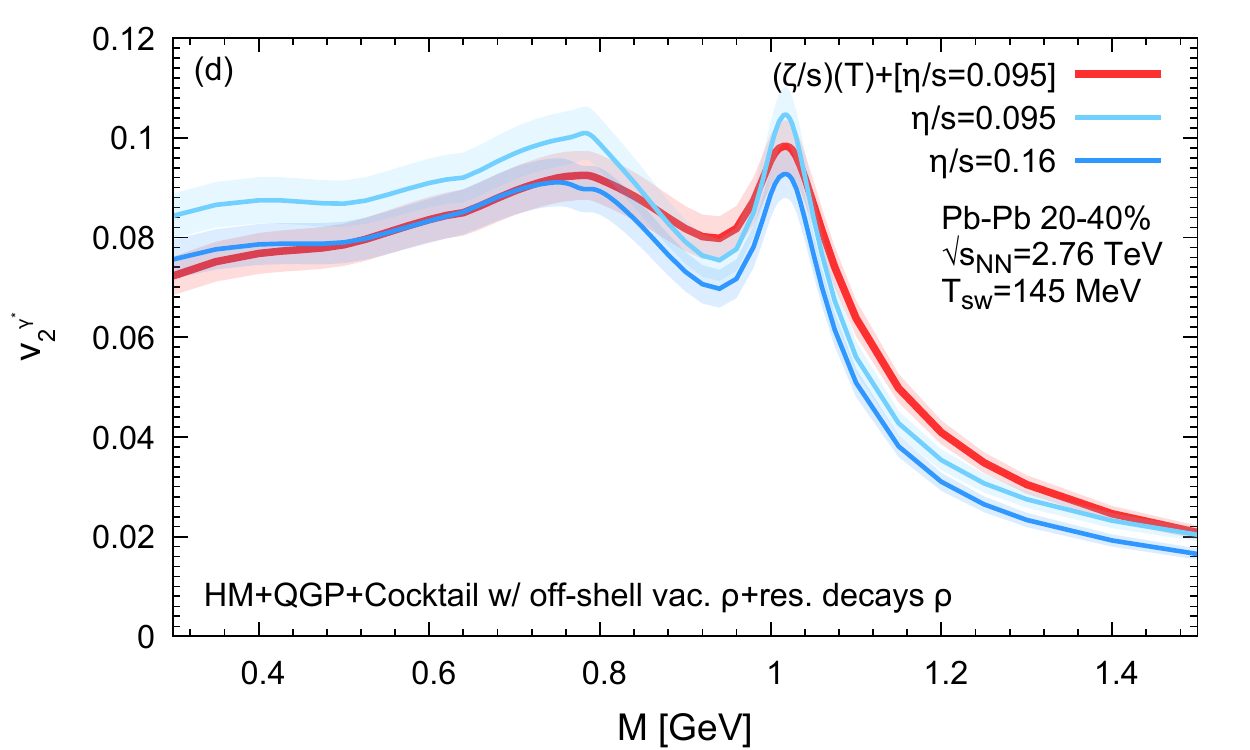}
\end{center}
\caption{(Color online) Invariant mass distribution of the dilepton $v_2$ including the contribution from the $\rho$ on the switching hypersurface at RHIC (a) and LHC (b) collision energies under the influence of bulk and shear viscosities. The contribution from resonance decays to the $\frac{p^0 d^3 N_\rho}{d^3 p}$ at RHIC and LHC is in (c) and (d), respectively.}
\label{fig:v2_M_cocktail_w_rho_RHIC_LHC}
\end{figure} 

Using the vacuum off-shell description of the $\rho$ emanating from the switching hypersurface, Figs.~\ref{fig:v2_M_cocktail_w_rho_RHIC_LHC}a,b show the effects of $\eta/s$ and $\zeta/s$ on the combined dilepton $v_2$. In Fig.~\ref{fig:v2_M_cocktail_w_rho_RHIC_LHC}c,d the calculation includes cocktail dileptons from $\rho$ mesons produced by the decays from higher-mass resonances. As the $\rho$ mesons emerging from resonance decays are on their mass shell in our calculation, we focus on the invariant mass window $0.3<M<1.1$ GeV. In that invariant mass window, including the $\rho$ into the cocktail does not substantially change the pattern that was observed in Fig.~\ref{fig:v2_M_cocktail_wo_rho_thermal_tot}. 

\begin{table}[!h]
\caption{The $\frac{v_2(M=0.9\, \mathrm{GeV})}{v_2(M=0.3\, \mathrm{GeV})}$ ratio, including the cocktail $\rho$ contribution, as a tool to measure the effects of bulk viscosity} 
\centering
\begin{tabular}{c | c | c | c || c | c | c}
LHC & $(\zeta/s)(T)+[\eta/s=0.095]$ & $\eta/s=0.095$ & $\eta/s=0.16$ & RHIC & $(\zeta/s)(T)+[\eta/s=0.06]$ & $\eta/s=0.06$ \\   
\hline
incl. off-shell vac. $\rho$ & 1.12 & 0.922 & 0.954 & incl. off-shell vac. $\rho$ & 1.18 & 0.859\\
\hline
incl. off-shell vac. $\rho$ &  &  &  & incl. off-shell vac. $\rho$ &  & \\
\& res. decay $\rho$ &  1.13 & 0.938 & 0.967 & \& res. decay $\rho$ & 1.23 & 0.920
\end{tabular}
\label{table:v2_diff_mass_w_CF_rho} 
\end{table}

Table \ref{table:v2_diff_mass_w_CF_rho} shows that the ratio $\frac{v_2(M=0.9\, \mathrm{GeV})}{v_2(M=0.3\, \mathrm{GeV})}$ continues to be sensitive to the effects of bulk viscosity even after resonance decay contributions are included. Thus, it is useful quantity to {\it highlight} the effects of bulk viscosity in our study. Additional studies will be necessary to clarify if this $v_2$ ratio is a robust observable to constrain the bulk viscosity of QCD. Nevertheless, the calculations presented herein show that the invariant mass distribution of $v_2$ exhibits a sensitivity to the presence of bulk viscosity. For the purposes of constraining bulk viscosity in data, a measurement of the entire invariant mass distribution of $v_2$ is needed. Moreover, combining hadron and dilepton anisotropic flow observables together, within the context of a Bayesian model to data comparisons, will put more robust constraints on bulk viscosity present inside hydrodynamical simulations. For this proposed study to yield the best possible outcome, a more precise measurement of dilepton $v_2$ is needed; such measurement are currently being planned \cite{Citron:2018lsq}.

\section{Conclusions}\label{sec:conclusion}
In the present study we explored the influence of bulk viscosity on dilepton production at RHIC and LHC collision energies. The total $v_2(M)$ in our calculations, composed of thermal and cocktail contributions, reacts similarly to bulk viscosity at RHIC and LHC collisions energies. Indeed, bulk viscosity affects most prominently the total dilepton invariant mass yield, through the increase in the spacetime volume occupied at lower temperatures --- thus, increasing the HM and cocktail dilepton yield, while leaving QGP dileptons yield essentially unaffected.\footnote{Note that this finding depends on the form of the bulk viscous modification $\delta n$ used in Eqs. (\ref{eq:bulk_deltan_1},\ref{eq:delta_n_bulk_hm}), and future studies will investigate how different parametrization of $\delta n$ affect dilepton production.} As the dilepton $v_2(M)$ is a yield-weighted average of the individual contributions, the effects of bulk viscosity also manifests itself in the total dilepton $v_2$, exhibiting similar features at RHIC and LHC collisions energies. Thermal dilepton $v_2$ is however different at RHIC and LHC collisions, owing to the different proportions of HM versus QGP dilepton yields present at those two collision energies.

Bulk viscous pressure has an interesting dynamical effect on the generation of the hydrodynamical momentum anisotropy ($\varepsilon_p$) as a function of temperature at both collision energies, and as a function of proper time $\tau$ at RHIC collision energy. Investigating the development of $\varepsilon_p$ as a function of temperature, starting from high temperatures and proceeding to lower temperatures, the medium with bulk and shear viscosities develops $\varepsilon_p$ faster than media without bulk viscosity, and reaches its maximum as the temperature approaches the peak in $\frac{\zeta}{s}(T)$. Once lower temperatures are reached, bulk viscosity reduces the amount of hydrodynamical momentum anisotropy. These features are mostly imprinted onto the HM dilepton radiation affecting RHIC and LHC dileptons differently, owing to the differently-sized spacetime volumes present at those two collision energies. For this reason, the enhancement in $\varepsilon_p$ has a modest increase on the $v_2(M)$ of HM dileptons at RHIC, while reducing the $v_2(M)$ of HM dileptons at the LHC. Given the non-linear nature of the hydrodynamical equations, the cause of these novel dynamics in $\varepsilon_p(T)$ necessitates a separate future investigation to inspect the role played by various transport coefficients governing bulk ($\Pi$) and shear ($\pi^{\mu\nu}$) sectors of viscous hydrodynamics. The role of these transport coefficients on the development of the expansion, shown in Fig. \ref{fig:later_latest_dynamics},  would also be interesting to determine. The more limited goal of the present investigation, however, was to perform a study that systematically investigates the effects of bulk viscosity on hadronic \cite{Ryu:2015vwa,Ryu:2017qzn} and electromagnetic \cite{Paquet:2015lta} probes using the same underlying hydrodynamical calculation, while also exploring the sensitivity of dileptons for highlighting new features in hydrodynamics in the presence of bulk viscous pressure. 

Since dileptons were sensitive to new features in hydrodynamics driven by bulk viscous pressure, the next goal was to investigate the consequences of these new features, and for this cocktail dileptons had to be included. Whether or not the effects of bulk viscosity on dileptons can be detected in experiment depends upon how well the cocktail dileptons can be calculated (and, therefore, potentially removed from) experimental measurements, to better expose thermal radiation. The contribution from semi-leptonic decays of open heavy flavor hadrons onto the dilepton spectrum needs to be removed as well. While the effects of open heavy flavor will be studied in an upcoming publication, that source can potentially be removed by using, for example, the Heavy Flavor Tracker installed in the STAR detector at RHIC. Hence, our investigation here concentrated more on dilepton production from thermal and cocktail sources, focusing on the invariant mass dependence of the dilepton yield and $v_2$, with a particular attention to the contribution of the $\rho$ meson inside the dilepton cocktail. Given its large width, it was found that the $\rho$ meson is sensitive to the invariant mass distribution assumed in the calculation of its cocktail contribution from the hypersurface of constant $T_{\rm sw}$. Therefore, if the cocktail $\rho$ is to be removed in experimental data in the process of isolating thermal dilepton radiation, its invariant mass distribution should be carefully taken into account. Combining all sources, our calculation has found that the ratio $\frac{v_2(M=0.9\, \mathrm{GeV})}{v_2(M=0.3\, \mathrm{GeV})}$ is useful to highlight the effects of bulk viscosity, while experimental measurement of $v_2(M)$ should be done at multiple invariant mass points to better constrain the effects of bulk viscosity. 

The upcoming dilepton calculation using SMASH will include various effects present in dynamical dilepton production from hadronic transport. However, before this investigation can begin, the equation of state must be modified to the updated lattice QCD equation of at high temperatures (e.g. \cite{Borsanyi:2013bia,Bazavov:2014pvz}) as well as include all the resonances present inside SMASH. The new components in the equation of state will certainly affect the speed of sound, by making it less sharply varying around the cross-over transition, thus affecting the evolution of the system. To match hadronic experimental observables, model parameters (e.g. $T_{\rm sw}$, $\eta/s$, $\zeta/s$, and so on) must be re-adjusted. Once this match is obtained, a study including SMASH allows for several effects to be investigated. First, it opens the possibility to assess through dileptons the effects of collisional broadening on the in-medium properties of the parent hadrons generating the lepton pairs. Collisional broadening effects on in-medium vector mesons are already included in our dilepton production from the hydrodynamical medium, and it would be intriguing to investigate how important those are inside of hadronic transport. Furthermore, different from collisional broadening, SMASH's long-range fields can leave interesting features on the in-medium properties of the parent hadrons, distinguishing their effects --- from collisional broadening via dileptons --- is an interesting avenue to explore. If found to be significant, those modifications would open a window to study the in-medium properties of hadrons inside a hadronic transport evolution. Moreover, calculating dilepton production dynamically through SMASH hadronic transport will generate additional anisotropic flow to that from the hydrodynamical simulation. Relying on the the free-streaming assumption, our current calculation of the dilpeton cocktail does not have this additional anisotropic flow.  

This study complements earlier investigations \cite{Vujanovic:2016anq,Vujanovic:2017psb} on the sensitivity of dileptons to various transport coefficients of hydrodynamical simulations. Together, they show the value of dileptons as probes of the strongly-interacting medium created in heavy-ion collisions. The simultaneous use of dileptons and hadronic observables will yield much better constraints on the properties of strongly interacting media than any of those observables alone. Indeed, degenerate parameter combinations in theoretical calculations, leading to the same hadronic anisotropic flow for example, often result in different dilpeton anisotropic flow \cite{Vujanovic:2016anq,Vujanovic:2017psb}. For this capacity of dileptons to be fully exploited, however, experimental measurements of dilepton elliptic flow (or even higher harmonics) are crucial.

\section*{Acknowledgments}
This work was supported in part by the Natural Sciences and Engineering Research Council of Canada, in part by the Director, Office of Energy Research, Office of High Energy and Nuclear Physics, Division of Nuclear Physics, of the U.S. Department of Energy under Contracts No. DE-AC02-98CH10886, DE-AC02-05CH11231, DE-SC0004286, DE-SC0013460, and DE-FG02-05ER41367, and in part by the National Science Foundation (in the framework of the JETSCAPE Collaboration) through award number ACI-1550233 and ACI-1550300. G. Vujanovic acknowledges support by the Natural Sciences and Engineering Research Council (NSERC) of Canada and the Fonds de Recherche du Qu\'ebec --- Nature et Technologies (FRQNT). Computations were performed on the Guillimin supercomputer at McGill University under the auspices of Calcul Qu\'ebec and Compute Canada. The operation of Guillimin is funded by the Canada Foundation for Innovation (CFI), the Natural Sciences and Engineering Research Council (NSERC) of Canada, NanoQu\'ebec, and the Fonds de Recherche du Qu\'ebec--- Nature et  Technologies (FRQNT). G.~S.~Denicol thanks Conselho Nacional de Desenvolvimento Cient\'ifico e Tecnol\'ogico (CNPq) for financial support.

\appendix
\section{Early time dynamics}\label{appdx:early-time}

\begin{figure}[!h]
\begin{center}
\begin{tabular}{cc}
\includegraphics[width=0.495\textwidth]{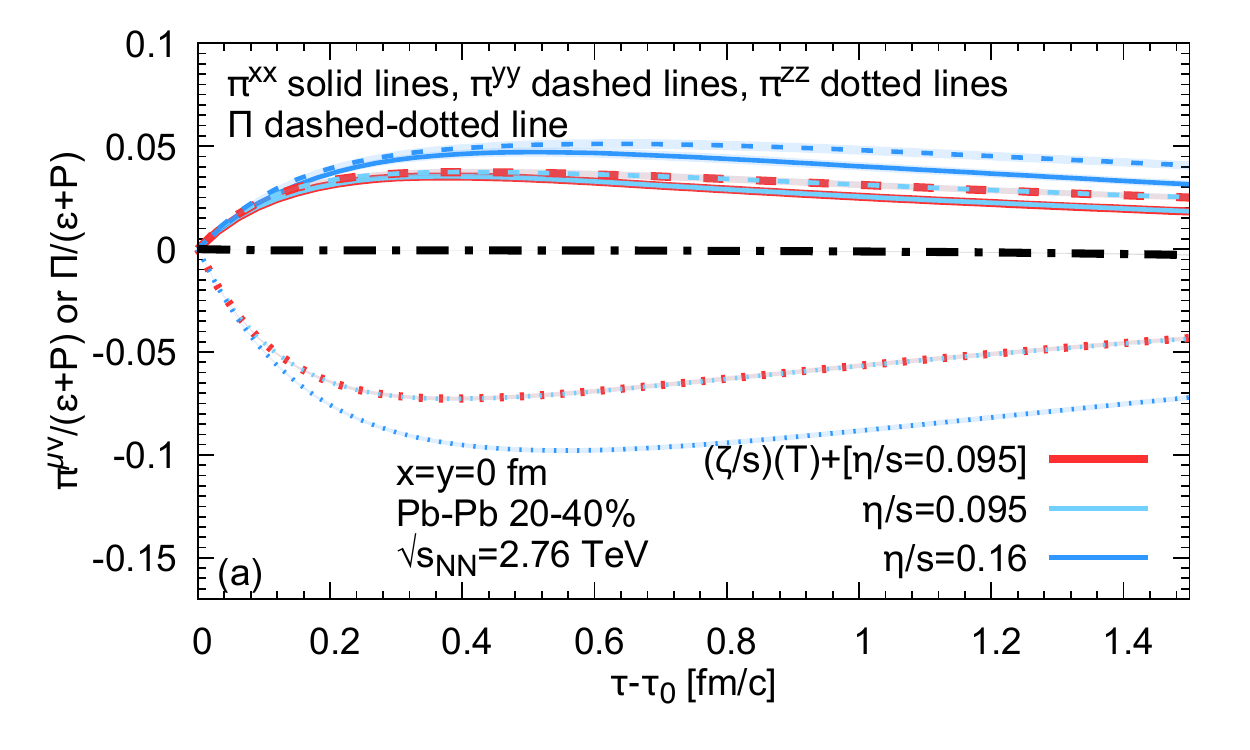} & \includegraphics[width=0.495\textwidth]{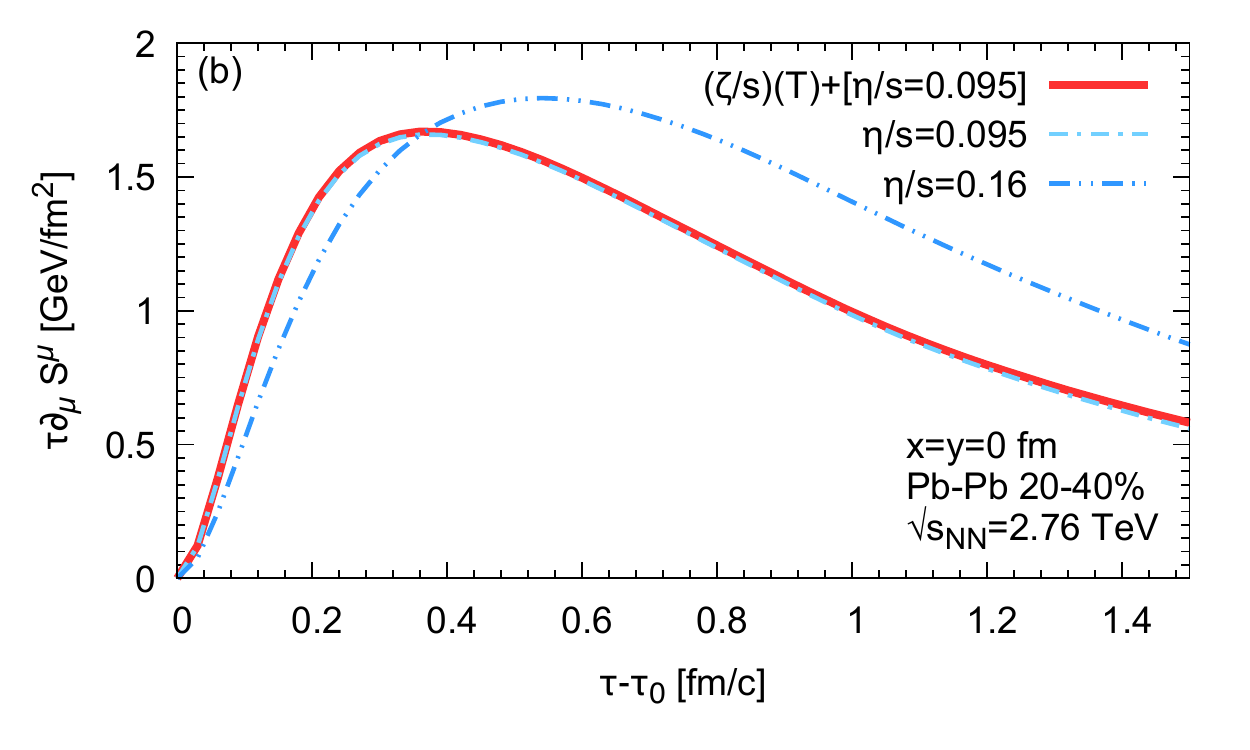}\\
\includegraphics[width=0.495\textwidth]{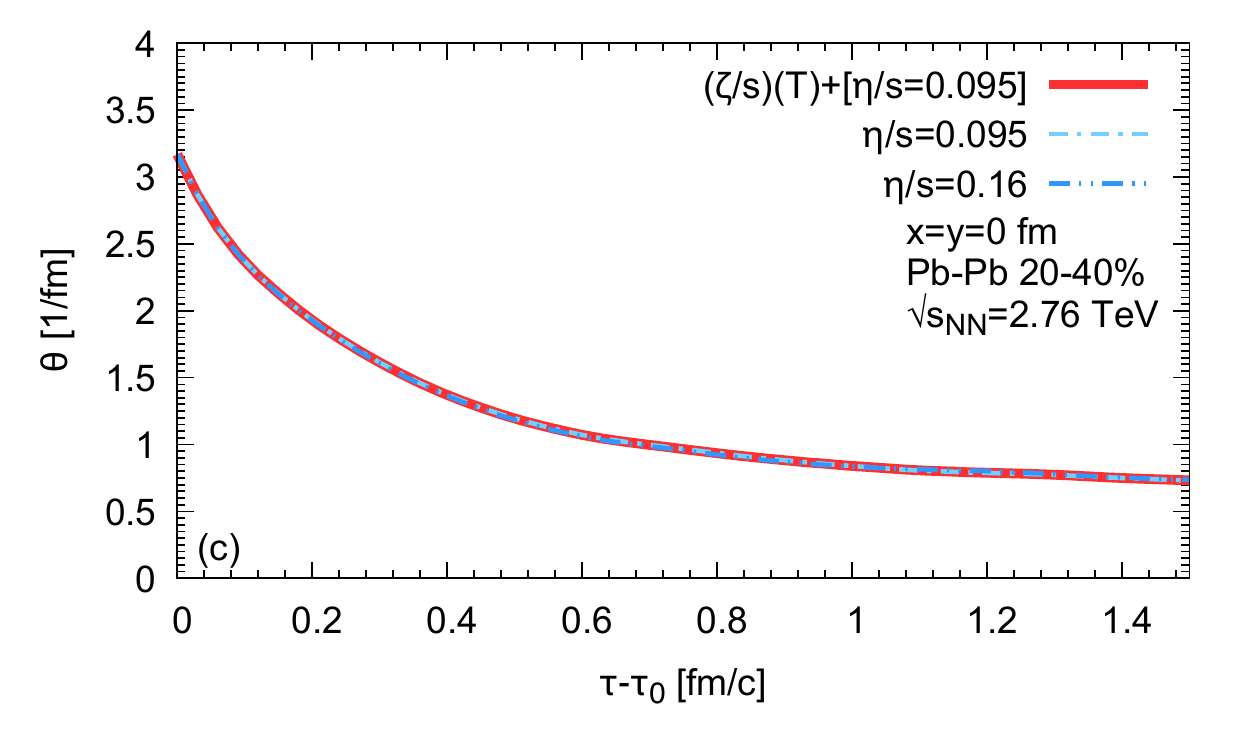}   & \includegraphics[width=0.495\textwidth]{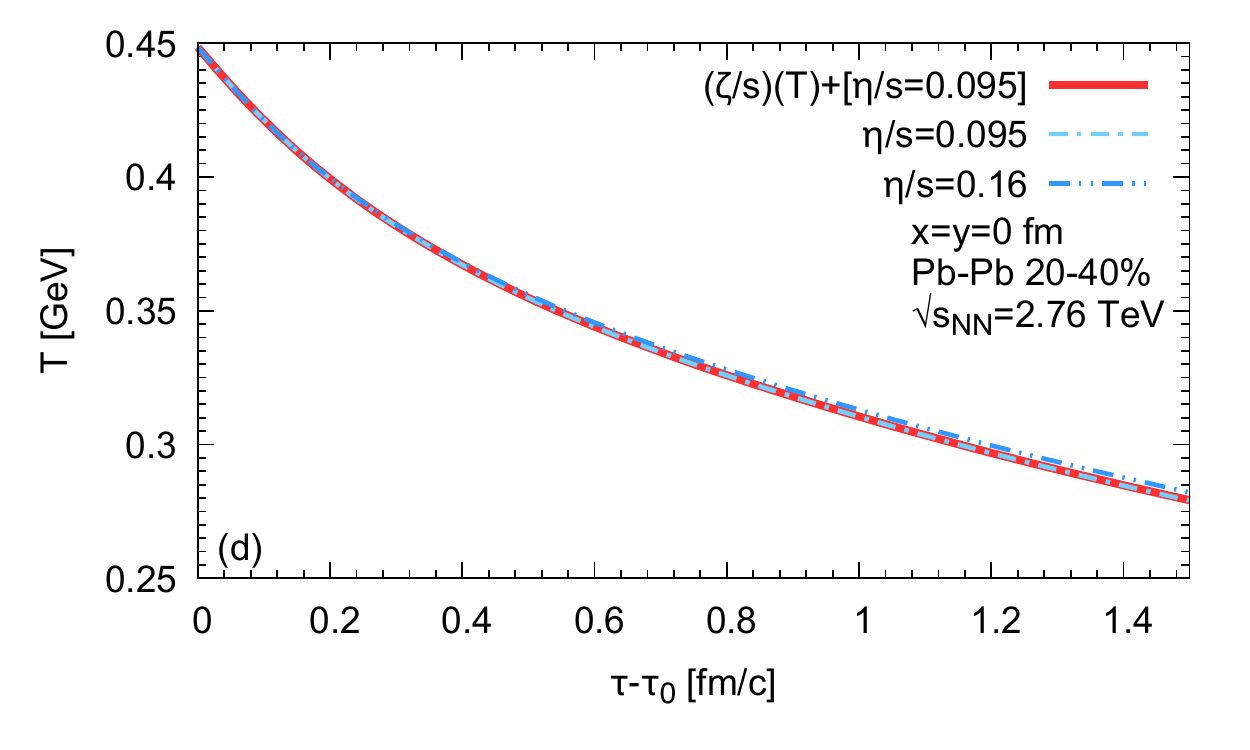}
\end{tabular}
\end{center}
\caption{(Color online) (a) Event-averaged enthalpy density normalized shear and bulk viscous pressure during the first few fm/$c$ of evolution. $\pi^{\mu\nu}/(\varepsilon+P)$ is evaluated in the local rest frame. (b) Event-averaged entropy production rate $\partial_\mu S^\mu$ rescaled by $\tau$ during the first few fm/$c$ of evolution. Note that the initial entropy density of this cell is 44.24 GeV/fm$^2$. (c) Event-averaged expansion rate $\theta$ during the first few fm/$c$ of evolution. (d) Event-averaged temperature for the central cell during the first $\tau-\tau_0\leq 1.5$ fm/$c$ of evolution.}
\label{fig:early_dynamics}
\end{figure}

To complete the results presented in Fig. \ref{fig:later_latest_dynamics}, the evolution of the medium at early times is displayed in Fig. \ref{fig:early_dynamics}. Given that the temperature dependence of $\zeta/s$ peaks at temperatures around 0.18 GeV, the early-time entropy production presented in Fig. \ref{fig:early_dynamics} is entirely dominated by the medium with largest shear viscosity --- i.e. $\eta/s=0.16$ --- which generates largest $\pi^{\mu\nu}$ and thus $\partial_\mu S^\mu$. The substantial Bjorken-like expansion rate $\left(\theta\propto \tau^{-1}\right)$ at early times drives the temperature evolution of the medium, dwarfing any entropy production, and resulting in a substantial temperature reduction by almost 0.17 GeV in $\tau-\tau_0\leq 1.5$ fm/$c$ regardless of what entropy is produced.   

\section{Exploring the $p_T$-dependence of viscous corrections to the dilepton rate}\label{appdx:dR_pT}

\begin{figure}[!h]
\begin{center}
\includegraphics[width=0.495\textwidth]{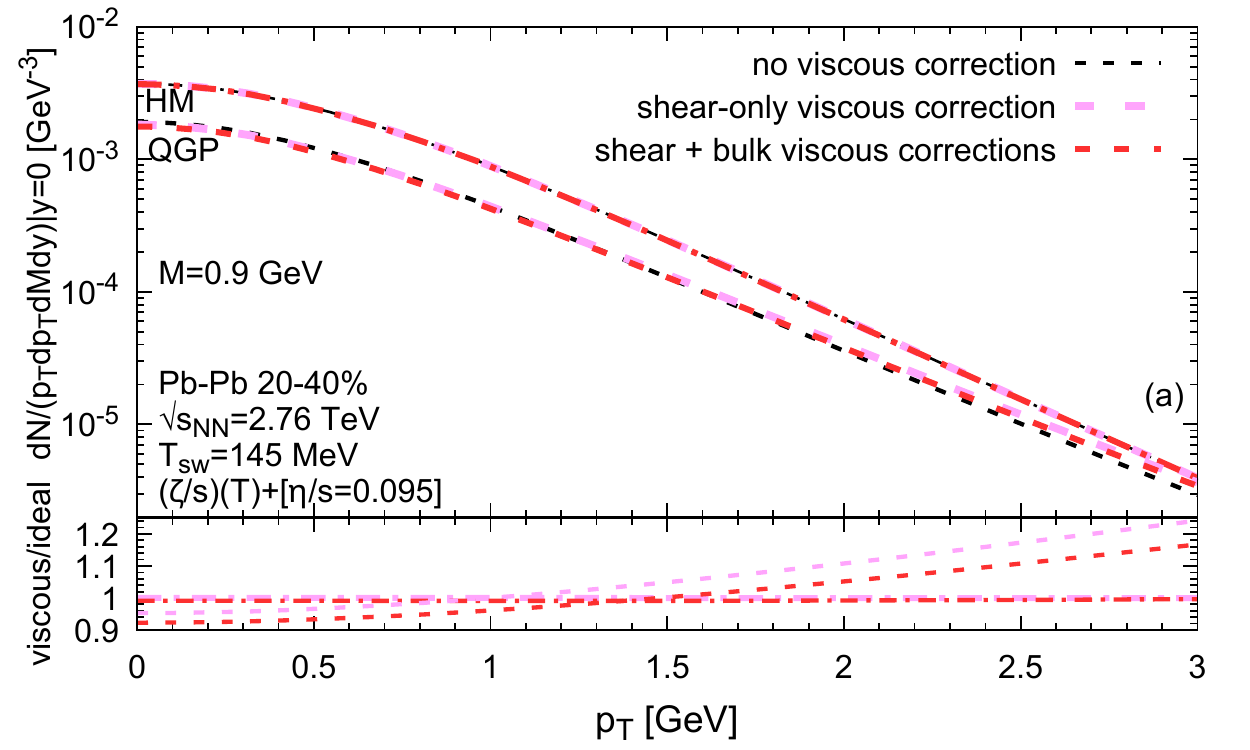}
\includegraphics[width=0.495\textwidth]{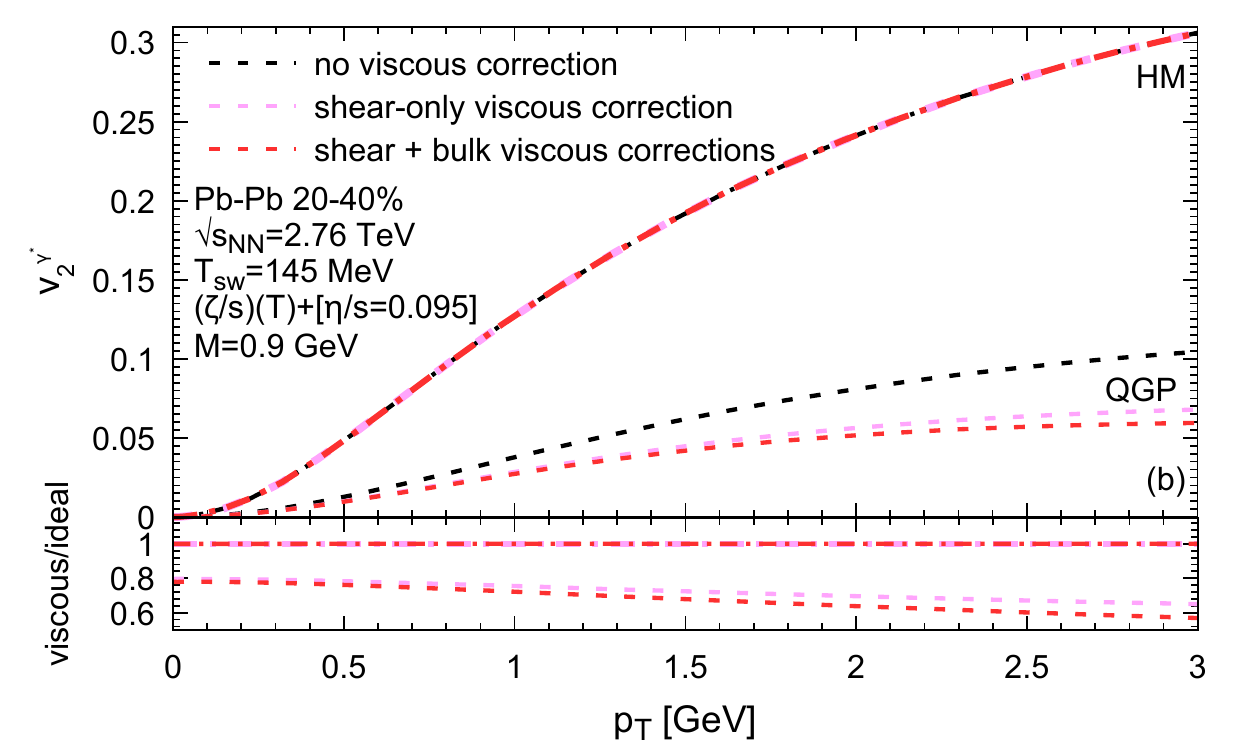}
\includegraphics[width=0.495\textwidth]{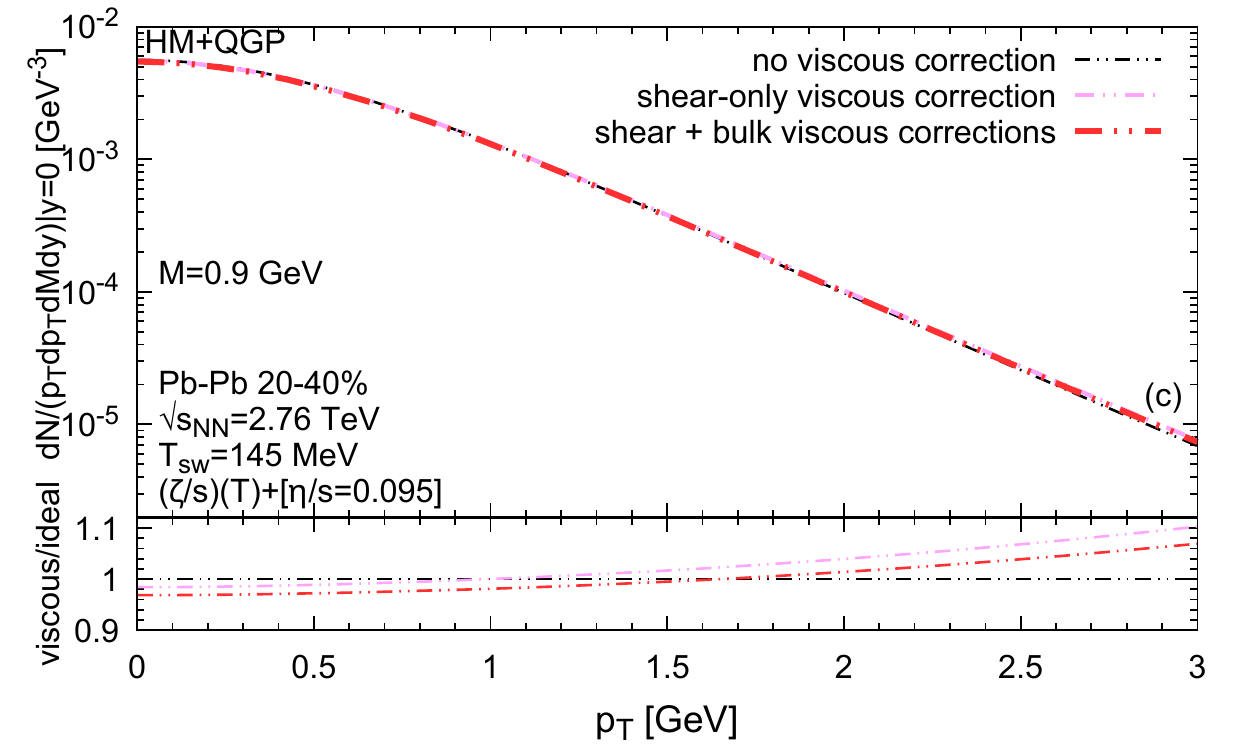}
\includegraphics[width=0.495\textwidth]{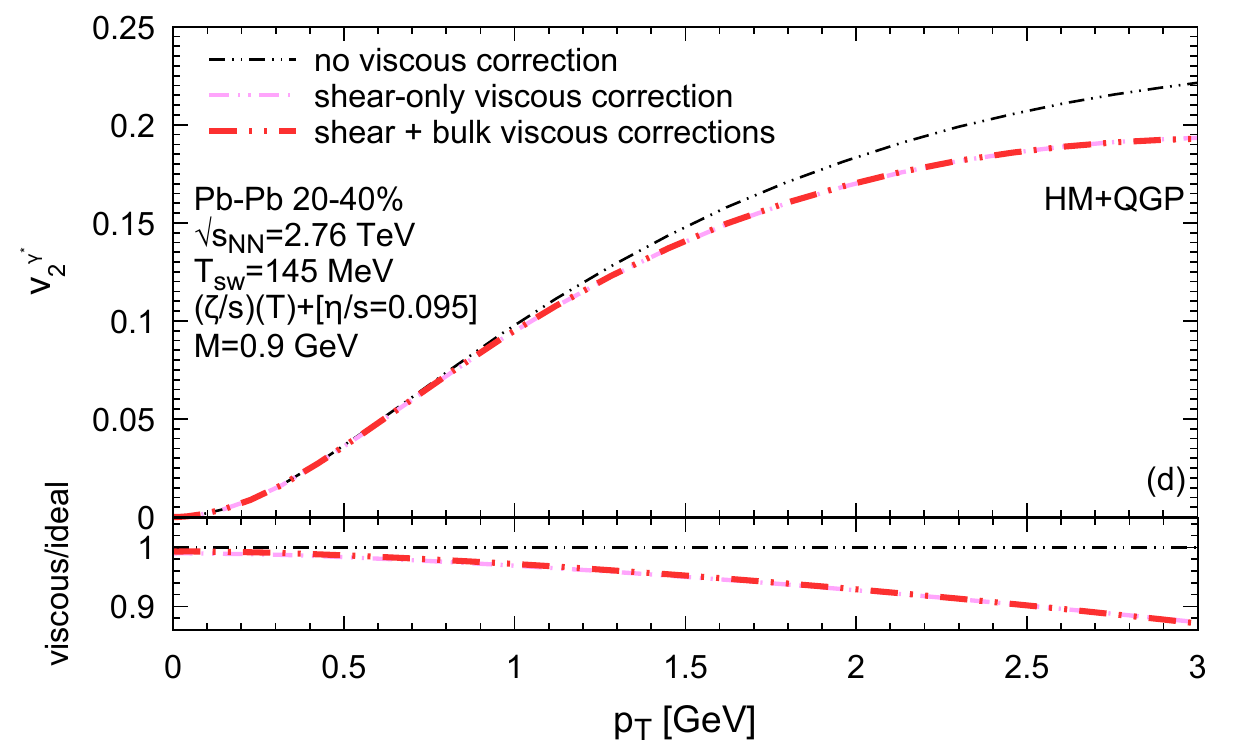}
\end{center}
\caption{(Color online) (a) Top panel: Effects of viscous corrections on the $p_T$-differential dilepton yield of HM (dash-dotted) and QGP (dashed) dileptons. Bottom panel: Ratios of the viscous dilepton yields over their respective inviscid (ideal) yields. (b) Top panel: Effects of viscous corrections on the $p_T$-differential $v_2$ for HM and QGP dileptons. Bottom panel: Ratio of the viscous over the ideal dilepton $v_2$, including both shear and bulk viscous corrections. (c) Top panel: Effects of viscous corrections on the $p_T$-differential thermal (HM+QGP) dilepton yield. Bottom panel: Ratios of the viscous dilepton yields over their respective inviscid (ideal) yields. (d) Top panel: Effects of viscous corrections on the $p_T$-differential $v_2$ for thermal (HM+QGP) dileptons. Bottom panel: Ratio of the viscous over the ideal dilepton $v_2$, including both shear and bulk viscous corrections. HM and QGP dileptons are produced in different temperature windows presented in Eq.~(\ref{eq:f_QGP}).}
\label{fig:dR_effects_on_M_0p7}
\end{figure}

In Fig.~\ref{fig:dR_effects_on_M_0p7}, we investigate the effects of bulk and shear viscous corrections on the dilepton emissions rates by looking at the $p_T$-differential yield and $v_2$ at low invariant mass. 

At $M=0.9$ GeV, the yield is dominated by radiation from the lower temperature medium --- i.e. hadronic medium (HM) dileptons. In Fig. \ref{fig:dR_effects_on_M_0p7}a, the yield of HM dileptons is essentially unaffected by our viscous corrections to the dilepton emission rate. The higher temperature partonic (QGP) dilepton yield is affected more significantly by viscous corrections to the rate, especially for $p_T\gtrsim 2$ GeV. To better appreciate the effects of viscous corrections, the bottom panel of Fig.~\ref{fig:dR_effects_on_M_0p7}a displays the ratio of viscous over inviscid (ideal) dilepton production. The thermal (HM+QGP) dilepton yield in Fig. \ref{fig:dR_effects_on_M_0p7}c is affected by viscous $\delta R$ corrections on the order of $\sim 10$\% for $p_T\gtrsim2.5$ GeV, as this is where QGP radiation becomes comparable with HM for $M=0.9$ GeV.  

As far as the $v_2$, our viscous correction to the lower temperature (HM) contribution leaves the $v_2$ unaffected. However, the $v_2$ of the higher temperature (QGP) region is affected by our viscous correction throughout the entire $p_T$-distribution, as seen in Fig.~\ref{fig:dR_effects_on_M_0p7}b and highlighted in the bottom panel. For thermal (HM+QGP) dileptons in Fig. \ref{fig:dR_effects_on_M_0p7}d, viscous emission rate corrections affect the entire $v_2(p_T)$, with dilepton $v_2$ being slightly more sensitive to the rate modifications than the yield. Since $v_2$ of thermal dileptons is a yield-weighted average of the low and high temperature (HM and QGP) sources, the high temperature (QGP) contribution becomes significant on the total $v_2$ at $p_T\gtrsim 1.5$ GeV: this is where viscous correction effects on $v_2$ can be more readily appreciated.  
\begin{figure}[!h]
\begin{center}
\includegraphics[width=0.495\textwidth]{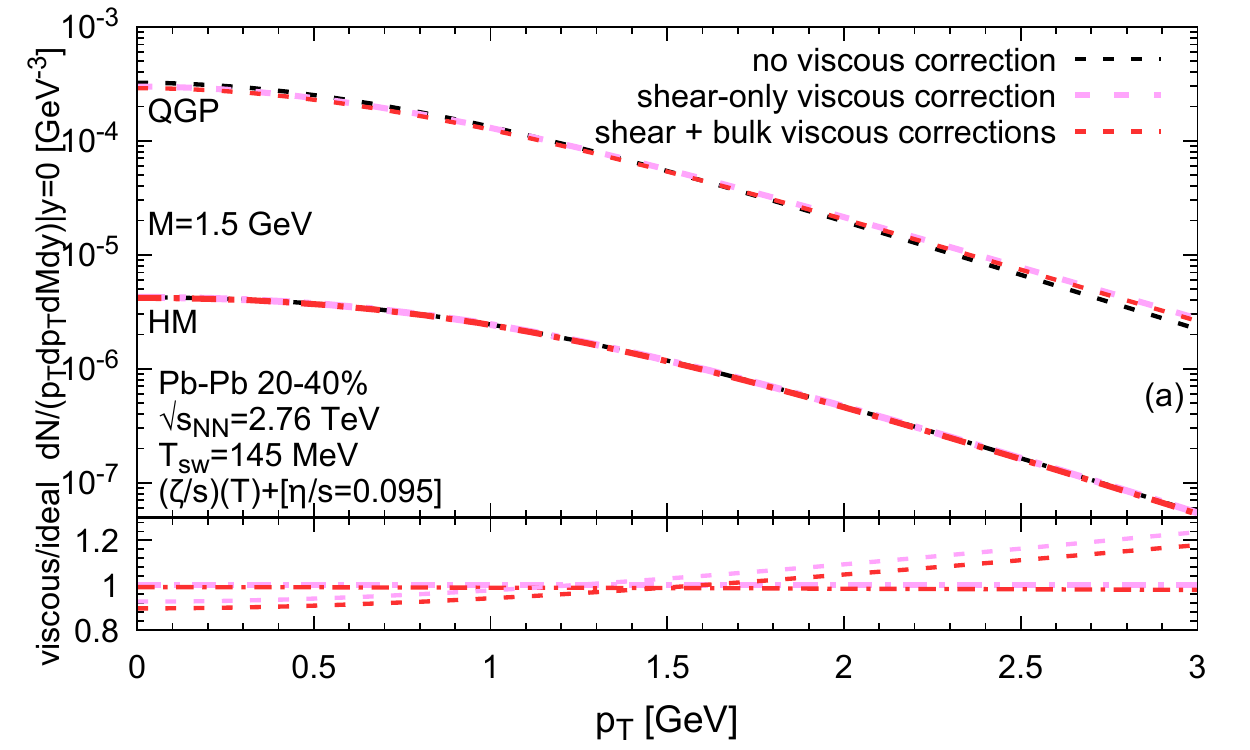}
\includegraphics[width=0.495\textwidth]{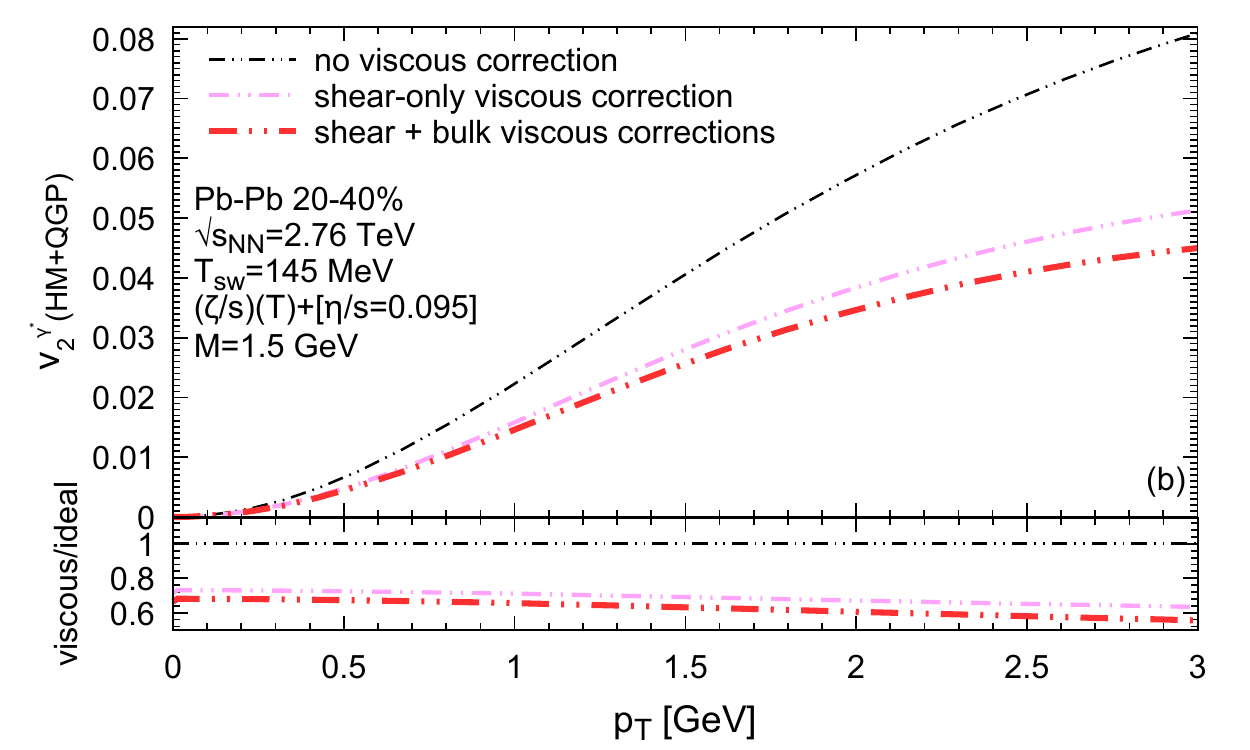}
\end{center}
\caption{(Color online) Similar to Fig. \ref{fig:dR_effects_on_M_0p7}, but for larger dilepton invariant mass $M=1.5$ GeV.}
\label{fig:dR_effects_on_M_1p5}
\end{figure}

At a higher invariant mass of $M=1.5$ GeV, the higher temperature (QGP) dilepton yield dominates any lower temperature (HM) contribution at all $p_T$, as shown in Fig.~\ref{fig:dR_effects_on_M_1p5}a. We only show the thermal $v_2$ in Fig.~\ref{fig:dR_effects_on_M_1p5}b as it closely follows radiations from higher (QGP) temperatures. At this invariant mass, the effects of viscous corrections can be seen, reducing the $v_2(p_T)$ of dileptons. Though measurements of the transverse momentum distribution of dilepton yield and $v_2$ at a fixed invariant mass are sensitive to the viscous corrections of the dilepton rates and the underlying hydrodynamical evolution, those measurements are also more difficult than the $p_T$-integrated invariant mass measurements, requiring higher statistics.    

\bibliography{references}
\end{document}